\documentclass[11pt]{article}
\usepackage{amssymb,epsf,graphics,graphicx}
\newcommand{\wtilde}{\widetilde}
\newcommand{\trisymb}{\wtilde}

\newcommand{\usa}[1]{ \left \{ #1 \right \} }
\newcommand{\ddt}{{d\over d\theta}}
\newcommand{\ubl}[1]{\{#1\}}
\newcommand{\phup}{^{\phantom{j_j}}}

\newcommand\toline[1]{--#1}
\newcommand{\fract}[2]{{\textstyle\frac{#1}{#2}}}
\newcommand{\fr}[2]{{\textstyle\frac{#1}{#2}}}
\newcommand{\ri}{\right}
\newcommand{\ep}{\varepsilon}
\newcommand{\lf}{\left}
\newcommand{\te}{\theta}

\newcommand\Zth{{\mathbb Z}_3}

\newcommand\ZN{{\mathbb Z}_N}
\newcommand\NN{{\mathbb N}}
\newcommand\Z{{\mathbb Z}}

\newcommand{\CM}{{\cal M}}
\newcommand{\tpsi}{ \psi}
\newcommand{\tphi}{\phi}

\newcommand{\B}{{\cal B}}

\newcommand{\CY}{ {\cal Y}}

\newcommand{\CL}{{\cal L}}

\newcommand\eq{\begin{equation}}
\newcommand\en{\end{equation}}
\newcommand\bea{\begin{eqnarray}}
\newcommand\eea{\end{eqnarray}}
\newcommand\nn{\nonumber}
\newcommand\ba{\(\begin{array}}
\newcommand\ea{\end{array}\)}

\newcommand{\resection}[1]{\setcounter{equation}{0}\section{#1}}

\newcommand{\fra}[2]{\ensuremath{\mbox{\footnotesize $\frac{#1}{#2}$}}}
\begin{document}
\begin{titlepage}
\vskip 0.5cm
\begin{flushright}
DCPT-02/53 \\
{\tt hep-th/0208202} \\
\end{flushright}
\vskip 1.5cm
\begin{center}
{\Large{\bf
Integrable aspects
of the scaling $q$-state Potts models II:\\[5pt]
finite-size effects
}}
\end{center}
\vskip 0.8cm
\centerline{Patrick Dorey%
\footnote{e-mail: {\tt p.e.dorey@durham.ac.uk}},
Andrew Pocklington$^2$%
\phantom{\footnote{e-mail: {\tt andrew@ift.unesp.br}}}
and Roberto Tateo%
\footnote{e-mail: {\tt roberto.tateo@durham.ac.uk}}}
\vskip 0.9cm
\centerline{${}^{1,3}$\sl\small Dept.~of Mathematical Sciences,
University of Durham, Durham DH1 3LE, UK\,}
\vskip 0.2cm
\centerline{${}^2$\sl\small IFT/UNESP, Instituto de Fisica Teorica, 
01405-900, Sao Paulo - SP, Brasil\,}
\vskip 1.25cm
\begin{abstract}
\noindent
We continue our discussion of
the $q$-state Potts models for $q\le 4$, in the
scaling regimes close to their critical and tricritical points.
In a previous paper, the spectrum and full S-matrix of the models 
on an infinite line were elucidated; here, we consider 
finite-size behaviour.  TBA equations are proposed for all cases 
related to $\phi_{21}$ and $\phi_{12}$
perturbations of unitary minimal models.
These are subjected to a variety of checks
in the ultraviolet and infrared limits,
and compared with results from a recently-proposed
nonlinear integral equation. A nonlinear integral equation is also
used to study the flows from tricritical to critical models, over the
full range of $q$.
Our results should also be of relevance to the study of the
off-critical dilute A models in regimes 1 and 2.
\end{abstract}
\end{titlepage}
\setcounter{footnote}{0}
\def\thefootnote{\fnsymbol{footnote}}
%
\resection{Introduction}
\label{intr}
As a second-order phase transition of a lattice model is approached,
its correlation length 
diverges. In the so-called `scaling region' near to the
transition, a continuum limit can be taken, and the model can then be
investigated using the techniques of continuum field theory. The $q$-state
Potts models \cite{potts,FK} illustrate this notion very nicely.
For $q\le 4$ the addition of vacancies allows them to be defined so as to
have both  critical and  tricritical points \cite{Nienh},
and thus to have
two distinct scaling regions, each with its own associated continuum 
field theory\footnote{Strictly speaking the {\em direction}
in which the critical
point is approached must also be specified - here, apart from in section
\ref{mless}, we only consider the 
(first) thermal direction, which is related to changes of temperature.}.
These are known as the scaling Potts models, with
the words critical or tricritical being added if there is a need to
be more precise.

This paper 
is the continuation of
a  companion paper~\cite{pottsI}, in which the exact
S-matrices of the scaling Potts models
were discussed. Our starting-point
there was some work by 
Chim and Zamolodchikov~\cite{CZ}, who, noting that the models
should be integrable, proposed a set of S-matrix elements
describing the scattering of elementary kink-like excitations.
Using the bootstrap technique and a `minimal'
hypothesis governed by  the presence 
or absence
of Coleman-Thun~\cite{CT}  type  explanations of S-matrix poles,  we 
were able to
close the  bootstrap  for all of the critical scaling Potts models,
and for the tricritical scaling Potts models with $4>q\ge 2$.

The scaling Potts models can be related  to $\phi_{21}$ and $\phi_{12}$ 
perturbations of $c<1$ conformal field theories, and in this context an 
alternative set of elementary S-matrix elements had previously been
proposed
by Smirnov \cite{Sm}, using a construction based on reductions of the
Izergin-Korepin $a_2^{(2)}$ S-matrix \cite{IK}. 
(Interesting features  of these scattering theories
have recently been discussed  in~\cite{KTW,TW}.) 
While the relationship between the two 
approaches is now being clarified \cite{paulandnick}, in \cite{pottsI}
we found it more 
convenient to work entirely within Chim and Zamolodchikov's framework -- 
the fundamental S-matrix elements and the formal vacuum structure  are then continuous functions of $q$, and
the connection with the Fortuin--Kasteleyn~\cite{FK} formulation of the
lattice model and its symmetries
is rather more direct.
Smirnov's approach is perhaps more natural if one wishes to discuss 
perturbations of specific minimal models, but, as we shall review below, the
identification of the scaling Potts models with such perturbations hides a
number of subtleties. Nevertheless, it does 
underline that a study via finite-size effects might be worthwhile, and
this topic forms the  main theme of the
present paper.
Taking as partial input the mass
spectrum and S-matrix elements found in \cite{pottsI}, we propose sets of
thermodynamic Bethe ansatz
(TBA) equations describing the ground-state energies of the
critical and tricritical scaling Potts models,
for the values of $q$ for which the
associated minimal model is unitary. These proposals are checked in
a variety of ways, and we also study aspects of
the finite-size behaviour of the
models using the so-called non-linear integral equation technique.

The plan of the paper is as follows. Section \ref{cftrev} discusses
the conformal field theory descriptions of the critical
and tricritical points. Section \ref{recap}
summarises the necessary S-matrix results obtained 
in~\cite{CZ} and~\cite{pottsI}. 
Section \ref{revsect}
contains a  short review  of previous work on
the TBA for $\phi_{21}$  and $\phi_{12}$ perturbations, 
and discusses some features of TBA systems in general. 
We also sketch some of the reasoning
which led to our main conjectures. 
The conjectures themselves are outlined in
sections~\ref{con21} and \ref{con12}, in the form of
sets of rules for the construction of
the eight new  families of massive TBA equations for $\phi_{21}$ and
$\phi_{12}$ perturbations of the minimal models $\CM_{p,p+1}$, related to 
the critical and tricritical
scaling  Potts models at the particular values $\sqrt{q}=2
\cos(\fract{\pi}{p+1})$ ($p=6,7,\dots$) and 
$\sqrt{q}=2
\cos(\fract{\pi}{p})$
($p=5$, $p=7,8,\dots$) respectively.
The first four sets of equations for the perturbed critical models
are given explicitly in section~\ref{con21}, while
the story for the tricritical models is illustrated in section~\ref{con12} 
by the  set of equations  for the
$\CM_{5,6}+ \phi_{12}$ theory. This is  related to the
tricritical branch of the Potts model at $\sqrt{q}=2 
\cos(\fract{\pi}{5})$.

In section~\ref{checks} our TBA systems
are subjected to a number of analytical and
numerical tests, all of which they pass. 
Further verification is provided in
section~\ref{NLIEQ}, where a variant of the non-linear
integral equation of \cite{DT} is proposed to describe the
finite-size ground-state energy at arbitrary values of $q$ and
its solutions are compared numerically with those of the TBA
equations. This section also shows how the nonlinear integral
equation technique can be used to study the interpolating flows
between the tricritical and critical models, by taking an
equation first introduced in \cite{Zam3} and tuning its parameters
to suitable values.

The full set of TBAs  related to  both $\phi_{21}$ and
$\phi_{12}$ perturbations of minimal unitary models,
and the associated sets of functional relations,
are given in four appendices.

We end this introduction with a remark on the possible wider relevance of
our results.
In the following, we have concentrated 
on the integrable quantum field theories 
associated to the continuum limits of
lattice Potts models near to their critical and tricritical points.
A renormalised field theory contains information which is universal 
in nature, and its relevance is not restricted to any specific member
of a universality class.
Other  lattice systems  associated with
the ``Potts''  universality classes
are the     
dilute $A$ models
of~\cite{WNS,Roche}. 
While these lattice models are only defined 
at discrete values of $q$, they have the advantage of being
soluble not only at but also away 
from the critical point, even on the lattice. The link with Potts models comes
via the identification
of their scaling limits with  the $\phi_{12}$ and $\phi_{21}$ 
perturbations of unitary minimal conformal models.
The exact correspondence is \cite{WPSN}:
\bea
\hbox{Dilute}~A_{p-1}\,, ~
\hbox{regime}~  1^{\pm } 
&\leftrightarrow& \CM_{p,p+1}+\phi_{21}~  \leftrightarrow ~
\sqrt{q}=2\cos(\fract{\pi}{p{+}1})\,,~\hbox{critical branch;}
\nn
\\
\hbox{Dilute}~A_p\,, ~~
\hbox{regime}~  2^{\pm } ~
&\leftrightarrow& \CM_{p,p+1}+\phi_{12}~ \leftrightarrow ~
\sqrt{q}=2\cos(\fract{\pi}{p})\,,~\hbox{tricritical branch.}
 \nn
\eea   
These are precisely the points at which we have been able to
conjecture TBA descriptions of the continuum models. Past experience
(see \cite{SuzE8,SuzEn} for examples directly relevant to the matter
in hand) suggests a close link between the underlying mathematical  
structures of the lattice and continuum models, when both are
integrable.
We therefore expect that many of the  results reported in this paper,
such as
the general forms of the  TBAs and the Y-systems,
will also   play a r{\^o}le  in the study of the off-critical dilute
$A$ models, at least in regimes 1 and 2.

\resection{The conformal field theory description of the critical
and tricritical points}
\label{cftrev}
If a model is placed precisely at a second-order transition, its correlation
length is infinite and it has no intrinsic length-scale. 
Its behaviour should therefore be
described by a conformal field theory \cite{BPZ}.
Some key features of
the conformal field theories relevant to the critical $q$-state Potts models
with $q\le 4$
were identified by Dotsenko~\cite{Dot} (see also \cite{DF}).
His work made use of previous 
predictions for certain critical exponents of the $q$-state Potts
models~\cite{dN,Na,N}, 
in particular
the following formula, first proposed by den Nijs \cite{dN}\,:
\eq
y\phup_T(q)=
\frac{6\lambda}{3+2\lambda}
\qquad\mbox{where}~~ 
\sqrt{q}= 2\sin\left(\frac{\pi}{3}\lambda\right)~~\mbox{and}~~ 
0\le \lambda\le 3/2~.
\label{dNf}
\en
Here, $y\phup_T$ is the renormalisation group 
eigenvalue for the energy operator $\epsilon$\,,
given in 
terms of the scaling dimension $x_{\epsilon}$ of $\epsilon$ by
$y\phup_T=
2-x_{\epsilon}$\,. It is related to the
specific heat exponent $\alpha$ as $\alpha=2-2/y_T$. (In fact, den
Nijs gave his formula in terms of $\mu=\pi/2-\pi\lambda/3$, with
$\sqrt{q}=2\cos\mu$, but the parametrisation in terms of $\lambda$
will be more convenient below.)

At the same time, conformal field 
theories with central charge $c<1$ can be parametrised, at least partially,
by a real number
$\xi>0$  such that
\eq
c(\xi)=1-\frac{6}{\xi(\xi{+}1)}~.
\label{cdef}
\en
For {\em any} (not necessarily rational) value of $\xi$, these
theories admit so-called `degenerate' primary fields,
which are special in that they have null fields in their sets of
descendants, causing their correlation functions to satisfy differential
equations~\cite{BPZ}. The possible
(left) conformal dimensions these fields
are\footnote{In~\cite{Dot,DF}, 
the conformal dimensions
are defined such that $\Delta_{r,s}$ becomes $\Delta_{s,r}$}
\eq
\Delta_{r,s}(\xi) = \frac{((\xi{+}1)r-\xi s)^2-1}%
{4\xi(\xi{+}1)}~
\qquad (r,s\in\NN)\,,
\label{ddef}
\en
with a similar formula 
for the right conformal dimensions $\bar\Delta_{r',s'}$\,.
The {\it spinless} degenerate 
primaries have $\Delta=\bar\Delta$ so that their scaling
dimensions are $x_{r,s}=2\Delta_{r,s}$\,; these fields will be denoted
$\phi_{rs}$\,.

In \cite{BPZ} and \cite{Dot}, the $2$- and $3$- state Potts models
were identified with $c<1$ conformal field theories with $\xi= 3$ and
$5$ respectively. In both cases, the energy operator
$\epsilon$ was shown to correspond to 
the primary field 
$\phi_{21}$\,.
In \cite{Dot}, Dotsenko
conjectured that the same should hold
for all $q\leq4$, implying the general relation
\eq
y\phup_T(\xi)=2-2\Delta_{2,1}(\xi)
= 
\frac{3(\xi{-}1)}{2\xi}
\en
and, comparing with (\ref{dNf}), 
\eq
\xi=\frac{3+2\lambda}{3-2\lambda}\qquad,\qquad
\lambda=
\frac{3}{2}\frac{(\xi-1)}{(\xi+1)}~.
\label{eq:qtopp'}
\en
An immediate check on this hypothesis comes via
the operator algebra $\phi_{21}\phi_{21}\sim I+\phi_{31}$, which
predicts that the critical exponent $y^{\phantom{l}}_{T_2}$ 
(the second thermal
exponent) should be given in terms of the scaling dimension
$x_{3,1}$ as
\eq
y\phup_{T_2}=2-x_{3,1}=-\frac{4}{\xi}
=-\frac{4(3-2\lambda)}{(3+2\lambda)}~.
\en
This matches the
value calculated by Nienhuis \cite{Na} using a mapping
onto a Coulomb gas.
Note, since $0\le\lambda\le 3/2$ for the Potts models, $\xi$ lies in
the range $[1,\infty]$ and $-2\le c(\xi)\le 1$.

So far so good; but one should beware that the value of $c$ does not
specify a conformal field theory uniquely. Consider the situation 
when $\xi$ is rational, and 
suppose that
\eq
\xi=\frac{p}{p'-p}
\en
with $p$ and $p'$ coprime integers with $p'>p$. Then 
(\ref{cdef}) and (\ref{ddef}) turn into the familiar formulae
\eq
c=1-\frac{6(p'-p)^2}{pp'}~~,\quad
\Delta_{r,s}=\frac{(p'r-ps)^2-(p'-p)^2}{4pp'}
\en
and the degenerate fields
can be identified in pairs as
\eq
\phi_{rs}=\phi_{p{-}r,\,p'{-}s}~.
\en
Operator product expansions between fields in the subset
\eq
\{\phi_{rs}\,|~r=1\dots p{-}1,\,s=1\dots p'{-}1\}~,
\en
and their descendants, then close amongst themselves. The conformal field
theory containing just these fields is called the diagonal, or `A'
series, minimal model
$\CM_{pp'}$\,.
Depending on the values of $p$ and $p'$, 
there may be other consistent truncations to other field theories
also containing finite numbers of degenerate primary fields. 
All are examples of
{\em rational} conformal field theories, and the full set
of possibilities for $c<1$ forms the famous
ADE classification of \cite{CIZ}.
However, apart from the special values 
$\xi=3$ and $5$ (corresponding to $q=2$ and $3$)
{\em none} of these is the conformal field theory of a 
$q$-state Potts model. This follows from the fact that the 
relevant torus partition functions do not coincide \cite{dFSZ}.

At first sight, this might seem to contradict
the claim of~\cite{CIZ} to have found a classification
of modular invariant partition functions with $c<1$. 
The Potts model partition functions
are certainly modular invariant, so how
can they escape this result?
The explanation is that
the proof in \cite{CIZ} made essential use of the requirement,
first emphasised in \cite{cardy}, that
all characters should appear in the partition function with non-negative
integer multiplicities. This must be true
for local quantum
field theories whose partition functions can be given as
traces of powers of
a transfer matrix, but it fails for the
Potts models at general values of $q$, for which no such transfer
matrix can be defined.

Another way to see the 
special nature of the Potts conformal field theories
is to consider the limit $q\to 4$,
or $\xi\to\infty$. To take this limit through minimal models
we must send
$p,p'\rightarrow\infty$, keeping $|p{-}p'|$ finite. The
scaling dimensions of the degenerate fields tend to
$x_{r,s}=(r-s)^2/2$,
as is reasonable for a theory with $c=1$. However, notice that
$x_{r,s}=x_{r+k,s+k}$ for any~$k$\,. Taking the limit $\xi\to\infty$
through a sequence of
minimal models therefore results in a theory in which 
each scaling dimension appears with an infinite degeneracy\footnote{This
limit needs some care.
Here we have implicitly focused on a finite set of
(`Kac'-like \cite{crdy}) operators, taken the limit $q\to 4$, and only
then allowed the number of operators to tend to infinity. Taking the
limit in the other way gives a theory even less like the 4-state Potts
model \cite{Runkel:2001ng}.}.
By contrast, the degeneracies of the scaling dimensions in the $4$-state 
Potts
model are {\em finite}.

The discussion so far has concentrated on the critical Potts models. 
For the tricritical models, it turns out that universal quantities
are still
given by formulae such as (\ref{dNf}) and (\ref{cdef}), 
with the {\em same} formula
$\sqrt{q}=2\sin(\frac{\pi}{3}\lambda)$ relating $q$ to $\lambda$ or 
$\xi=(3{+}2\lambda)/(3{-}2\lambda)$, but with
$\lambda$ now required to lie in the range $3/2\le\lambda\le 3$
\cite{Nienh}. At fixed $q$ this is
achieved by sending $\lambda$ to $3-\lambda$, or $\xi$ to $-2-\xi$.
Under this continuation, $c(\xi)$ is mapped to $c(\xi{+}1)$, and
$\Delta_{21}(\xi)$ to $\Delta_{12}(\xi{+}1)$.
This is in accord with the observation of
\cite{DF}, that the exponents for the tricritical models \cite{dN,Na}
can be recovered by the same style of argument as followed for the
critical models, if the energy operator
is identified not with $\phi_{21}$, but rather
with $\phi_{12}$.
However,
just as for the critical models, it is only when $q$ is an integer that 
the partition function of a tricritical Potts model coincides with that of
a minimal model.

For both the critical and the tricritical models, the conformal
dimension $\Delta$ of the energy operator $\varepsilon$ is
\eq
\Delta(\lambda)=-\frac{1}{2}+\frac{9}{2(3{+}2\lambda)}~.
\label{Delform}
\en
Note that $\Delta(0)=1$, while $\Delta(3)=0$\,: the range $[0,3]$
for $\lambda$ is precisely the range for which the energy operator is
both relevant and of positive conformal dimension.

We have stressed the fact that the critical and tricritical
Potts model partition functions do not 
generally coincide with those of minimal models, even at points where the
central charges match, because a knowledge of the operator content as encoded 
in the partition function is crucial for the
proper interpretation of finite-size
effects. 
Imagine that the theory is defined on a 
cylinder of circumference $R$. 
If it is conformal, then the only scale
is provided by the system size itself, and so $E(R)$, the ground state energy,
must be proportional to $1/R$. The precise relation is \cite{BCN,Aff1}
\eq
E(R)\,=\,-\frac{\pi}{6 R}\,c_{\rm eff}
\label{eformc}
\en
where $c_{\rm eff}$, the `effective central charge', is equal to 
$c-12x_{\rm min}$\,, with $x_{\rm min}$ is the smallest scaling dimension
in the model. This is the dimension of some field $\phi_{\rm vac}$ which 
generates the vacuum, or ground, state 
for the conformal field theory on the cylinder.
For a unitary theory, there are no negative-dimension
fields and $\phi_{\rm vac}=I$, the identity operator. Thus $x_{\rm min}=0$,
and $c_{\rm eff}=c$. However, if the model is non-unitary
then negative-dimension operators are to be expected, and as a result
$c_{\rm eff}$ is larger than $c$. Whether the Potts models should be 
considered as unitary or not is perhaps a moot point, but an examination
of the partition functions in 
\cite{dFSZ} shows that all the scaling dimensions there are non-negative,
and hence, for {\em all} $0\le q\le 4$, for both the critical and the
tricritical models, we have $c_{\rm eff}=c$.

We are interested in the scaling regions around the critical and
tricritical points.
In the continuum limit, these should be described by
perturbed conformal field theories of the form~\cite{Zam1}:
\eq
{\cal A}_{q,\tau}={\cal A}_{CFT} +
\tau\int\epsilon(x)\,d^2x
\label{action}
\en
where ${\cal A}_{CFT}$ is the action at the critical or tricritical
point, $\tau$ measures the (scaled) deviation from the critical
temperature, and
the energy density operator $\epsilon(x)$ can be identified with
$\phi_{21}(x)$  or $\phi_{12}(x)$ for the critical or tricritical
points respectively. The dimensionful coupling $\tau$ introduces an
independent length scale $m\propto \tau^{1/(2{-}2\Delta)}$, where
$\Delta=\Delta_{21}$ or $\Delta_{12}$
is the conformal dimension of $\epsilon$, given in terms of $\lambda$
by (\ref{Delform}). The ground state energy
$E(R)$ can still be expressed as in (\ref{eformc}), but now the
effective central charge can depend on the dimensionless quantity
$r\equiv mR$, and we should also allow for the possibility of a `bulk' 
term in $E(R)$, proportional to the system size:
\eq
E(R)\,=\,{\cal E}(\tau)R-\frac{\pi}{6 R}\,c(r)\,.
\label{eform}
\en
The `scaling function' $c(r)$
(for brevity, we shall omit the
subscript `eff')
encodes a great deal
of information about the off-critical model, and will be the main
topic of this paper.

\resection{Mass spectrum and S-matrix data} \label{recap}
\label{SpecD}
An important input to our conjectures is the infrared information
provided by
the mass spectrum and
S-matrix of the model on an infinite line. To make this paper
relatively self-contained, in this section we
summarise the relevant data from \cite{pottsI}.
The full spectrum contains both kink and breather states, but since
only the diagonal S-matrix elements (those
involving at least one breather) 
enter directly into the TBA systems, only these will be quoted here.
The S-matrix elements in \S \ref{p0}, \S \ref{4part} and \S \ref{6part}
depend on
the  parameter $\lambda$ introduced in the last section.
For rational values of $\lambda$, our results will be equally
applicable to perturbations of minimal models 
$\CM_{p,p'}$, modulo the issues of choice of vacuum state $\phi_{\rm vac}$
discussed above. (For the unitary cases $p'=p{+}1$ which form the main
concern below, the ground states agree and these do not arise anyway.)
The relation between $\lambda$ and $p$ and $p'$ is
\bea
\lambda &=& \fract{3 p}{p'} - 
\fract{3}{2}~~,~~~(\phi_{21}~\mbox{perturbations})  \\
\lambda &=& \fract{3 p'}{p} - \fract{3}{2}~\,,~~~(\phi_{12}~\mbox{perturbations}) 
\eea 
\subsection{Perturbed critical models: $0<\lambda \le 1$   (one particle)}

In this regime there is only the fundamental kink $K \equiv K_1$. 
The  physical--strip poles in the kink S-matrix elements 
 at $\te= \fract{2i\pi}{3}$ and
 $\te=  \fract{i\pi}{3}$
are due to the fundamental kink itself.

\subsection{Perturbed critical models: $1<\lambda<3/2$ (two particles)}
\label{p0}
A pole at $\te= i\pi(1 {-}1/\lambda)$ enters the physical strip as
$\lambda$ passes $1$, and is
interpreted as neutral bound state $B \equiv B_1$ with mass
$m_{B_1}=2 m_K \cos( \frac{\pi}{2}-\frac{\pi}{2\lambda})$.
This extra particle brings with it two new scattering amplitudes,
${\cal S}_{B_1K_1}$ and ${\cal
S}_{B_1B_1}$\,:
\begin{eqnarray}
{\cal S}_{B_1K_1} & = & [\fra{1}{2}+\fra{1}{2\lambda}]
[\fra{1}{6}+\fra{1}{2\lambda}]~, \\[3pt]
{\cal S}_{B_1B_1} & = & [\fra{2}{3}][\fra{1}{\lambda}]
[\fra{1}{\lambda}-\fra{1}{3}]~,
\end{eqnarray}
where
\eq
[a] =
(a)(1-a)\ \ \ \ \mathrm{and}\ \ \ \
(a)=\frac{\sinh\left(\fra{\theta}{2}+\fra{i\pi a}{2}\right)}
{\sinh\left(\fra{\theta}{2}-\fra{i\pi a}{2}\right)}~.
\label{bldef}
\en
Even though ${\cal S}_{B_1K_1}$ and ${\cal S}_{B_1B_1}$ contain
further poles, for
values of $\lambda$ in this region they can all be explained \cite{pottsI}
by invoking a variant of the Coleman-Thun \cite{CT,CDS} mechanism.
\subsection{Perturbed tricritical 
models: $3/2<\lambda\le 2$ (four particles) }
\label{4part}
In addition to $K_1$ and $B_1$\,, two
extra particles, a kink $K_2$ and a neutral breather $B_3$, now
enter the spectrum.
Their masses are
\eq
m_{K_2}= 2 m_K \cos( \fract{\pi}{3} - \fract{1}{2 \lambda})~~,~~
m_{B_3}=2m_{B_1}\cos(\fra{\pi}{2\lambda}-\fra{\pi}{6}) ~~\, .
\en
No further
particles are
needed to explain the pole structure up to $\lambda = 2$. The
new diagonal S-matrix elements are as follows:
\bea
{\cal S}_{B_1K_2}
& = &
[\fra{1}{2}][\fra{5}{6}][\fra{1}{6}+\fra{1}{\lambda}]
[\fra{1}{\lambda}-\fra{1}{6}] \nn\\
{\cal S}_{B_3K_1}
& = & [\fra{1}{3}]^2[\fra{1}{\lambda}]
[\fra{1}{\lambda}+\fra{1}{3}]\nn \\
{\cal S}_{B_3K_2}
& = & [1-\fra{1}{2\lambda}]^2
[\fra{1}{3}+\fra{1}{2\lambda}]^3[\fra{2}{3}+\fra{1}{2\lambda}]
[\fra{3}{2\lambda}-\fra{1}{3}][\fra{3}{2\lambda}] \nn\\
{\cal S}_{B_3B_3}
& = & [\fra{2}{3}]^3[\fra{1}{\lambda}]^3
[\fra{4}{3}-\fra{1}{\lambda}]^2[\fra{1}{3}+\fra{1}{\lambda}]
[\fra{2}{\lambda}-\fra{2}{3}][\fra{2}{\lambda}-\fra{1}{3}] \nn\\
{\cal S}_{B_3B_1}
& = & [\fra{1}{6}+\fra{1}{2\lambda}]^2
[\fra{1}{2}+\fra{1}{2\lambda}][\fra{7}{6}-\fra{1}{2\lambda}]
[\fra{3}{2\lambda}-\fra{1}{2}][\fra{3}{2\lambda}-\fra{1}{6}] 
\label{Smat0}
\eea

\subsection{Perturbed tricritical models: $2<\lambda < 9/4$  (six particles)}
\label{6part}
For $2<\lambda<\frac{9}{4}$\,,
two more breathers,
$B_2$ and  $B_5$, enter the  spectrum. The
scattering amplitudes for the new particles
become increasingly
complicated, and
greater reliance is placed on the Coleman-Thun mechanism to explain the
pole structure. The new S-matrix elements in this region are:
\begin{eqnarray}
{\cal S}_{B_2B_1} & = & [1-\fra{1}{2\lambda}]
[\fra{2}{3}-\fra{1}{2\lambda}][\fra{3}{2\lambda}]
[\fra{3}{2\lambda}-\fra{1}{3}] \nn\\
{\cal S}_{B_2B_2} & = & [\fra{2}{3}][\fra{2}{3}-\fra{1}{\lambda}]
[\fra{1}{\lambda}-\fra{1}{3}][\fra{2}{\lambda}]
[\fra{2}{\lambda}-\fra{1}{3}][1-\fra{1}{\lambda}]^2 \nn\\
{\cal S}_{B_2B_3} & = & [\fra{1}{2}][\fra{5}{6}]
[\fra{2}{\lambda}-\fra{1}{6}][\fra{7}{6}-\fra{1}{\lambda}]^2
[\fra{2}{\lambda}-\fra{1}{2}][\fra{5}{6}-\fra{1}{\lambda}]^2 \nn\\
{\cal S}_{B_2K_1} & = & [\fra{1}{2}][\fra{1}{6}]
[\fra{1}{2}+\fra{1}{\lambda}][\fra{1}{6}+\fra{1}{\lambda}] \nn\\
{\cal S}_{B_2K_2} & = & [\fra{1}{2}-\fra{1}{2\lambda}]^2
[\fra{1}{6}+\fra{1}{2\lambda}]^2[\fra{1}{6}+\fra{3}{2\lambda}]
[\fra{3}{2\lambda}-\fra{1}{6}] \nn\\
{\cal S}_{B_5B_1} & = &
[\fra{1}{3}]^2[\fra{4}{3}-\fra{1}{\lambda}]
[\fra{1}{3}+\fra{1}{\lambda}][\fra{2}{\lambda}-\fra{2}{3}]
[\fra{2}{\lambda}-\fra{1}{3}][1-\fra{1}{\lambda}]^2 \nn\\
{\cal S}_{B_5B_2} & = & [\fra{4}{3}-\fra{3}{2\lambda}]^2
[1-\fra{3}{2\lambda}]^2[\fra{2}{3}+\fra{1}{2\lambda}]
[\fra{1}{3}+\fra{1}{2\lambda}]^3[1-\fra{1}{2\lambda}]^2
[\fra{5}{2\lambda}-\fra{1}{3}][\fra{5}{2\lambda}-\fra{2}{3}] \nn\\
{\cal S}_{B_5B_3} & = & [\fra{7}{6}-\fra{1}{2\lambda}]
[\fra{1}{6}+\fra{3}{2\lambda}][\fra{3}{2}-\fra{3}{2\lambda}]^2
[\fra{5}{2\lambda}-\fra{5}{6}][\fra{1}{2}+\fra{1}{2\lambda}]^3
[\fra{3}{2\lambda}-\fra{1}{6}]^3[\fra{5}{2\lambda}-\fra{1}{2}]
[\fra{5}{6}-\fra{1}{2\lambda}]^4 \nn\\
{\cal S}_{B_5B_5} & = &
[\fra{5}{3}-\fra{2}{\lambda}]^2[\fra{2}{3}]^5
[\fra{3}{\lambda}-1][\fra{3}{\lambda}-\fra{2}{3}][\fra{1}{\lambda}]^5
[\fra{1}{3}+\fra{1}{\lambda}]^3[\fra{4}{3}-\fra{2}{\lambda}]^3
[\fra{2}{\lambda}][\fra{4}{3}-\fra{1}{\lambda}]^2 \nn\\
{\cal S}_{B_5K_1} & = & [\fra{1}{2}-\fra{1}{2\lambda}]^2
[\fra{5}{6}-\fra{1}{2\lambda}]^2[\fra{1}{6}+\fra{3}{2\lambda}]
[\fra{3}{2\lambda}-\fra{1}{6}] \nn\\
{\cal S}_{B_5K_2} & = & [\fra{5}{6}]^2[\fra{1}{2}]^2
[\fra{7}{6}-\fra{1}{\lambda}]^2[\fra{5}{6}-\fra{1}{\lambda}]^3
[\fra{1}{2}+\fra{1}{\lambda}][\fra{2}{\lambda}-\fra{1}{6}]
[\fra{2}{\lambda}-\fra{1}{2}]
\label{Sel}
\end{eqnarray}
The complete mass spectrum up to $\lambda=9/4$ is:
\eq
\begin{array}{ll}
m_{K_1}  =  m  ~&(\lambda >0) \nn\\[3pt]
m_{B_1}  =  2m\cos(\fra{\pi}{2}-\fra{\pi}{2\lambda}) ~&(\lambda >1) \nn\\[3pt]
m_{K_2}  =  2m\cos(\fra{\pi}{3}-\fra{\pi}{2\lambda})  ~&(\lambda>3/2) \nn\\[3pt]
m_{B_3}  =  4m\cos(\fra{\pi}{2}-\fra{\pi}{2\lambda})
\cos(\fra{\pi}{2\lambda}-\fra{\pi}{6})  ~&(\lambda>3/2) \nn\\[3pt]
m_{B_2}  =  2m\cos(\fra{\pi}{2}-\fra{\pi}{\lambda})~&(\lambda>2)\nn\\[3pt]
m_{B_5}  =  4m\sin(\fra{\pi}{\lambda})
\cos(\fra{\pi}{3}-\fra{\pi}{2\lambda})  ~&(\lambda>2)
\end{array}
\en
\resection{The thermodynamic Bethe ansatz}
\label{revsect}
Our aim is to probe the models (\ref{action}) via their
properties in finite spatial volumes. The main tool will be the
thermodynamic Bethe ansatz (TBA) approach, a technique which allows
the finite-size ground state energy of a theory to be obtained via 
the solution of a collection of coupled nonlinear integral equations.

If the set of TBA equations is to be finite, then we would
expect that the unperturbed theory should be associated
with a minimal model in some way (as recalled above,
for non-integer values of $q$ the critical or tricritical $q$-state
Potts model is {\em never} precisely the same as a minimal model, but 
the operator subalgebras generated by the energy operator 
agree, and so the ground-state energies based on the `conformal'
vacuum state $I$ should
coincide.) The
precise form of the equations will depend on the particular model
being perturbed, and only a limited number of cases have been
studied to date. Before describing our new conjectures, 
we shall give
a brief summary of the earlier work.

\subsection{Earlier work}
\label{earlierTBA}
The situation is summarised in table~\ref{t:tba};
the cited papers can be consulted for further
explanations.
Many of these results can, sometimes
with hindsight, be labelled according to the `$g\diamond g'$'
scheme put forward
in~\cite{Rav}, where $g$ and $g'$ indicate a pair of diagrams of
ADET type. Since the $A_1$ incidence matrix is zero, the TBA for the
$A_1\diamond A_1$ case is trivial, reflecting the fact that the
field theory of $\CM_{3,4}+\phi_{13}$, the thermally-perturbed Ising 
model, is free. Two other cases, labelled $(G_2)$ and $(F_4)$, also
have a Lie-algebraic interpretation, though 
the fact that these algebras are not simply-laced 
places them outside the set of $g\diamond g'$ systems.
However, their form is in line with a more general set
of Lie algebraic TBA systems discussed
in \cite{KN,FusH}. As was remarked in \cite{pottsI}, there is
an intriguing coincidence between the sequence
$\{A_1,A_2,G_2,D_4,F_4,E_6,E_6,E_7,E_8\}$ which arises  naturally in
connection with the Potts models, and the `exceptional series' of Lie
algebras making up the last line of the extended Freudenthal
magic square, as discussed by Deligne, Cohen and de Man, 
Cvitanovic and others \cite{Del,Cohman,DelII,Cvit,Macf}.
In \cite{DelII}, Deligne and de Man 
identified 
an extra member of the series, namely
the superalgebra $OSp(1|2)$. From our point of view this corresponds
to $\CM_{7,10}+\phi_{21}$, where the unperturbed CFT is the 
$OSp(1|2)^{(1)}\times OSp(1|2)^{(1)}/OSp(1|2)^{(2)}$ coset conformal
field theory \cite{pottsI}. The relevant TBA is the $T_1\diamond
D_3$ system, the $n=3$ case of the third line of table~\ref{t:tba}.

However, it is important to realise that all of the
infinite sets of systems listed in table \ref{t:tba}
describe  the
ground states of perturbations of non-unitary minimal 
models, generated by negative-dimension operators and {\em not} the
identity.
These TBAs are therefore 
not directly  relevant for the $q$-state Potts models,
although one might hope that 
the addition of a suitable chemical
potential would allow them to describe the  Potts vacuum state as well
(see for example \cite{M0,F0,RST}). We will not discuss this
possibility any further here, but it would be an interesting avenue to
explore.
\begin{table}[ht]
\hspace{-25pt}
\begin{tabular}{|l|l|c|c|c|c|l|l|}
\hline
 && & & &&&\\[-4pt]
Minimal & Perturbing &
$\lambda$ &$c$ &$c_{\rm eff}(0)$ & $\phi_{\rm vac}$&
 TBA& Ref. \\
model &field&&&&&system& \\[5pt]
\hline
&& & &&&&\\[-4pt]
$\CM_{n,2n-1}$&$\phi_{21}$& $\fra{3}{4n{-}2}$&
$-\fra{(n-2)(4n-3)}{n(2n-1)}$ &
$\fra{(n-2)(2n+3)}{n(2n-1)}$& $\phi_{12}$ & {}\,$T_1\diamond
A_{2n-4}\!$ &
\cite{RST,BP}$\!$\\[3pt]
$\CM_{2n+1,4n}$&$\phi_{21}$&
$\fra{3}{4n}$& $-\fra{(2n-3)(4n-1)}{2n(2n+1)}$  & 
$\fra{4n^2+2n-3}{2n(2n+1)}$ &$\phi_{n,2n-1}$& {}\,$T_1\diamond T_{2n-1}$ &
\cite{MelzerSUSY}\\[3pt]
$\CM_{2n+1,4n-2}\!$&$\phi_{21}$& $\fra{3}{2n-1}$& 
$-\fra{4(n -1)(2n -7)}{(2n -1)(2n +1)}$ & $\fra{4 n^2 -4}{4 n^2 -1}$ &
$\phi_{\left[\fra{n+1}{2}\right] , 2 \left [\fra{n+1}{2} \right]-1}\!$
& {}\,$T_1\diamond D_{n}$ &
\cite{DDT}\\[3pt]
$\CM_{3,4}$&$\phi_{21}({=}\phi_{13})$&$ \fra{3}{4}$ & 
$\fra{1}{2}$ & $\fra{1}{2}$  & $\phi_{11}$ & \,$A_1\diamond A_1$ & \\[3pt]
$\CM_{4,5}$&$\phi_{21}$&$ \fra{9}{10}$ & 
$\fra{7}{10}$ & $\fra{7}{10}$  & $\phi_{11}$ & ~ &\cite{BP} \\[3pt]
$\CM_{5,6}$&$\phi_{21}$& {\footnotesize $1$} 
& $\fra{4}{5}$ & $\fra{4}{5}$ & $\phi_{11}$ 
& \,$A_2\diamond A_1$&\cite{AlZam1} \\[3pt]
$\CM_{9,10}$&$\phi_{21}$&$\fra{6}{5}$ & $\fra{14}{15}$ & $\fra{14}{15}$&
  $\phi_{11}$&{}~$(G_2)$   &\cite{Ta1} \\[3pt]
$(c=1)$&$\phi_{21}{=}\phi_{12}$& 
$\fra{3}{2}$& {\footnotesize $1$} & {\footnotesize $1$}
 &{\small $I$}& {}\,$D_4\diamond A_1$ & \cite{KM}\ \\[3pt]
$\CM_{10,11}$&$\phi_{12}$& $\fra{9}{5}$ & $\fra{52}{55}$ &
$\fra{52}{55}$ & $\phi_{11}$  &{}~$(F_4)$  & \cite{Ta1}\ \\[3pt]
$\CM_{6,7}$&$\phi_{12}$& {\footnotesize $2$} & $\fra{6}{7}$ & $\fra{6}{7}$ &
$\phi_{11}$  & {}\,$E_6\diamond A_1$ & \cite{KM}\ \\[3pt]
$\CM_{4,5}$&$\phi_{12}$& $\fra{9}{4} $& $\fra{7}{10}$ & $\fra{7}{10}$ &
$\phi_{11}$  & {}\,$E_7\diamond A_1$ & \cite{KM}\ \\[3pt]
$\CM_{3,4}$&$\phi_{12}$& $\fra{5}{2}$&$\fra{1}{2}$&$\fra{1}{2}$
 &$\phi_{11}$& {}\,$E_8\diamond A_1$ & \cite{KM}\ \\[3pt]
$\CM_{2,5}$&$\phi_{12}$& {\footnotesize $6$} &$-\fra{22}{5}$ &$\fra{2}{5}$&
$\phi_{12}$ & {}\,$T_1\diamond A_1$   & \cite{AlZam1}\ \\[3pt]
$\CM_{2,2n+1}$&$\phi_{12}$& {\footnotesize $3n$}& $-\fra{2(n-1)(6n-1)}{2n +1}$
&$\fra{2n-2}{2n+1}$& $\phi_{1n}$ & {}~ & \cite{MK}\\[3pt]
\hline
\end{tabular}
\caption{TBA systems for minimal models perturbed by
$\phi_{21}$ and $\phi_{12}$. For the systems of $g\diamond g'$ type,
the mass in the magnonic system should be placed on
the first node (that furthest from the tadpole or fork for
$T_n$ or $D_n$). $[x]$ is the integer part of $x$. } \label{t:tba}
\end{table}


The observation that a TBA system related to $G_2$
might describe the $\phi_{21}$ perturbation of
$\CM_{9,10}$ ($\lambda=\frac{6}{5}$)
was made in \cite{Ta1}. This point lies in the region
where the analysis of \cite{pottsI} suggests two particle
types in the model, and indeed that is precisely
what the TBA system of \cite{Ta1}
predicts. This case will be discussed along with the new
TBA systems below, and for now we merely stress that, just as for
almost all the above examples,
this TBA system was first obtained as a
conjecture, only subsequently being checked to describe the
claimed model.

A more deductive approach can be found in a paper by Bazhanov and
Ellem~\cite{BP}, who 
discussed the $\phi_{21}$ perturbations of
$\CM_{4,5}$ ($\lambda=\frac{9}{10}$) and $\CM_{3,5}$
($\lambda=\frac{3}{10}$), taking
Smirnov's RSOS description of the massive scattering 
theories~\cite{Sm} as their starting-point. 
Subject to some mild assumptions, they
derived sets of TBA equations for these two cases.
Crucial was the fact~\cite{Zamtr}  that for these models the RSOS
restriction on the 
allowed 
vacuum 
states coincides with the
restriction imposed on adjacent spins in the hard hexagon
model~\cite{Bax}. Unfortunately  this equivalence  holds only for
a subset of the whole family of models related to $a_2^{(2)}$ (as
mentioned in~\cite{BP}, another example, to which we shall return
in \S\ref{con12} below, is the theory $\CM_{5,6}+\phi_{12}$). Thus,  the
extension of the approach of \cite{BP} to further models faces
considerable obstacles. 

One other piece of work should be mentioned at this stage,
even though its ultimate conclusions appear to
run counter to the findings
that will be reported below.
In the course of a detailed study of
character and polynomial identities associated with general
$\phi_{21}$ perturbations,
Berkovich, McCoy and Pearce \cite{BMP} discussed
possible TBA equations 
for all unitary minimal models $\CM_{p,p+1}$, which reduced to the
result of \cite{BP} for $p=4$.
They were only able to
specify the general form of these equations, and certain
(`kernel') functions necessary for a complete conjecture were left
undetermined. Nevertheless, at least on a na\"{\i}ve reading, the
equations of \cite{BMP} appear to entail just a single massive
kink in the model for all values of $p$. This is at variance with
the results of \cite{pottsI}, which for $\phi_{21}$ perturbations
predict that there should be a kink
and a breather for all $p>5$.
In later sections
we shall propose some new TBA systems which {\em are} consistent
with this spectrum. There 
may be room to
accommodate both our results and those of \cite{BMP} -- the closing
remarks of~\cite{BM} indicate one possibility -- but we suspect
that the final resolution will reside in some modification to the
considerations of \cite{BMP} to bring their TBA systems into line
with ours, at least insofar as they concern $\phi_{21}$ perturbations
of unitary minimal models\footnote{In this context, it is worth 
noting  that there is strong evidence
that further novel sets  of   character and 
polynomial identities can be found using the  TBA equations proposed in this 
paper. We would like to thank Ole Warnaar for performing an initial
check of this.}.

For now we shall leave this question unresolved, and
proceed with a discussion of the general form that we would expect
the TBA equations to take, assuming for the time being that the
results of \cite{pottsI}
do indeed provide a reliable guide.

\subsection{The general structure of TBA systems}
When all scattering is diagonal, the
integral equations of the TBA
follow directly from the bulk S-matrix.
The ground state energy of the
model on a circle of
circumference $R$ is
written in terms of `dressed'
single-particle energies $\ep_a(\theta)$
(pseudoenergies)\cite{AlZam1}.
These pseudoenergies solve a
system of  non-linear integral equations of the following form:
\eq
\ep_{a}(\theta) = R m_a \cosh \theta  -  \sum_{c=1}^{N}
\Phi_{ac}*L_{c}(\theta) \,.
\label{TBA1}
\en
Here
$
L_{c}(\theta)=\ln(1+e^{-\ep_{c}(\theta)}),
$
$*$ denotes the  convolution
\eq
f*g(\theta)= { 1 \over 2 \pi} \int_{- \infty}^{\infty} f(\theta')
g(\theta- \theta') d \theta'\,,
\en
and the
$2\to 2$ S-matrix elements ${\cal S}_{ac}(\theta)$ influence the equations
through the kernel functions $\Phi_{ac}(\theta)$:
\eq
\Phi_{ac}(\theta)= -i \ddt\ln {\cal S}_{ac}(\theta)\,.
\en
The number of pseudoenergies  coincides with the number $N$ of
particle types  in the original scattering theory.
{}From (\ref{TBA1}), they have the large
$\theta$ asymptotics
\eq
\ep_{a}(\te)|_{\theta \rightarrow \infty} \sim
R m_a \cosh \te~~,~~ (a=1,2,\dots,N~).
\label{as1}
\en
If off-diagonal scattering is
also involved  the complexity of the method increases significantly, as it is
necessary to perform an extra `diagonalisation' step before the final equations
can be written down (see, for example, \cite{BP} for further discussion
of this point).
But once the dust has settled, one generally finds that
the parts of the TBA equations associated with diagonal scattering are as
before, while
the contribution of each kink multiplet $K$ is
split into two parts.
The first is described by a single
pseudoenergy $\ep_{K}(\te)$ with an
asymptotic of the type~(\ref{as1}), with $m_K$ the common mass
of all kinks in the multiplet.
In addition,
diagonalisation
results in the introduction of
a (possibly infinite) number of auxiliary pseudoenergies.
These behave as
\eq
\ep_{a}(\theta)|_{\theta \rightarrow \infty} \sim
\mbox{constant} ~~,~~a=N+1,N+2,\dots~~,
\en
and can be associated with fictitious `particles'
transporting  zero  energy and zero momentum
(note that  $m_a \cosh \te$ in~(\ref{as1}) is the single-particle energy
on the infinite line).  These  new  particles are
often called  `magnons', and can be thought of as constructs
introduced to get the counting of states right.
The diagonalisation is by no means trivial, and in the following we will
instead make a conjecture, using the above considerations as our
guide and borrowing elements from some
previously encountered TBA systems.
\subsection{The steps towards the main conjectures}
\label{stepsec}
\begin{figure}[ht]
\vskip -70pt
\[
\begin{array}{cc}
\refstepcounter{figure}
\label{f1.a}
\epsfxsize=.25\linewidth
\epsfbox[0 0 288 338]{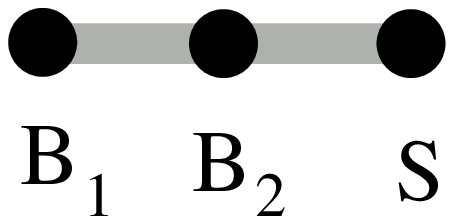}
&
\refstepcounter{figure}
\label{f1.b}
\epsfxsize= .25\linewidth
\epsfbox[0 0 288 338]{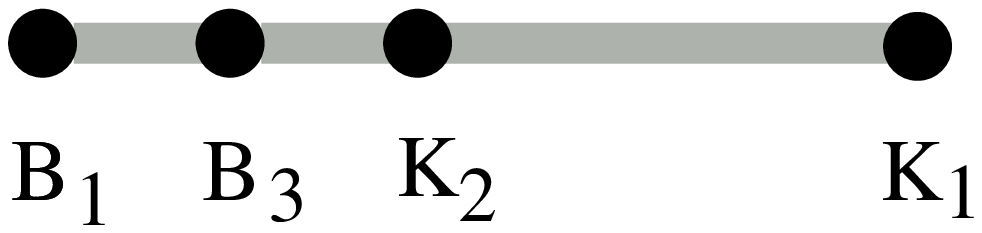}
\\[5pt]
\parbox[t]{.50\linewidth}{\small\raggedright%
Figure \ref{f1.a}: Sine-Gordon TBA at $\fract{\beta^2}{8 \pi}= \fract{1}{4}$ }
&
\parbox[t]{.34\linewidth}{\small\raggedright%
Figure \ref{f1.b}: $\CM_{6,7}+\phi_{12}$} \\[-30pt]
\refstepcounter{figure}
\label{f2.a}
\epsfxsize=.25\linewidth
\epsfbox[0 0 288 350]{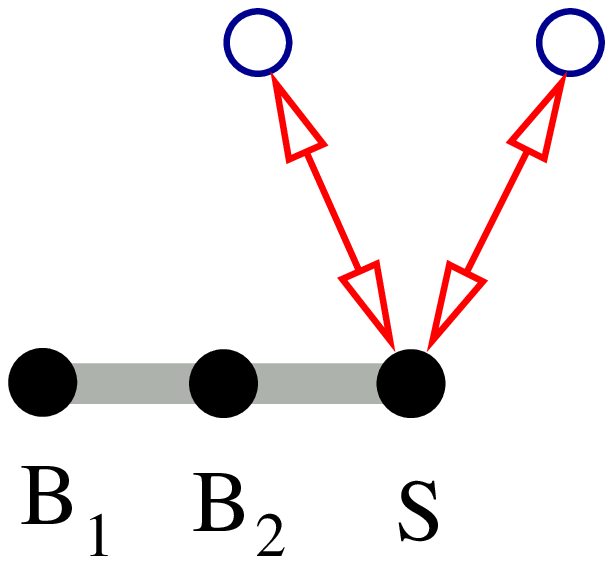}
&
\refstepcounter{figure}
\label{f2.b}
\epsfxsize= .25\linewidth
\epsfbox[0 0 288 350]{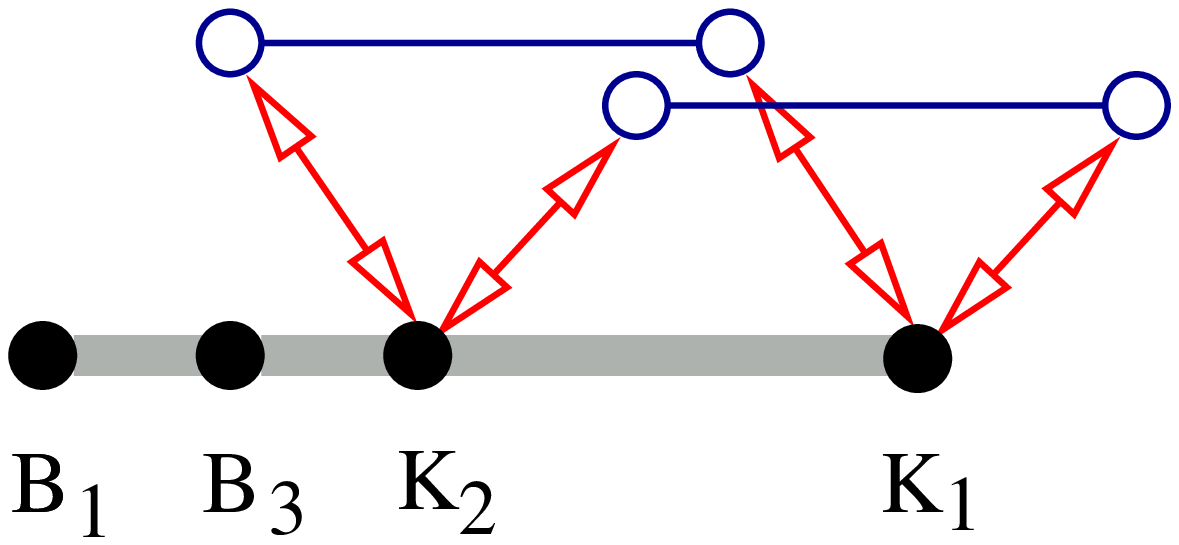}
\\[5pt]
\parbox[t]{.50\linewidth}{\small\raggedright%
Figure \ref{f2.a}:  Sine-Gordon TBA at $\fract{\beta^2}{8 \pi}= \fract{2}{7}$ } 
&
\parbox[t]{.34\linewidth}{\small\raggedright%
Figure \ref{f2.b}: $\CM_{10,11}+\phi_{12}$ } \\[30pt]
\refstepcounter{figure}
\label{f3.a}
\epsfxsize=.25\linewidth
\epsfbox[0 0 288 350]{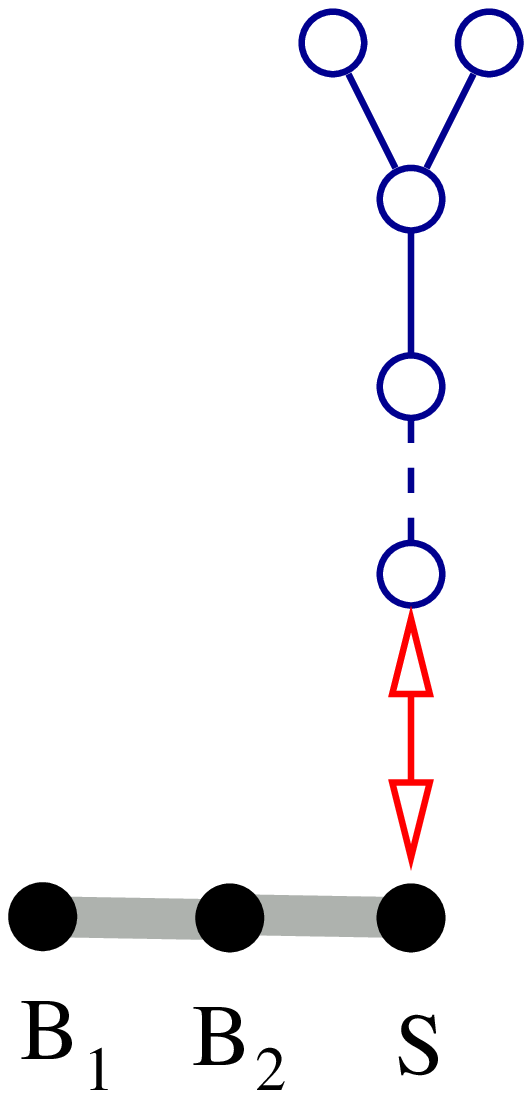}
&
\refstepcounter{figure}
\label{f3.b}
\epsfxsize= .25\linewidth
\epsfbox[0 0 288 350]{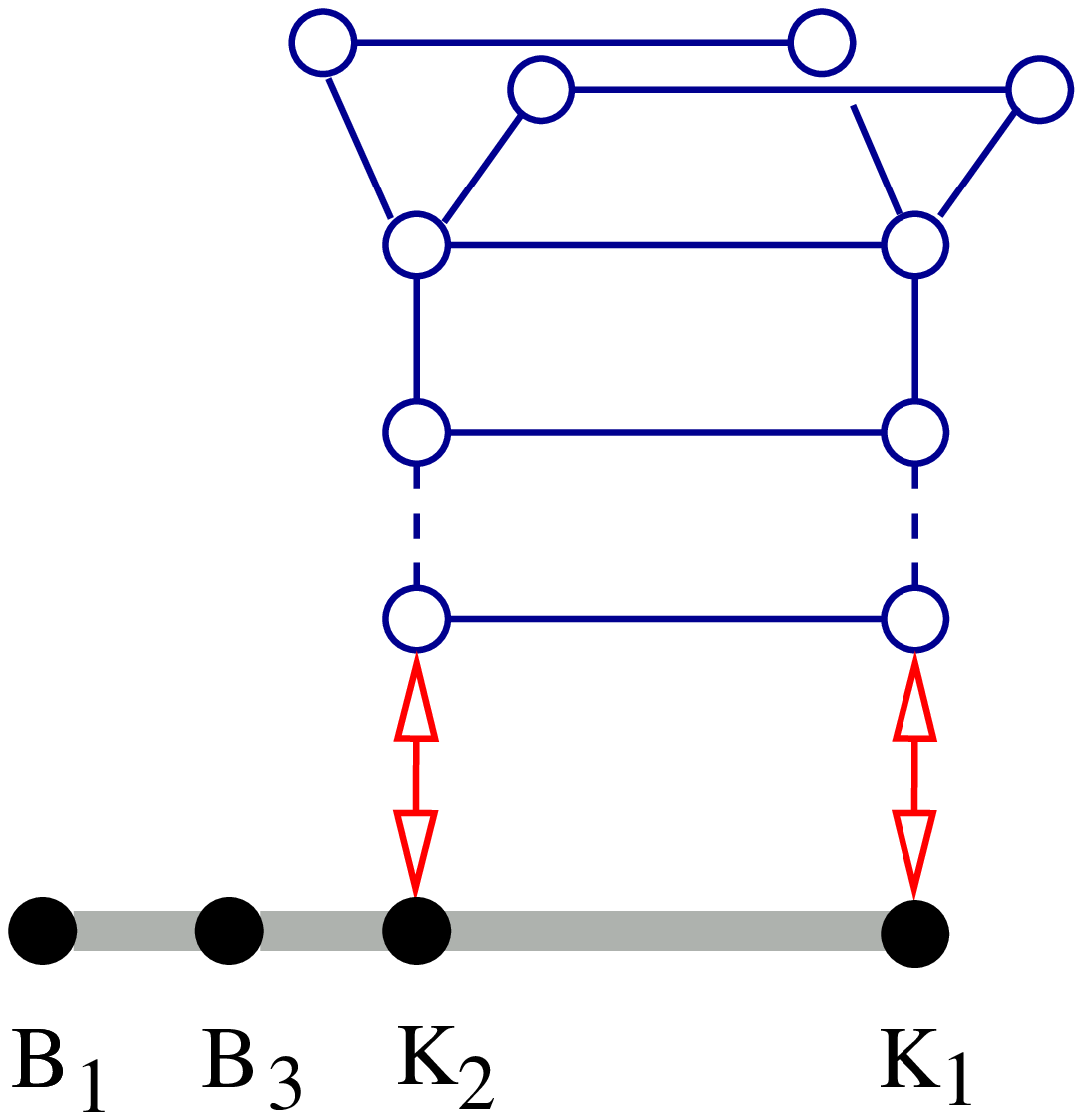}
\\[5pt]
\parbox[t]{.50\linewidth}{\small\raggedright%
Figure \ref{f3.a}:   Sine-Gordon TBA at $\fract{\beta^2}{8 \pi}=
 \fract{n-1}{3n-2}$ 
 }
&
\parbox[t]{.34\linewidth}{\small\raggedright%
Figure \ref{f3.b}:  $\CM_{4n+6,4n+7}+\phi_{12}$   } 
\end{array}
\]
\end{figure}

Our TBA conjectures were found via a link 
with the TBA equations for perturbed $\Z_N$ systems
of \cite{DTT}. 
Originally this emerged from a study of four previously-known  
$\phi_{12}$-related  cases. 
These were:
$\CM_{3,4}+\phi_{12}$  (related to $E_8$),  
$\CM_{4,5}+\phi_{12}$  (related to  $E_7$), $\CM_{6,7}+\phi_{12}$  
(related to  $E_6$) and 
$\CM_{10,11}+\phi_{12}$ (related to $F_4$).
No  hint of a common
structure came  by  just looking at the first three models,
but
a consideration of
 $\CM_{6,7}+\phi_{12}$ and 
$\CM_{10,11}+\phi_{12}$ alone revealed
some striking  analogies with systems related to $a_1^{(1)}$.
To see this, let us compare them  with the sine-Gordon TBAs 
at $\fract{\beta^2}{8 \pi}= \fract{1}{4}$ and 
$\fract{\beta^2}{8 \pi}=\fract{2}{7}$.
The four 
systems are depicted in figures \ref{f1.a}, \ref{f1.b}, \ref{f2.a} and
\ref{f2.b}. (Figure \ref{f1.b} is the $E_6$-related system: the distinction
between solitons and antisolitons is special  
to this, the case of the tricritical Potts
model, and so the $E_6$ diagram symmetry has been used 
to superimpose each pair of soliton-antisoliton nodes, resulting in the
diagram shown. Likewise, figure \ref{f1.a} is the $D_4$-related point
of the sine-Gordon model,
and the $D_4$ diagram symmetry has been used 
to superimpose the  soliton and the antisoliton.) 
There is a formal similarity between the pairs of  
diagrams in figures  \ref{f1.a} and \ref{f1.b} and in
figures \ref{f2.a} and
 \ref{f2.b}, the  
main differences being  
related to the fact  that there is just a single
soliton-antisoliton pair in the sine-Gordon
model, and correspondingly
just a single massive `soliton' node in the 
$a_1^{(1)}$-related 
TBA systems.

The sine-Gordon TBA systems of figures \ref{f1.a} and \ref{f2.a} are the first 
members of a series of models at 
$\fract{\beta^2}{8 \pi}=\fract{n-1}{3n -2} $, represented in 
figure \ref{f3.a},
which have  an  $A_1 \diamond D_{n-1}$ magnonic structure.
This fact made it natural  to generalise the
$\CM_{6,7}+\phi_{12}$ and $\CM_{10,11}+\phi_{12}$ TBA systems
in a similar manner, by `nesting'
an $A_2 \diamond D_{n-1}$ magnonic 
system on top of the two  $a_2^{(2)}$ soliton nodes. Graphically the 
result is shown in figure \ref{f3.b}.

Working first at the level of associated sets of functional equations
called Y-systems, 
checks on periodicities and central charges were used to 
fix the details.
We finally  converged to the  proposal of appendix~\ref{TBAY12} 
(Case $\trisymb{D}$), which covers the family of  models  
$\CM_{4n+6,4n+7}+\phi_{12}$. This gave 
a clear signal that the general
$\phi_{12}$ perturbations of unitary minimal models share the
`modulo 4' property of the $\ZN$-related systems found in \cite{DTT}, 
an observation which enabled us to obtain the remaining cases by replacing the  
$A_2 \diamond D_{n-1}$ part with the other  families of systems from 
\cite{DTT}. 
Full TBA equations were  then obtained by Fourier transforming
the Y-systems.

Once it was seen  that the systems in \cite{DTT} were playing a 
central r\^ole, it was possible, following similar reasoning, to 
conjecture the  TBAs for the $\phi_{21}$ perturbations. 
In these cases elements of the S-matrix were first  used to fix 
the `diagonal' part of the TBA equations, with the systems of 
ref.~\cite{DTT}
then being adapted to describe  the magnonic 
part\footnote{The two $O(N)$-related set of equations, i.e. those 
connected  to the $\ZN$ TBAs with $N$ even, are  very similar 
to the equations   for the perturbed    
$O(N)_3/O(N-1)_3$ coset models  proposed by  Paul Fendley in~\cite{Fsig}.
The main difference between  the two systems 
can be, naively speaking,  traced back to the dissimilarity 
in the neutral bound-state  sector.}.
The Y-systems were then derived  by using standard methods, paying
attention to the pole structure of the kernel functions
(see appendix~\ref{kersec} and, for example, \cite{Rav}).   
As should be
clear from this discussion, the process was one of educated
guesswork; however the detailed checks that we report below leave us
in no doubt that the final systems are correct.

\resection{TBA conjectures: $\phi_{21}$ perturbations of unitary
minimal models}
\label{con21}
In this section we shall be concerned with thermal perturbations of
critical Potts models at points where the central charges of the
unperturbed models coincide with those of the unitary minimal
models $\CM_{p,p+1}$, $p\ge 3$, which means that we set
$\sqrt{q}=2\cos\frac{\pi}{(p{+}1)}$\,, and
$\lambda=\frac{3}{2}(p{-}1)/(p{+}1)$\,, $\xi=p$\,. As explained earlier, 
the fact that the related minimal models are unitary means that we
will equally be describing the behaviour of the ground states of
perturbations of minimal models, the perturbing operator being
$\phi_{21}$ in all cases.

For $p=3$, $4$ and $5$, $\lambda\le 1$ and the spectrum consists of
the fundamental kink alone. TBA systems are already known for these
cases -- the first is the (trivial) example of the thermally-perturbed
Ising model, the second was treated in \cite{BP}, and the third is
the thermally-perturbed 3-state Potts model, for which the TBA was
written in \cite{AlZam1}.
Therefore we will suppose that $p\ge 6$, which ensures that the bulk
spectrum has one kink, and one breather~\cite{pottsI}.
The form of the `non-magnonic' part of the TBA is then
fixed almost completely
by the results of \S \ref{SpecD}:
\bea
\ep_{B}(\theta) &=& R m_B \cosh \theta  -  \sum_{c \in \{ B,K \}}
\Phi_{Bc}*L_{c}(\theta)~, \nn \\
\ep_{K}(\theta) &=& R m_K \cosh \theta - \sum_{c \in \{ B,K \}}
\Phi_{Kc}*L_{c}(\theta)
+ (\hbox{magnonic terms})~,
\label{TBA0}
\eea
with the  effective central charge given by
\eq
c(r)= {3 \over \pi^2} \sum_{c \in \{B ,K \}}
\int_{-\infty}^{\infty} d \theta~m_c R~\cosh \theta L_{c}(\theta)
{}~~~,~~~(r=m_KR)~.
\label{ceffe}
\en
{}The kernel functions involving diagonal S-matrix elements are:
\eq
\Phi_{BB}(\theta)= -i \ddt \ln {\cal S}_{BB}(\theta)
{}~~~~,~~~~ \Phi_{KB}(\theta)=\Phi_{BK}(\theta)=
-i \ddt \ln {\cal S}_{KB}(\theta)~.
\label{diagk}
\en
The last term in (\ref{TBA0}) indicates
the presence of the extra contributions
from the as-yet unknown number of magnons.
In addition,  equations  involving
magnons and  kink pseudoenergies are required.
It is at this stage that our conjectures
begin, with
most of the deduction proceeding, as explained in the last section,
by analogy with the models
studied in~\cite{Ta1,DTT}.
We start with the kernel $\Phi_{KK}(\te)$. Since the scattering of two kinks is
not generally diagonal, this is not expected to be a simple logarithmic
derivative of an S-matrix element. However, if we set
\eq
\rho= \left ({1 \over 2} -  {1 \over 2  \lambda} \right)~,
\en
then we  observe
that  the mass relation
\eq
m_B=2\cos(\pi\rho)m_K
\label{massrel}
\en
is accompanied by the following S-matrix identity:
\eq
{\cal S}_{aB}(\te)={\cal S}_{aK}(\te+ i {\pi \rho}) 
{\cal S}_{aK}(\te- i {\pi \rho})~~,~~~(a=B)
\label{boot}
\en
or, equivalently
\eq
\Phi_{aB}(\theta)=
\Phi_{aK}(\theta+ i \pi  \rho)+\Phi_{aK}(\theta- i \pi  \rho)~~,~~~(\te \ne 0,a=B)~.
\label{ker}
\en
The constraint $\te \ne 0$ is due to the fact that $\Phi_{BK}(\te)$
has poles at
$\te= \pm i\pi \rho$. Care is needed when the identity is
integrated, either in Fourier transforms or
in convolutions.
Relations between the kernels such as~(\ref{massrel}) and
(\ref{ker}), involving
every   pair of particle species, are needed
in order to convert a
TBA equation into a Y-system, and so
we shall assume that~(\ref{ker})  holds for $a=K$ too.
Assuming   also that   $\Phi_{KK}(\te)$ is  free of
singularities  in the strip
$|\Im m \te| \le \pi \rho$, we obtain
\eq
\Phi_{KK}(\theta)=
\chi_{\rho}*\Phi_{KB }(\theta)~~,~~\chi_{\rho}(\theta)=
{1 \over 2 \rho  \cosh(\theta/ 2 \rho)}~.
\label{phikk}
\en

For the reasons outlined in the last section,
we looked to the TBA systems
which had previously arisen in the context of 
self-dual perturbations of $\Z_N$-symmetric conformal field
theories for the remaining equations.
It turns out 
that for the $\phi_{21}$ perturbations of the unitary minimal models
$\CM_{p,p+1}$ with $p>5$, the equations proposed in \cite{DTT}
can be adapted to provide an appropriate set of `magnonic' equations.
The equations in \cite{DTT} were obtained via
a `doubling' of the sine-Gordon TBA systems at certain
special points $\beta^2=32\pi/N$, $\xi=1/(N/4-1)$.

The recipe goes as follows.
Consider first the continued-fraction expansion of
\eq
{6 \rho}=
\lf( {p-5 \over p-1} \ri)~.
\label{sixrho}
\en
For the unitary models $\CM_{p,p+1}$
this results in four families of cases:
\bea
A)~~~p=4n+2~~:~~ {6 \rho}&=& { 1 \over 1+ {1 \over {n-1 + {1 \over 4}}}} 
\nn \\
B)~~~p=4n+3~~:~~ {6 \rho}&=& { 1 \over 1+ {1 \over {n-1 + {1 \over 2}}}} 
\nn \\
C)~~~p=4n+4~~:~~ {6 \rho}&=& {1  \over 1+ {1 \over
        {n-1 + {1 \over {1+ {1 \over 3}} } } }}  
\label{fourc}
\\
D)~~~p=4n+5~~:~~ {6 \rho}&=& { 1 \over  1+ {1 \over n} }  \nn
\eea
matching the four families of cases seen in \cite{DTT}.
The second `nested' continued-fraction decomposition of $6\rho$ in
(\ref{fourc}),
i.e.\ the decomposition of $\rho_2=1/(6\rho)-1$,
matches the special values $\xi=1/(N/4-1)$ observed in connection with the
$\Z_N$ models, if we identify $N=p-1$.
The fact that the match is at a nested level reflects the
idea that
these extra pseudoenergies are supposed to be of magnonic type, and
for this reason   we also trade  the driving  terms of
\cite{DTT} for
\eq
\nu_j^{(\alpha)}(\te)= \delta_{\alpha 1} \tpsi_{1j}*L_{K}(\te)~~,
\en
and set
\eq
(\mbox{magnonic terms}) =   \sum_{j}
\tpsi_{1j}*L^{(1)}_{j}(\te)\,,
\en
where $j=1, \dots, n$ for cases
A, B and C and $j=1,\dots, n+1$  for case D.
Finally, the variables  $\te$, $\te'$  used in~\cite{DTT} should be
rescaled. This can again be interpreted as a consequence of the fact that
the match is at
a  nested level and  the r\^ole of $1/h=2/(N-4)$
of~\cite{DTT} is now played by the quantity
$1/g=\rho \rho_2=4 \rho/(p-5) = 2 \rho/h$ (i.e. $g=3(p-1)/2$).
This is equivalent to a replacing
the kernels
$\phi_{ij}^{[\Z_N]}(\te)  $ , $\psi_{ij}^{[\Z_N]}(\te)  $  and
$\phi_k^{[\Z_N]}(\theta)  $  defined in appendix A of~\cite{DTT}
with
\footnote{Notice that
the fact that the kernels involved in different `nested zones'
just differ by a rescaling of the arguments of the type:
$\te/\xi_1 \rightarrow \te/(\xi_1 \xi_2) \rightarrow \te/(\xi_1 \xi_2 \xi_3)
\dots$, with
\eq
\xi_1=\xi~~~,~~ \xi_i={ 1 \over { \mbox{Int}[1/\xi_i] + \xi_{i+1}}}
\en
is typical of the $a_1^{(1)}$-related systems~\cite{TS} (see also \cite{Ta2}).}
\eq
\phi_{ij}(\te) = { 1 \over 2 \rho} \phi_{ij}^{[\Z_N]}( {\te \over 2\rho})~~,
~~
\psi_{ij}(\te) =  {1 \over 2 \rho}
\psi_{ij}^{[\Z_N]}({\te \over 2 \rho})~~,~~
\phi_k(\te) =   {1 \over 2 \rho} \phi_k^{[\Z_N]}({ \te \over 2 \rho})~~.
\label{magkerdefa}
\en
In  appendix~A the resulting  set of TBA equations is written
explicitly, while in \S\ref{checks} we report some analytical and
numerical evidence for their correctness.

{
Before that, we will give the explicit forms of
the TBA systems for the first four cases,
$p=6,7,8,9$. These are the first unitary models for which
the extra particle $B \equiv B_1$ is expected to appear,
and at the same time their
magnonic structures are still relatively simple.
In each case, the TBA system is conjectured to describe the perturbation of
$\CM_{p,p{+}1}$ by $\phi_{21}$.

\bigskip
\noindent
A) ${\bf p=6}$
(n=1, $\lambda={15 \over 14}$):
\bea
\ep_{B}(\te) &=& R m_B \cosh \te  -  \sum_{c \in \{ B,K \}}
\Phi_{Bc}*L_{c}(\te) \nonumber \\
\ep_{K}(\theta) &=& R m_K \cosh \te - \sum_{c \in \{ B,K \}}
\Phi_{Kc}*L_{c}(\te)
-\tphi_2*L^{(1)}(\te)  \nn\\
\ep^{(\gamma)}(\theta) &=& - \sum_{\beta=1}^{6} l^{[E_6]}_{\gamma \beta}
\tphi_2*L^{(\beta)}(\theta) -\delta_{\gamma 1} \tphi_2*L_{K}(\te)
~~~~(\gamma=1,\dots,6)
\eea
In this case there are six  magnons and
$l^{[E_6]}_{\alpha \beta}$ is the incidence matrix of the $E_6$ Dynkin
diagram with nodes labelled as in figure \ref{figure9.a}.

\bigskip
\noindent
B) ${\bf p=7}$
(n=1, $\lambda={9 \over 8}$):
\bea
\ep_{B}(\theta) &=& R m_B \cosh \theta  -  \sum_{c \in \{ B,K \}}
\Phi_{Bc}*L_{c}(\theta) \nonumber \\
\ep_{K}(\theta) &=& R m_K \cosh \theta - \sum_{c \in \{ B,K \}}
\Phi_{Kc}*L_{c}(\theta)
-\tphi_1*L^{(1)}(\theta)  \nn \\
\ep^{(\gamma)}(\theta) &=&
- \sum_{\beta=0}^{4} l^{[A_5]}_{\gamma \beta}
\tphi_1*L^{(\beta)}(\theta) -\delta_{\gamma 1} \tphi_1*L_K(\theta)
~~~~~(\gamma=0,\dots,4)
\eea
Here, there are five  auxiliary pseudoenergies and
$l^{[A_5]}_{\gamma \beta}$ is the incidence matrix of the $A_5$ Dynkin 
diagram with
nodes labelled
as in figure \ref{figure9.b}. 

\bigskip
\noindent
C) ${\bf p=8}$
(n=1, $\lambda={7 \over 6}$):
\bea
\ep_B(\theta) &=& R m_B \cosh \theta  -  \sum_{c \in \{ B,K \}}
\Phi_{Bc}*L_c(\theta) \nn \\
\ep_K(\theta) &=& R m_K \cosh \theta - \sum_{c \in \{ B,K \}}
\Phi_{Kc}*L_c(\theta)
-\tphi_3*L^{(1)}(\theta) \nn  \\
\ep^{(1)}(\theta) &=&
-\tphi_3*(L_K(\theta)+ L^{(6)}(\theta)+L^{(4)}(\theta))
-\tphi_4*L^{(3)}(\theta)-
\tphi_5*L^{(5)}(\theta)
\nn \\
\ep^{(3)}(\theta) &=&
\tphi_2*(K^{(5)}(\theta)-L^{(1)}(\theta)) \nn \\
\ep^{(5)}(\theta) &=&
\tphi_2*(K^{(3)}(\theta)+K^{(6)}(\theta)+K^{(4)}(\theta)) \nn   \\
\ep^{(6)}(\theta) &=&
\tphi_2*K^{(5)}(\theta) \nn \\
\ep^{(4)}(\theta) &=&
\tphi_2*(K^{(5)}(\theta)-L^{(2)}(\theta)) \nn  \\
\ep^{(2)}(\theta) &=&
-\tphi_3*(L^{(6)}(\theta)+L^{(3)}(\theta)) -\tphi_4*L^{(4)}(\theta)-
\tphi_5*L^{(5)}(\theta)
\label{ccase}
\eea
where
$
K^{(c)}(\theta)= \ln(1+e^{\ep^{(c)}(\theta)})$\,.

\bigskip
\noindent
D) ${\bf p=9}$ (n=1, $\lambda={6\over 5}$):
\bea
\ep_B(\theta) &=& R m_B \cosh \theta  -  \sum_{c \in \{ B,K \}}
\Phi_{Bc}*L_c(\theta) \nn \\
\ep_K(\theta) &=& R m_K \cosh \theta - \sum_{c \in \{ B,K \}}
\Phi_{Kc}*L_c(\theta)
+\tpsi* \lf(  L^{(1)}(\theta) +L^{(3)}(\theta) \ri) \nn\\
\ep^{(\gamma)}(\theta) &=&  \sum_{\beta=1}^{2} l^{[A_2]}_{\gamma \beta}
\tpsi*L^{(\beta)}(\theta) + \delta_{\gamma 1}\tpsi*L_{K}(\theta)
~~~~~~~(\gamma =1,2)  \nn \\
\ep^{(\gamma)}(\theta) &=&  \sum_{\beta=3}^{4} l^{[A_2]}_{\gamma \beta}
\tpsi*L^{(\beta)}(\theta) + \delta_{\gamma 3} \tpsi*L_K(\theta)
~~~~~~~(\gamma =3,4)
\eea
where $\tpsi(\theta) \equiv \tpsi_{11}(\theta) = \tpsi_{22}(\theta)$.
In this case there are four  auxiliary pseudoenergies and
$l^{[A_2]}_{\alpha \beta }$ is the incidence matrix of the $A_2$ Dynkin
diagram with
nodes labelled as in figure \ref{figure9.d}.
It was noted in \cite{Ta1} that this TBA can be mapped into a particular
case of the $G_2$-related Y-systems of \cite{KN}.
Notice also that the pseudoparticle  part factorises into a pair
of $A_2$-type TBAs
while a single one was found for the model $\CM_{4,5}$ studied in~\cite{BP}.
This can be  explained by observing that the vacuum incidence diagram
($k=8$ in~\cite{Ko})  is (orbifold) equivalent
to the 
product of two ``hard hexagon''  diagrams ($k=3$ in~\cite{Ko}).
Thus also in this case a simple
variant of the  analysis of~\cite{BP} should  lead to a more rigorous
derivation of  the system, though this remains to be done.

\bigskip
\bigskip

\[
\begin{array}{cc}
\refstepcounter{figure}
\label{figure9.a}
\epsfxsize=.25\linewidth
\epsfbox[0 50 288 338]{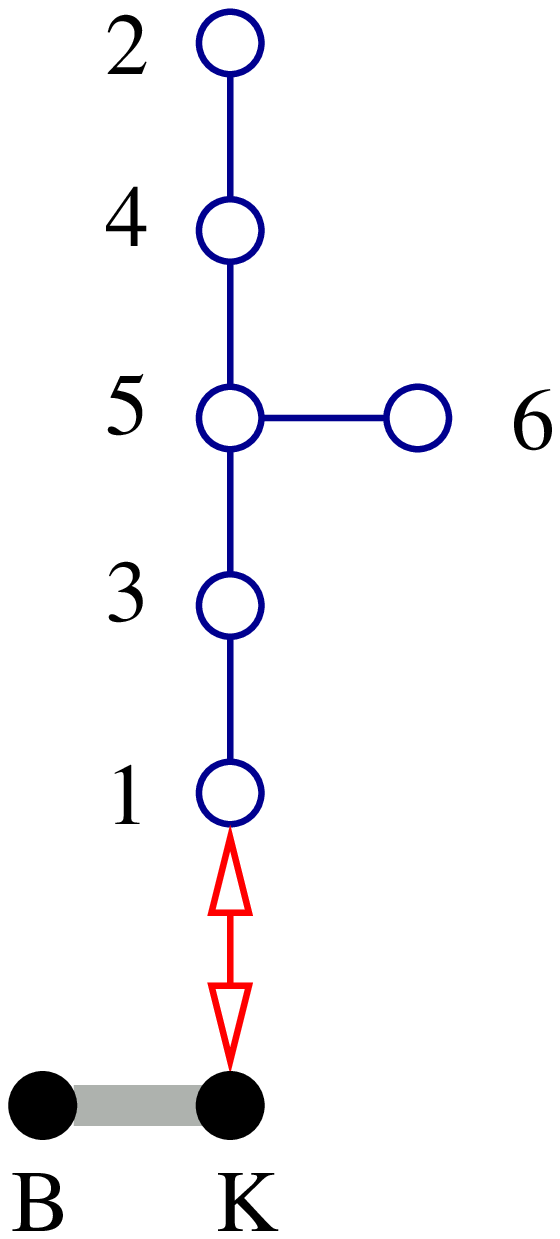}
&
\refstepcounter{figure}
\label{figure9.b}
\epsfxsize= .25\linewidth
\epsfbox[0 50 288 338]{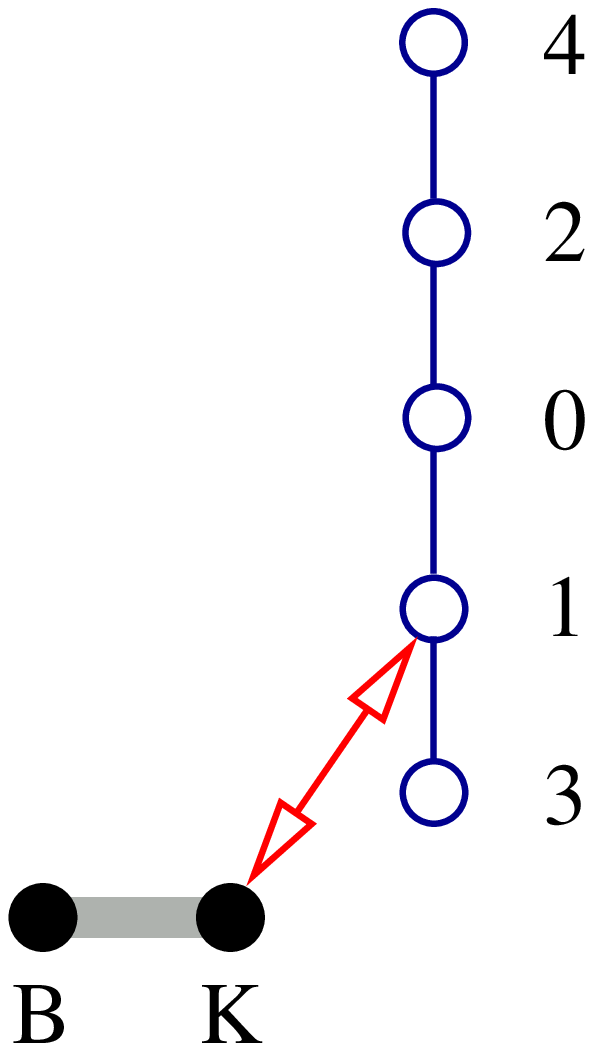}
\\
{} \\
{} \\
{} \\
\parbox[t]{.35\linewidth}{\small\raggedright%
Figure \ref{figure9.a}: $\CM_{6,7}+\phi_{21}$
}
&
\parbox[t]{.35\linewidth}{\small\raggedright%
Figure \ref{figure9.b}: $\CM_{7,8}+\phi_{21}$
} 
\end{array}
\]

\[
\begin{array}{cc}
\refstepcounter{figure}
\label{figure9.c}
\epsfxsize=.25\linewidth
\epsfbox[0 50 288 350]{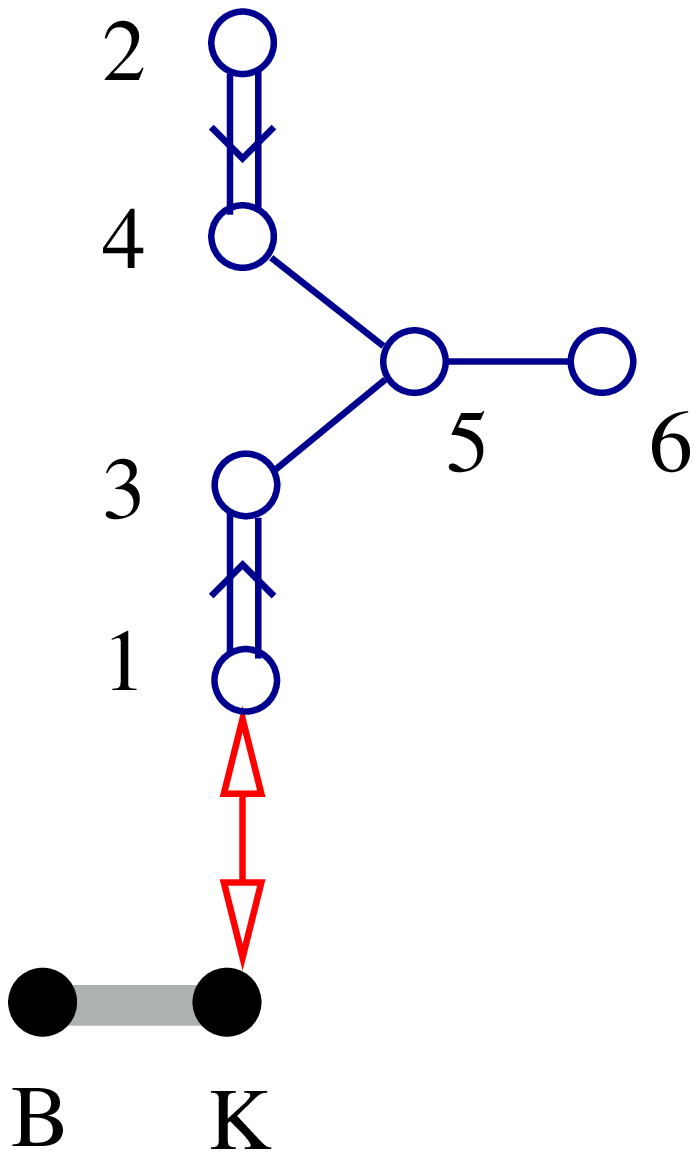}
&
\refstepcounter{figure}
\label{figure9.d}
\epsfxsize= .25\linewidth
\epsfbox[0 50 288 350]{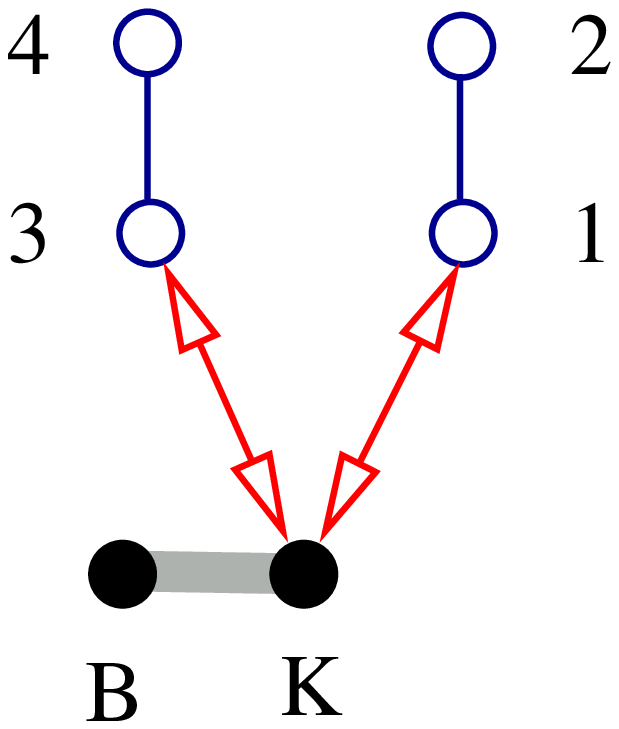}
\\
{} \\
{} \\
{} \\
\parbox[t]{.35\linewidth}{\small\raggedright%
Figure \ref{figure9.c}:  $\CM_{8,9}+\phi_{21}$
}
&
\parbox[t]{.35\linewidth}{\small\raggedright%
Figure \ref{figure9.d}:  $\CM_{9,10}+\phi_{21}$
}
\end{array}
\]
}
\resection{TBA conjectures: $\lambda>3/2$}
\label{con12}
We were also able to construct the TBA equations describing the 
ground-state energies of the unitary $\CM_{p,p+1}+\phi_{12}$
models. 
As in the cases related to  $\phi_{21}$\,, we split
the set of unitary minimal models into four families, which
we label $\trisymb{A}\dots\trisymb{D}$. For
$p \ge 7$, the spectrum consists of just
four particles (two  neutral particles $B_1$ and $B_3$ and two
kinks $K_1$ and $K_2$). Here the magnonic sector
resembles again the  $\Z_{N}$-related structure described in~\cite{DTT}.
A pictorial representation of the TBA systems  for the family
$\CM_{4n+3,4n+4}$  is given in figure~\ref{figtba}.
Note that at  $n=0$, what for $n>0$ had been a  magnonic  $E_6$-type sector
in the  TBA
naturally generates  the extra six neutral particles needed to
reconstruct the full $E_8$-related  mass spectrum of the
$\phi_{12}$-perturbed Ising model.
\begin{figure}[ht]%
\begin{center}
\resizebox{0.8\linewidth}{!}%
{\includegraphics{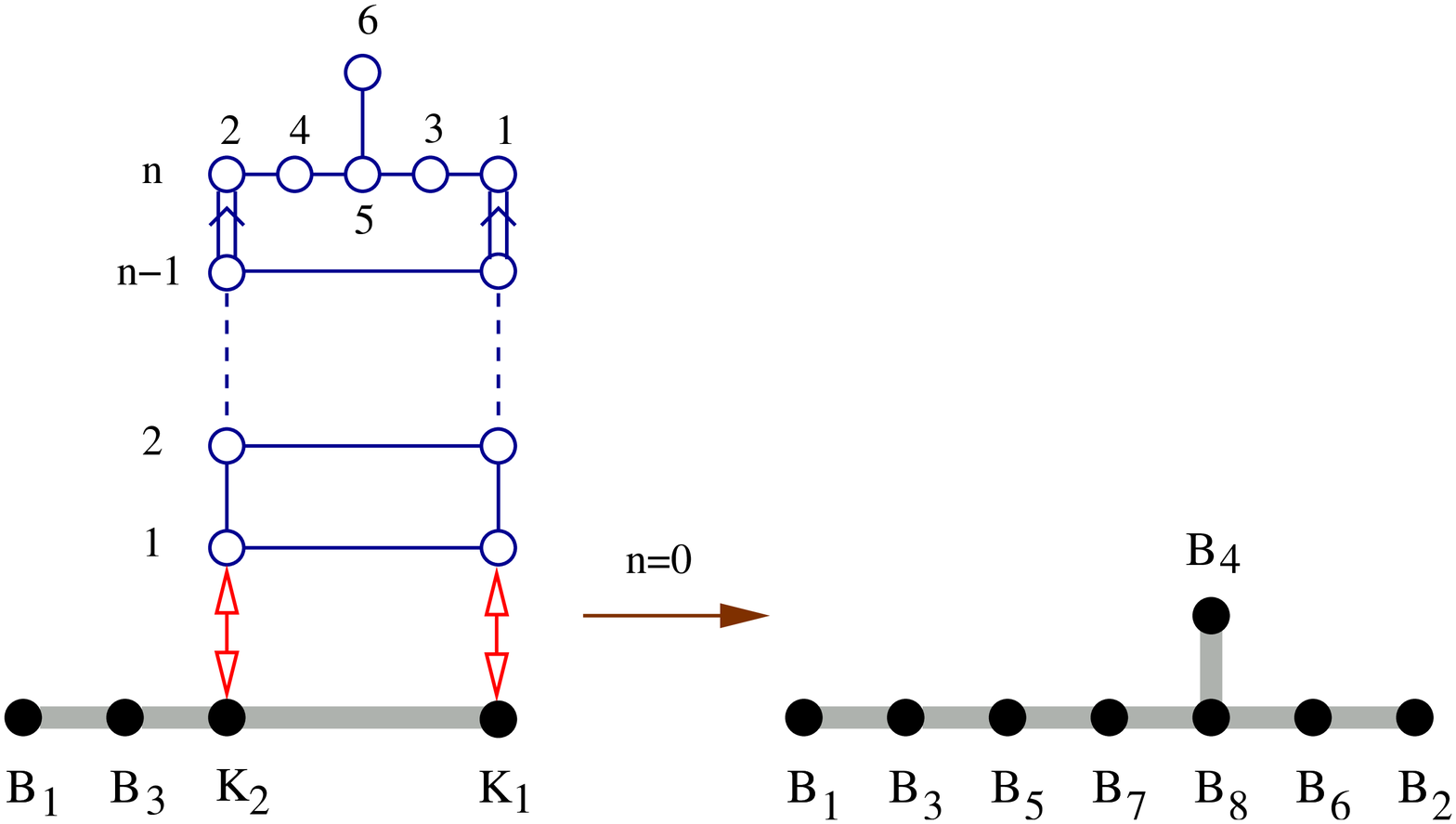}}
\caption{ \protect{\footnotesize
Diagrammatic representation for the $\CM_{4n+3,4n+4}+\phi_{12}$
TBA systems}}
\label{figtba}
\end{center}
\end{figure}%
The full sets of TBA equations and Y-systems are given
in appendix~\ref{a12}.

\begin{figure}[ht]%
\begin{center}
\resizebox{0.38\linewidth}{!}%
{\includegraphics{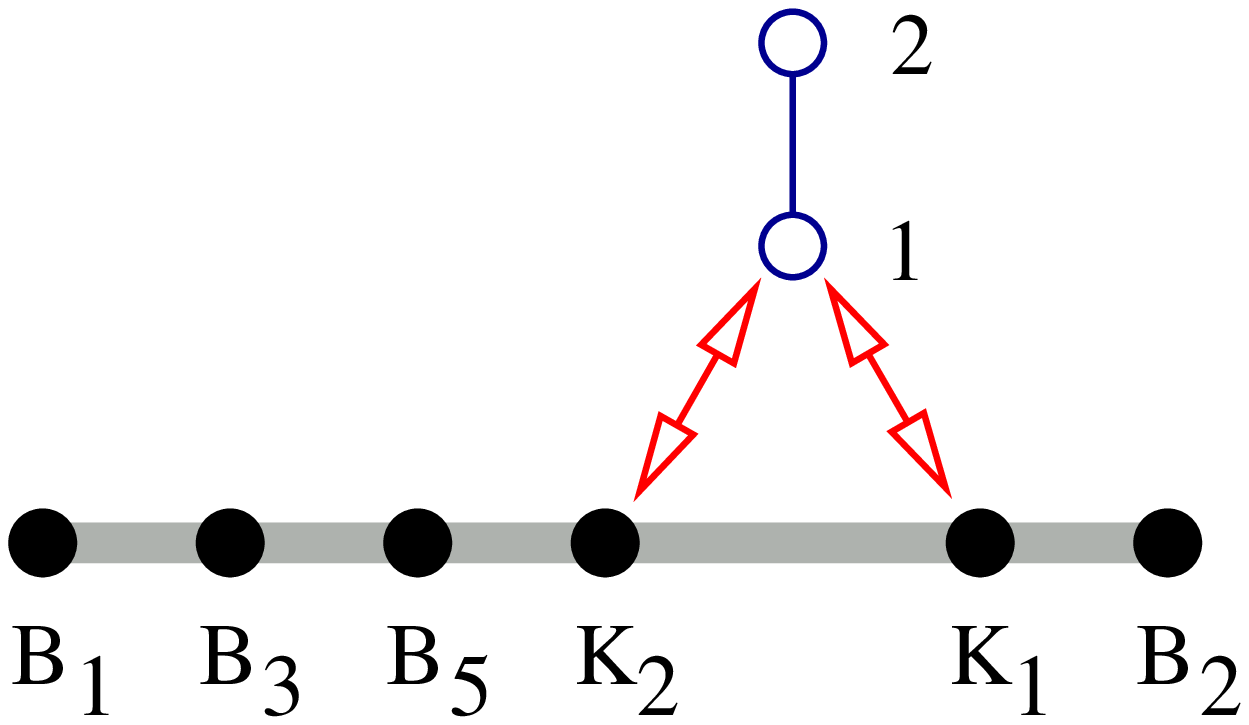}}
\caption{ \protect{\footnotesize
Diagrammatic representation for the $\CM_{5,6}+\phi_{12}$
TBA system}}
\label{figtba0}
\end{center}
\end{figure}

We shall illustrate these proposals with
the simplest new system, corresponding to
the model    $\CM_{5,6}+\phi_{12}$.
For this case, $\lambda=\fract{21}{10}$ and so, referring back to 
\S \ref{recap},
we expect the spectrum to contain six particles, 
$B_1$, $B_2$, $B_3$, $B_5$, $K_1$ and $K_2$.
In addition, we conjecture that
two magnonic pseudoenergies, $\ep^{(1)}$ and $\ep^{(2)}$, should be
introduced to deal with the non-diagonal nature of the kink scattering.
The equations are:
\bea
\ep_{b}(\theta) &=& R m_b \cosh \theta  -  \sum_{c}
\Phi_{bc}*L_{c}(\theta)-(\delta_{b,K_1}+\delta_{b,K_2})
\phi*L^{(1)}(\theta) \nonumber \\
\ep^{(1)}(\theta) &=& -
\phi*(L^{(2)}(\theta)+L_{K_1}(\theta)+L_{K_2}(\theta))  \\
\ep^{(2)}(\theta) &=& - \phi*L^{(1)}(\theta)
\nonumber
\eea
with $b,c \in \{B_1,B_2,B_3,B_5,K_1,K_2 \}$. The kernels are:
\eq
\Phi_{bb'}(\theta)= -i \ddt  \ln {\cal S}_{bb'}(\theta)
{}~~~~,~~~~ \Phi_{kb}(\theta)=\Phi_{bk}(\theta)=
-i \ddt  \ln {\cal S}_{kb}(\theta)
\en
where $b, b' \in \{ B_1,B_2, B_3 ,B_5 \}$, $k \in \{K_1 ,K_2 \}$ and
\eq
\Phi_{K_1 K_1}(\theta)= \phi*\Phi_{K_1 B_2}(\theta)
{}~~,~~
\Phi_{K_2 K_2}(\theta)= \phi*\Phi_{K_2 B_5}(\theta) \, ,
\en
\eq
\Phi_{K_1 K_2}(\theta)= \Phi_{K_2 K_1}(\theta)=
\phi*\Phi_{B_5 K_1}(\theta)
{}~~,~~\phi(\theta) = \frac{21}{\cosh(21\,\theta)}\,.
\en
A diagrammatic representation is shown in figure \ref{figtba0}.

Again the $A_2$-type pseudoparticle sector reflects the relationship
between this theory and the  hard hexagon model~\cite{BP}.

\resection{Analytical and numerical checks}
\label{checks}
This section gathers together some
evidence, analytical and numerical, in favour of the conjectured TBA
systems.
The value of the UV central charge is obtained through standard
manipulations of the TBA equations~\cite{AlZam1,KM}, which allow
$c_{\mbox{\tiny UV}} \equiv c(0)$ to be expressed as a sum of
Rogers dilogarithm functions. The
arguments of the dilogarithms  are expressed in terms of the set of
stationary  solutions of the TBA at
$r=0$ and $r=\infty$.
We did not  recognise these sums among any of
the standard relations for the Rogers function (see for example~\cite{Kir}),
so we  relied on  numerical work
to check  that, up to $p=13$ and 
to about $15$ digit accuracy, we have
the expected result
\eq
c_{\mbox{\tiny UV}} \equiv c(0)=1-{6 \over p (p+1)}~.
\label{cuv}
\en

Next, we discuss the corrections to $c(0)$ which make up
the UV expansion of $c(r)$. 
A key idea is to find a set of functional relations satisfied by the
exponentiated pseudoenergies, called a Y-system. These generally imply
a periodicity property for the pseudoenergies under a certain
imaginary shift in $\theta$, which in turn can be used to deduce
information about the $r$-dependence of $c(r)$.
This phenomenon was first noticed by   
Al.B.~Zamolodchikov in~\cite{Zam2}, for Y-systems related to the   
TBA equations
described in ~\cite{KM} and encoded by a single ADE Dynkin diagram. 
(The periodicity for the A and D cases was subsequently proved
in  \cite{GT,FSz,CGT}, while a general proof also
covering the E cases was only given recently, in 
\cite{fomin}.)

The Y-systems pertinent to $\phi_{21}$ perturbations are collected
together in appendix A.
We did not attempt to {\em prove} their periodicity 
properties, but numerically verified
the following result:
\eq
Y_a(\te + i \pi P)=Y_a(\te),~~~~  P= {8 p \over 6 (p-1)}
\en

Following the arguments of \cite{Zam2}, periodicity 
implies an
expansion for $c(r)$ in powers of $r^{4/P}$ and fixes  the 
dimension of an odd perturbing operator to
\eq
\bar{\Delta}=\Delta=1-{1 \over P} = {p+3 \over 4 p}\,.
\en
This value coincides with the conformal dimension of 
the operator $\phi_{21}$.

A numerical
study of $c(r)$ for the $\phi_{21}$ TBA systems
also indicates the presence of
a single irregular  `anti-bulk' term:
\eq
c(r)=c- {\cal B}\,r^2 + \sum_{n=1}^{\infty} c_n r^{ n (3p-3)/p} \, .
\label{exp}
\en
This term is related to the bulk free energy of (\ref{eform}) by
${\cal B}=-6 {\cal E}/(\pi m_K^2)$\,;
it can be  calculated analytically from the TBA by
adapting   the method  of~\cite{AlZam1,KM}.
For $\te \rightarrow -\infty$ we have
\eq
\Phi_{ab}(\te) \sim - \Phi_{BB}^{(1)} \lf({ m_a \over m_B} \right)
 \lf({ m_b \over m_B} \right)  e^{\te} + \dots ~,~~~(a,b \in \{B,K
\})\,,
\en
with
\eq
\Phi_{BB}^{(1)}= 4 \lf (\sin(\frac{2 \pi}{3})+
\sin(\frac{\pi}{\lambda})+ 
\sin(\frac{\pi}{\lambda} -\frac{\pi}{3}) \ri)\,.
\en
Arguing as in \cite{AlZam1,KM}, we therefore have
\eq
{\cal B}={3 \over  \pi \Phi_{BB}^{(1)} } \lf({m_B \over m_K} \ri)^2
= \frac{\sqrt{3}\,}{2\pi}
  \,\frac{\sin( {\pi \over 3}{p +1 \over p-1})
   }{\sin( {2 \pi \over 3} {p \over p-1} )}~~.
\en
This  matches  the result given in~\cite{Fat}.

Next, the coefficients $c_n$ in the expansion (\ref{exp}) can be 
checked against the results of conformal perturbation theory (CPT).
This predicts an expansion 
\eq
c_{\scriptstyle\rm{CPT}}(r) \equiv  -{ 6 R \over \pi}  E(\tau,R)=
{  c } +  
\sum_{m=2}^{\infty} B_m t^m~~,~~t= - \pi \tau (R/2 \pi)^{2-2 \Delta}\, ,
\label{expcft}
\en
which should match (\ref{exp}) save for the absence of the anti-bulk term 
$-{\cal B}\,r^2$.
The CPT coefficients $B_m$ are given in terms of connected correlation
functions on the plane. For unitary cases,
\eq
B_m=  {24 \over \pi^{m-1} m!}
 \int \langle \ep(1,1) \ep(z_1, \bar{z}_1)
\dots \ep(z_{m-1}, \bar{z}_{m-1} ) \rangle_c \prod_{k=1}^{m-1}
{d^2 z_k \over (z_k \bar{z}_k)^{1-\Delta}}~,
\en
with conformal invariance fixing
\eq
B_2= 12  {\gamma^2(\Delta) \over \gamma(2 \Delta)}~~,~~
\gamma(x)= {\Gamma(x) \over \Gamma(1-x)} ~ .
\en

The high-low temperature duality symmetry  of the model is reflected in the
fact that correlators with an odd  number of $\ep$ operators
vanish identically
($B_{m}=0$ for odd $m$).
Comparing (\ref{exp}) with (\ref{expcft}) confirms that our effective central
charge
behaves as that of a minimal conformal field theory
perturbed by  $\phi_{21}$
($
\bar{\Delta}_{21}= \Delta_{21}={ p+3 \over 4 p}
$).
On dimensional grounds there must be a relation between $\tau$ and $m_K$
(the  mass of the kink $K \equiv K_1$
appearing in (\ref{TBA0}) and (\ref{ceffe})) of the form
$\tau =\kappa m_K^{2-2 \Delta}$,
where $\kappa$ is a dimensionless constant whose exact value was
calculated by other methods in~\cite{Fat}.
Table~2 compares numerical results from the TBA with
the exact formula.
%
\begin{table}[htb]
\begin{center}
\vskip 3mm
\begin{tabular}{c  | l | l   }
Model &~~~$\kappa^2$~~(TBA)
~~~~ &~~~~~$\kappa^2$~~(Exact)
\\ \hline
\rule[0.2cm]{0cm}{2mm}
$\CM_{6,7}+\phi_{21}$   & 0.0259899061737        & 0.02598990617395    \\
$\CM_{7,8}+\phi_{21}$   & 0.02506361950681       & 0.02506361950686    \\
$\CM_{8,9}+\phi_{21}$   & 0.0242806346097        & 0.02428063460990    \\
$\CM_{9,10}+\phi_{21}$  & 0.02362954340274       & 0.02362954340277    \\

\end{tabular}
\caption{TBA data versus the  exact results of~{\protect \cite{Fat}}}
\end{center}
\vspace{-0.2cm}
\end{table}

Similar checks can be performed for the $\phi_{12}$ perturbations.
The relevant
functional relations are given in appendix B,
and we verified
that in these cases the periodicities implied by the Y-systems are
$P= \frac{8(p+1)}{6(p+2)}\,$. The resulting conformal 
dimensions  $(p-2)/(4p+4)$ match those of the
$\phi_{12}$ operator.
  
Finally, we turn to the large-$R$ behaviour, restricting attention to
the $\phi_{21}$ systems for brevity. 
The system~(\ref{TBA0}) implies the following asymptotic
for the ground-state energy $E(m_K,R)$:
\eq
E(m_K,R)= - { \pi \over 6 R} c(r) \sim {\sigma m_K \pi} K_1(r) \, ,
\en
where
$\sigma$ can be evaluated in terms of the quantities $\Upsilon_i^{(1)}$
in appendix C of~\cite{DTT}. Setting  $\eta=\pi/(p+1)$ and
$n=\mbox{Int}[(p-2)/4]$ we have
\bea
\sigma &=&
\left(1 + { 1 \over \Upsilon^{(1)}_n } \right )^{\! 1/2}\,
\prod_{i=1}^{n-1} \lf(1+  {1 \over \Upsilon_i^{(1)}} \ri) \nn \\
&=&
\left(1 + { \sin(2 \eta) \sin((2 n+2) \eta)
\over \sin^2(n \eta )} \right )^{\! 1/2}\,
\prod_{i=1}^{n-1} \lf(1+  {\sin( 2 \eta)  \sin(\eta) \over
\sin((i+3) \eta) \sin(i \eta)} \ri) \nn \\
&=&4 \cos^2 \eta -1= q-1
\eea
for cases A,B and C and
\bea
\sigma &=&
\left(1 + { 1 \over \Upsilon_n^{(1)} } \right )^{\! 1/2}\,
 \left(1 + { 1 \over \Upsilon_{n+1}^{(1)} } \right )^{\! 1/2}\,
\prod_{i=1}^{n-1} \lf(1+  {1 \over \Upsilon_i^{(1)}} \ri) \nn \\
&=&
\left(1 + {2  \sin(\eta) \cos (( n+1) \eta)
\over \sin(n \eta )} \right )
\prod_{i=1}^{n-1} \lf(1+  {\sin( 2 \eta)  \sin(\eta) \over
\sin((i+3) \eta) \sin(i \eta)} \ri) \nn \\
&=&4 \cos^2 \eta -1= q-1
\eea
for case D.
The value $\sigma=q-1$  matches the prediction of an
argument given in \cite{DTT}, extending the instanton ideas
of  \cite{fir0,fir1}, that
$\sigma$ should be the  Perron-Frobenius  eigenvalue
of the (in this case $S_q$-symmetric)
incidence matrix $I^{(q)}_{ab}$ encoding the number
of light kinks which join each pair of vacua. Note that the argument is a
little formal here, since in general the number of vacua is not an integer.
For these unitary cases this worry can be averted by shifting to
the RSOS approach of \cite{Sm};
either way, the TBA systems meet our expectations.

\resection{Results from the nonlinear integral equation approach}
\label{NLIEQ}
The TBA is not the only exact technique on the market for the study of
finite-size effects. Another class of methods is known as
the nonlinear integral equation (NLIE) approach. These equations
generally
depend smoothly on their parameters, and so are in some
senses better-suited to a description of the $q$-state Potts
models. The downside is that the full particle content is encoded
in a much more implicit way, and so they have less to say on the
issues of bootstrap closure discussed in \cite{pottsI}.

\subsection{Massive flows}
\label{mnliesec}
A non-linear integral equation describing
general $\phi_{12}$, $\phi_{21}$ and $\phi_{15}$ perturbations within
a unified framework
was proposed in~\cite{DT}, 
similar in spirit to the equations related to
$\phi_{13}$ perturbations introduced in
\cite{NLIE,Destri}. In contrast to the  TBA  equations discussed
earlier, in \cite{DT} the set of pseudoenergies
$\{\,\varepsilon_a\,\}$ is traded for a single unknown function
$f(\theta)$, which solves the following equation:
\eq
f(\te)=
 i \pi \alpha  -i r\sinh\theta
+\!\int_{{\cal
C}_1}\!\!\varphi(\te{-}\te')\ln(1{+}e^{f(\te')})\,d\te'
-\!\int_{{\cal
C}_2}\!\!\varphi(\te{-}\te')\ln(1{+}e^{-f(\te')})\,d\te'\,,
\label{pnlie}
\en
where $r=m_KR$ with $m_K$ the mass of the fundamental kink.
Given $f(\theta)$, the effective central charge can be evaluated as
\eq
c_{\rm eff}(r)= {3 i r \over \pi^2}
 \lf(\int_{{\cal
C}_1} \sinh\theta\, \ln(1{+}e^{f(\te)})\,d\te-
\int_{{\cal
C}_2}\sinh\theta\,\ln(1{+}e^{-f(\te)})\,d\te \ri)\,.
\label{mnlie}
\en
The  contours ${\cal C}_1$
and ${\cal C}_2$ run from $-\infty$ to $+\infty$, just below and just
above the real $\te$-axis, and the kernel $\varphi(\theta)$ depends
continuously on the parameter 
$\lambda$ as
\eq
\varphi(\te)=-\int_{-\infty}^{\infty}
\frac{e^{i k\te} \sinh(\fr{\pi}{3}k) \cosh( \fr{\pi}{6} 
(1{-}\fr{3}{\lambda})k)}
{\cosh( \fr{\pi}{2} k) \sinh(\fr{\pi}{2\lambda}k)}%
\frac{d k}{2\pi}
=-\frac{i}{2\pi}\frac{d}{d\theta}\log\,S(\theta)~.
\label{krnl}
\en
As before, $\sqrt{q}=2\sin(\fr{\pi}{3}\lambda)$\,, while $S(\theta)$
is the scalar part of the fundamental kink-kink scattering amplitudes as 
given in \cite{pottsI}.
(A scale-invariant version of this system had previously appeared in
the context of lattice models; see \cite{WBN}.)
In \cite{DT}, we showed how to use (\ref{pnlie}) to describe 
perturbations of
unitary and non-unitary minimal models, tuning  the twist parameter
$\alpha$ so as to account for the fact that the
ground state may be generated by a negative-dimension field.
For current purposes, we are more interested the $q$-state Potts
models and their tricritical variants, for which, as discussed
earlier,
the ground state {\em always} comes from the identity operator. 
For these theories, 
$\alpha$  should be related to $\lambda$ as
\eq
\alpha=\frac{2\lambda}{3}-1\,.
\en
The reader should note that the NLIE~(\ref{mnlie}) is  simpler and, 
not being restricted to special points,  of wider applicability 
than  the TBA equations proposed in \S \ref{con21} and \S \ref{con12}
above. It is also much more appropriate for discussions of the $q\to
1$ limit relevant to percolation.
However,  it does not encode, in any direct way,   
information about the bound-state content of the associated 
field theory. 

In Table~3 we  compare results from the NLIE and TBA
approaches. The agreement is very good, ranging from 10
to 14  significant digits.
\begin{table}[htb]
\begin{center}
\vskip 3mm
\begin{tabular}{c  c | l l   }
Model & \rule[-0.2cm]{0cm}{2mm}~$r$~~~&~~~~~TBA  ~ &~~~~~NLIE
\\ \hline
\rule[0.2cm]{0cm}{2mm}
$\CM_{10,11}+\phi_{21}$ & 0.2    &  0.9328551607515   & 0.9328551607514  \\
            & 0.3    &  0.9185246910885   & 0.9185246910884  \\ \cline{1-4}
\rule[0.2cm]{0cm}{2mm}
$\CM_{11,12}+\phi_{21}$ & 0.2    &  0.9421554960427   & 0.9421554960424  \\
            & 0.3    &  0.9279952931412   & 0.9279952931410  \\ \cline{1-4}
\rule[-0.2cm]{0cm}{2mm}
$\CM_{12,13}+\phi_{21}$ & 0.2    &  0.949321521251    & 0.949321521253  \\
            & 0.3    &  0.935305379646    & 0.935305379648 \\ \cline{1-4}
\rule[0.2cm]{0cm}{2mm}
$\CM_{13,14}+\phi_{21}$ & 0.2    &  0.954961250612733   &  0.954961250612730 \\
            & 0.3    &  0.941068318324234   &  0.941068318324237 \\ \cline{1-4}
\rule[0.2cm]{0cm}{2mm}
$\CM_{5,6}+\phi_{12}$ & 0.2  &   0.79241547642     & 0.79241547640 \\
            & 0.3  &   0.78313881742     & 0.78313881742 \\
\end{tabular}
\caption{\label{selection} Comparison between TBA  and the NLIE proposed
in~~{\protect \cite{DT}}}
\end{center}
\vspace{-0.2cm}
\end{table}
\subsection{Massless flows
from tricritical  to critical  models}%
\label{mless}
\begin{figure}[ht]%
\begin{center}
\resizebox{0.8\linewidth}{!}%
{\includegraphics{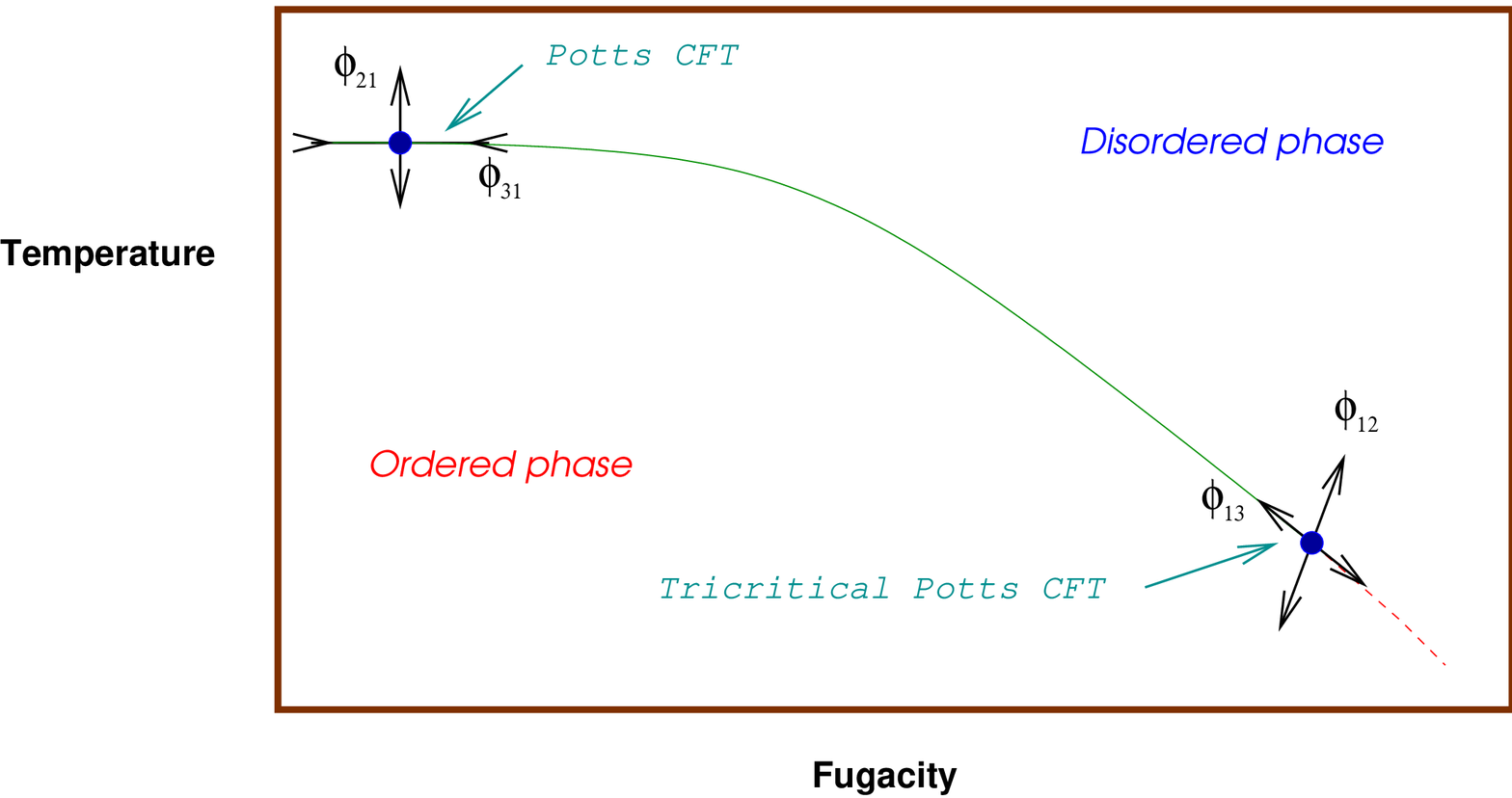}}
\caption{ \protect{\footnotesize
The phase diagram}}
\label{phasediag}
\end{center}
\end{figure}%
In addition to the perturbations discussed so far,
the concentration of vacancies in a tricritical Potts model can be
adjusted so as to drive the theory either onto a line of first-order
transitions, or else down via a massless flow to the corresponding
critical model. The relevant operator turns out to be $\phi_{13}$,
which is always in the spectrum of the tricritical models.
The full picture is illustrated in figure \ref{phasediag}; 
it is most convenient to use the parameter $\xi$ for these flows,
related to $q$ by
$\sqrt{q}=2\cos\pi/(\xi{+}1)$ with $\xi\in [1,\infty]$. The
massless flow is then
\eq
c_{\rm tricrit}(\xi)=
1-\frac{6}{(\xi+1)(\xi+2)}~~~\to~~~
 c_{\rm crit}(\xi)=
1-\frac{6}{\xi(\xi+1)}~.
\en
For $\xi=p\in\Z$, this is the well-known massless $\phi_{13}$ flow
from $\CM_{p{+}1,p{+}2}$ to $\CM_{p,p{+}1}$~\cite{Zamolodchikov:ti,%
Ludwig:gs}
for which equations of TBA type were proposed in
\cite{Zamolodchikov:vx}. 
To describe the flows at general values of $\xi$,
we instead adapt an equation proposed by
Zamolodchikov in \cite{Zam3} for the
study of flows in the imaginary-coupled 
sine-Gordon model from $c=1$ to $c=1$. 
In~\cite{DDT}, it was shown that the twist parameters in 
Zamolodchikov's equation could be chosen so as to
describe flows between both unitary and non-unitary minimal models,
revealing a striking non-monotonicity of $c(r)$ at
intermediate scales in the non-unitary cases.
Here, we shall propose a further modification to capture the
general flows between tricritical and critical Potts 
models. 
As in \cite{Zam3},
introduce two analytic functions
$f_R(\te)$ and $f_L(\te)$, and couple them together via
\bea
f_{R}(\te)&=& - i \frac{r}{2}e^{\te}  + i   \pi \alpha' 
\nn \\
&&+ \int_ {{\cal C}_1}\!\phi(\te-\te')
\ln(1+e^{f_{R}(\te')}) ~ d\te' 
- \int_ {{\cal C}_2}\! \phi(\te-\te')
\ln(1+e^{-f_{R}(\te')} ) ~ d\te'  
\nn \\
&&+  \int_ {{\cal C}_1}\!\chi(\te-\te')
\ln(1+e^{-f_{L}(\te')}) ~ d\te' 
- \int_ {{\cal C}_2}\! \chi(\te-\te')
\ln(1+e^{f_{L}(\te')} ) ~ d\te'\quad
\nn
\\[2pt]
f_{L}(\te)&=&-i \frac{r}{2} e^{-\te} - i   \pi \alpha' 
\nn \\
&& + \int_ {{\cal C}_2}\!\phi(\te-\te')
\ln(1+e^{f_{L}(\te')}) ~ d\te' 
- \int_{{\cal C}_1}\! \phi(\te-\te')
\ln(1+e^{-f_{L}(\te')} ) ~ d\te'  
\nn \\
&& + \int_{{\cal C}_2}\!\chi(\te-\te')
\ln(1+e^{-f_{R}(\te')}) ~ d\te' 
- \int_ {{\cal C}_1}\! \chi(\te-\te')
\ln(1+e^{f_{R}(\te')} ) ~ d\te'\quad
\label{zmikRL}	
\eea
where 
\eq
\phi(\te)=\int_{-\infty}^{\infty}
\frac{e^{i k\te} \sinh( k (\xi-1)\fr{\pi}{2})} 
{2 \cosh( \fr{\pi}{2} k) \sinh( k \fr{\pi}{2}\xi )}%
\frac{d k}{2\pi}~,
\label{nkrnl}
\en
\eq
\chi(\te)={-}\int_{-\infty}^{\infty}%
\frac{e^{i k\te}\sinh(\fract{ \pi k}{2} )}%
{2 \cosh(\fract{\pi k}{2})  \sinh(\fract{\pi k }{2}\xi)  }%
\frac{d k}{2\pi}~.
\label{chikrnl}
\en
(In \cite{Zam3}\,, the kernels were given in terms of a parameter
$p\equiv \xi{+}1$\,.)
We have
$\sqrt{q}=2\cos\pi/(\xi{+}1)$ with $\xi\in [1,\infty]$, and,
for the interpolating Potts flows, $\alpha'$ must be
chosen as
\eq
\alpha'= {1 \over \xi}~.
\en
In terms of the solutions to these equations, the effective central charge
is
\bea
c_{\rm eff}(r)&=& \frac{3 i r }{2 \pi^2} \biggl[ \int_{{\cal
C}_1}\!\!e^{\te} \ln(1{+}e^{f_R(\te)})\,d\te
-\int_{{\cal
C}_2}\!\!e^{\te} \ln(1{+}e^{-f_R(\te)})\,d\te
\nn\\[2pt]
&&{}~~~{}+\int_{{\cal
C}_2}\!\!e^{-\te} \ln(1{+}e^{f_L(\te)})\,d\te
-\int_{{\cal
C}_1}\!\!e^{-\te} \ln(1{+}e^{-f_L(\te)})\,d\te
 \biggr]~.
 \label{zmnliec}
\eea
\begin{figure}[ht]
\[
\begin{array}{c}
\includegraphics[width=0.8\linewidth]{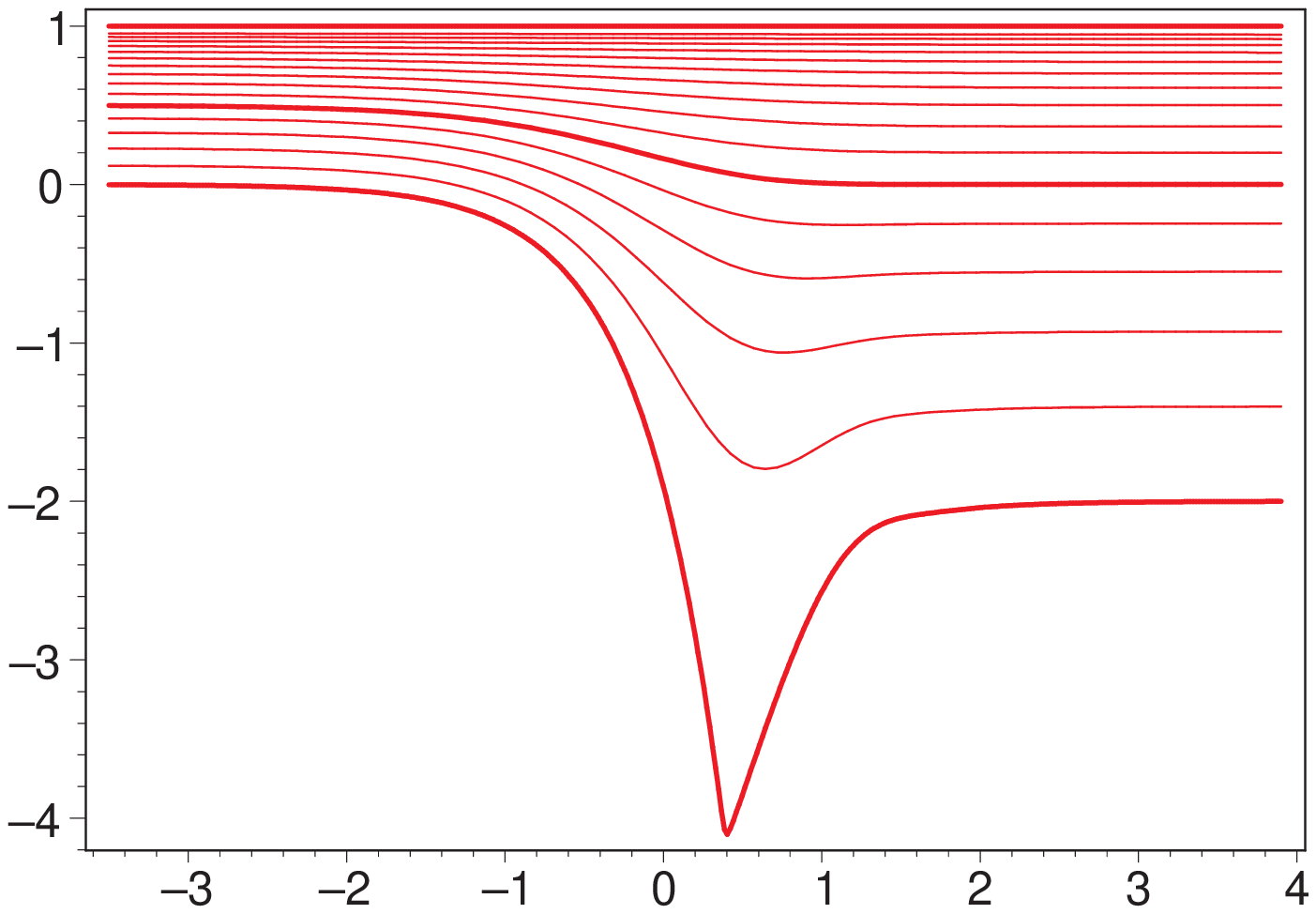}
\end{array}
\]
\caption{Flows of $c(r)$ as obtained from the NLIE
(\ref{zmikRL}), plotted against $\log(r/2)$.
Highlighted are the flows 
from $c=1/2$ to $c=0$
and
from $c=0$ to $c=-2$.
}
\label{cflows}
\end{figure}
Figure \ref{cflows} shows our numerical results. 
As long as the IR 
destination has a positive central charge, the flows are monotonic, even
at points corresponding to non-unitary minimal models. This is in
sharp contrast to the behaviour of the $\phi_{13}$-perturbed 
minimal models themselves, for which a number of non-monotonic flows of
$c(r)$ were exhibited in \cite{DDT}.
To some extent, this reflects
the already-mentioned fact that the ground states of Potts and minimal
models do not coincide away from unitary points, but it is nonetheless
surprising that
the switch to the identity vacuum state manages to eliminate 
{\em all} of the non-monotonicity while $c_{IR}$ remains positive. 

\begin{figure}[ht]
\[
\begin{array}{c}
\includegraphics[angle=270,width=0.7\linewidth]{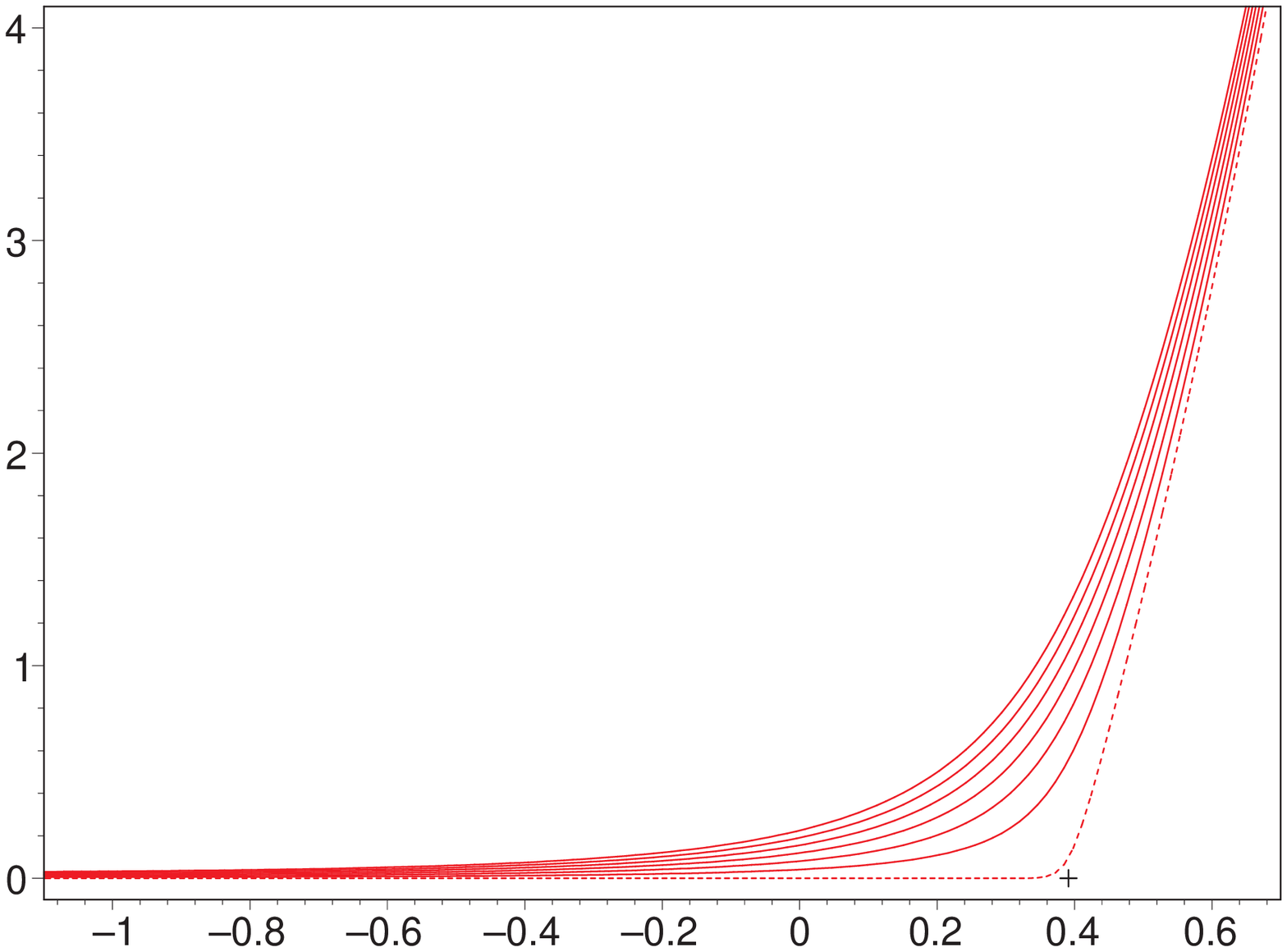}
\end{array}
\]
\caption{Flows of $c_{\mbox{\tiny CPT}}(r)$
 for $\xi=1.03, 1.025,
\dots 1.005$ as obtained from the NLIE, plotted against $\log(r/2)$.
Also shown is the
extrapolated curve for
$\xi=1$ (dotted), and the location of the level-crossing
predicted in \cite{Fendley:1993wq}.}
\label{cflows2}
\end{figure}

The other notable feature of figure~\ref{cflows} is
the behaviour of the flow from $c=0$ to $c=-2$, which exhibits
a cusp, suggestive of a level-crossing, at an intermediate 
length-scale. In fact, this curve deserves a
second glance for another reason: in \cite{Fendley:1993wq},
Fendley, Saleur and Zamolodchikov pointed out that the effective
central charge along this particular flow should be protected by
supersymmetry, at least for small enough values of $r$, and hence should be
identically zero. They also, by an indirect argument,
predicted that this picture would be changed by
a level-crossing at $r\approx 2.95708396$.
The apparent contradiction of their first claim with our results has
a neat resolution: one has to remember that, just as for the
TBA equations discussed earlier, the quantity $c(r)$
produced by the NLIE includes an `anti-bulk' piece, non-analytic
in the coupling constant, which ensures that at large values of
$r$, $c(r)$ tends to a constant. Thus, just as in equations (\ref{exp})
and (\ref{expcft}) above, we expect
\eq
c(r)=c_{\mbox{\tiny CPT}}(r) - {\cal B}\,r^2 
\en
where $c_{\mbox{\tiny CPT}}(r)$ is the physical, unsubtracted
quantity, while $c(r)$ is the quantity which is found directly from the NLIE.
The `$\mbox{Bulk}$' constant ${\cal B}$ was found exactly in
\cite{Fendley:1993xa} by
considering a related theory in a magnetic field; it can also be obtained
directly from equations (\ref{zmikRL}) and (\ref{zmnliec}), 
using an argument described for a similar
equation in \cite{DTT}. Either way, the answer is
\eq
{\cal B} =-\frac{3}{2\pi}\frac{1}{\cos(\frac{\pi}{2}(\xi{+}1))}~.
\en
Specialising to $\xi=1$, we see that before the bottom curve of
figure~\ref{cflows} is compared with the proposals of \cite{Fendley:1993wq},
the quantity $\frac{3}{2\pi}r^2$ should be added to it.
Numerically, it is hard to do this directly due to instabilities in
the equations near $\xi=1$, so in
figure~\ref{cflows2} we show a sequence of plots of the
appropriately-adjusted functions $c_{\mbox{\tiny CPT}}(r)=c(r)+{\cal
B}(\xi)r^2$, 
for $\xi=1.03$ down to
$\xi=1.005$, together with an extrapolated curve for $\xi=1$. (For the
extrapolation, numerical data down to $\xi=1.001$ was used.)
Not only does the revised curve meet general supersymmetric expectations;
the point at which supersymmetry was
predicted in \cite{Fendley:1993wq} to be  
spontaneously broken via a level-crossing is also reproduced.
We suspect that there is more to be said on this matter, and the nice
agreement between our results and those of \cite{Fendley:1993wq}
certainly deserves to be understood at a deeper and more analytical
level. For now, we leave this
for future investigations.

\resection{Conclusions}
\label{conc}

In the past decade there has been a common belief that
$A_n$--, $D_n$-- and  $E_n$-- related  integrable   
systems should exhibit a certain uniformity of structure.
As we remarked in \cite{pottsI}, 
the bulk, S-matrix, description of 
the real-coupling simply-laced
affine Toda field theories (and their minimal variants)
illustrate this philosophy well:
all the scattering data can be encoded using
the root systems of the associated Lie algebras \cite{Patrick}. 
A first sign of a breakdown in this pattern in other contexts came
with the attempt to make the coupling imaginary in these same Toda
models: the $A_1$ Toda Hamiltonian 
remains Hermitian, 
turning into the sine-Gordon Hamiltonian, 
while all the others become 
complex. Despite this crucial difference, it was initially thought
that
the relatively simple $A_1$-related
sine-Gordon spectrum might be the first instance of some unified
description valid for all Lie algebras~\cite{Ho,Gandenberger:1995gu}.
Smirnov's
work~\cite{Sm} had already suggested that matters were likely to be much 
more complicated, and subsequent work, in particular
\cite{SalKau} and \cite{pottsI}, 
has confirmed this expectation, to the
extent that there is still no complete understanding of the bulk
scattering
theory associated with any imaginary-coupled Toda theory beyond the
$A_1$ case.

On side of lattice models and the Bethe ansatz, a longstanding
hope  has been to find
generalisations of the
construction of
Takahashi and Suzuki~\cite{TS}, who many years ago found the
correct `string hypothesis' describing the
Bethe ansatz solutions relevant to
the finite volume corrections-to-scaling  of the  $XXZ$ model
at arbitrary values of the anisotropy.
The form taken by the
Bethe ansatz equations for this model are the $A_1$ cases of
a general set of ADE-related Bethe ansatz equations, given just in terms
of Lie algebra parameters in \cite{BA}.
The symmetry of equations in nature is often broken  by their  solutions, 
and for  three decades all attempts to generalise  the construction by 
Takahashi and Suzuki to the other lattice models 
at arbitrary values of the coupling constant have failed.
(The difficulties even for the $SU(3)$ case were
highlighted in \cite{SalKau}.)

Building on a relationship with the $q$-state Potts models,
in this paper we have conjectured and checked TBA equations related to
$a_2^{(2)}$  at points related to the  unitary series of minimal models
perturbed by their
$\phi_{21}$ and  $\phi_{12}$ operators. Although we have proposed    
a continued-fraction decomposition, {\it a la} Takahashi-Suzuki,  
that  governs 
these  systems,  
 we are currently unable to  extend the   analysis to the  other 
rational points. 
Yet, the models described, 
being unitary, are physically  the most interesting, and we hope 
that our work  will motivate a more rigorous derivation of 
these equations. 
In this respect we feel that  the functional approach described
in, for example, \cite{JKS,KSS},
and already applied by J.~Suzuki
to particular examples of $\phi_{12}$ perturbations in \cite{SuzE8,SuzEn}, 
is likely to be the most effective. 
For the Potts models, the TBA equations that we have been able to
find serve to confirm the mass
spectrum found by bootstrap techniques in \cite{pottsI}, at least at
the points $\sqrt{q}=2\cos(\pi/(p{+}1))$.

Finally, we mention two further open problems. 
First, it would be nice, along the lines of \cite{ex1,ex2,ex2a,ex3},
to adapt both the NLIE and the TBA equations to describe excited 
states.
(The recent  paper \cite{exnew} discusses TBA-like equations for
excited states in one particular $\phi_{21}$-related model.) 
In particular, we note that the ground-state NLIE of \S\ref{mnliesec}
works without problems even in the region $\lambda\in (\frac{9}{4},3]$
where we encountered problems closing the bootstrap in \cite{pottsI}.
Having  control over the full finite-size spectrum using 
the NLIE technique should  help us to
resolve this  open  question.
Second, we have introduced many new sets of TBA equations and
Y-systems in this paper. It seems important to elucidate the
associated character identities\footnote{See the introductory section
of \cite{BMP} for a nice review of this topic.}, 
to find the T-systems~\cite{KSS} and to
give to the T functions a spectral interpretation
in the spirit of  the ODE/IM correspondence \cite{DT}.
Relations with (generalised) quantum KdV 
equations~\cite{BLZa,FRS,BHK, FR} and with perturbed  
boundary conformal field theory~\cite{BLZa,BHK} might also be
explored.

\bigskip

\noindent
{\bf Acknowledgements -- }
We would like to thank
John Cardy,
Aldo Delfino,
Vladimir  Dotsenko,
Clare Dunning,
Davide Fioravanti,
Philippe di Francesco,
Paul Fendley,
Barry McCoy,
Bernard Nienhuis, 
Marco Picco,
Hubert Saleur,
Gabor Takacs,
Ole Warnaar,
Gerard Watts
and
Jean-Bernard Zuber
for useful discussions and correspondence.
We are especially greatful to Junji Suzuki for many detailed comments
and suggestions about the matters discussed in this paper.
In addition, PED thanks the Yukawa Institute for Theoretical
Physics and Shizuoka University
for hospitality.
Visits of PED to the
YITP and Shizuoka were partially
funded by a Daiwa-Adrian Prize and a Royal Society / JSPS
Anglo-Japanese Collaboration grant, title `Symmetries and
integrability'.
RT thanks the EPSRC for an Advanced Fellowship,
and AJP thanks the JSPS and the FAPESP for postdoctoral fellowships.
%
\appendix
%
\resection{TBA and Y-systems ($\phi_{21}$ perturbations)}
In this appendix we list the TBA equations and 
Y-systems for the $\CM_{p,p+1}+\phi_{21}$ models.
As mentioned in  \S\ref{con21} above, the continued-fraction 
expansion of
the parameter 
$
{6 \rho}=\lf( {p-5 \over p-1} \ri)~
$
allows  us to  distinguish  four families of systems:
\bea
A)~~~p=4n+2~~:~~ {6 \rho}&=& { 1 \over 1+ {1 \over {n-1 + {1 \over 4}}}} 
\nn \\
B)~~~p=4n+3~~:~~ {6 \rho}&=& { 1 \over 1+ {1 \over {n-1 + {1 \over 2}}}} 
\nn \\
C)~~~p=4n+4~~:~~ {6 \rho}&=& {1  \over 1+ {1 \over
        {n-1 + {1 \over {1+ {1 \over 3}} } } }}  
\label{fourc1a}
\\
D)~~~p=4n+5~~:~~ {6 \rho}&=& { 1 \over  1+ {1 \over n} } 
\nn
\eea
\subsection{The kernels}
\label{kersec}
The kernels functions related to the  diagonal S-matrix elements 
were defined  in  (\ref{diagk}), and we do not repeat them here.

There are also kernels associated with the magnonic pseudoenergies,
which were given in equation (\ref{magkerdefa}) of
the main text in terms of those of \cite{DTT}. 
For completeness, we give them here explicitly. 
By analogy with a notation used for the affine Toda
theories, define the blocks
\eq
(x)(\theta)=
{\sinh\bigl({\theta\over 2}+{i \pi \over 2} x\bigr)\over
        \sinh\bigl({\theta\over 2}-{i\pi \over 2} x\bigr)}\quad,
\quad\ubl{x}= \left(x- \fract{1}{h} \right) \left( x+ \fract{1}{h} \right)~.
\label{blocks} 
\en
Then
\eq
\hat{S}_{jk}=
 \prod^{ \atop j+k-1}_{|j-k|+1\atop {\rm
step~2}} \usa{ \fract{l}{h}} \usa{ 1- \fract{l}{h}} \, ,
\qquad (j,k=1\dots n{-}1)~,
\label{Sjk}
\en
and
\eq
\hat{S}_{kn}=\hat{S}_{k,n{+}1}=
(-1)^k\prod^{ \atop h/2+k-1}_{h/2-k+1 \atop {\rm step~2}}
\usa{ \fract{l}{h}} \, ,
\qquad (k=1\dots n{-}1)~,
\label{Sks}
\en
where $h=(p-5)/2$.
Some of the magnonic kernels can now be expressed in terms of the
logarithmic derivatives of the  $\hat{S}$ functions:
\eq
\tphi_{jk}(\theta)=-i \ddt \ln \hat{S}_{jk}(\theta/2\rho)\quad,\quad
\tpsi_{jk}(\theta)=-i \ddt \ln \hat{T}_{jk}(\theta/2 \rho)\,.
\label{kernels}
\en
In the definition of $\psi_{jk}$, the function $\hat{T}_{jk}$ is
obtained by replacing each
block $\ubl{x}$ in (\ref{Sjk})--(\ref{Sks}) by $(x)$.
In (\ref{kernels}), $j$ and $k$ run from $1$ to $n{-}1$, with $n$
depending on $p$ as in (\ref{fourc1a}).

For the D-type  models $h=2n$ and the definition can be extended immediately to
cover the remaining cases when both $j$ and $k$ take the values
$n$ or $n{+}1$.
The remaining functions needed are given by
\bea
\llap{n~{\rm even}}~:~~\hat{S}_{nn}=\hat{S}_{n{+}1,n{+}1}=
\prod^{\atop 2n-3}_{l=1\atop {\rm step~4}}\usa{\fract{l}{h}}&,&
\hat{S}_{n,n{+}1}=
\prod^{\atop 2n-1}_{l=3\atop {\rm step~4}}\usa{\fract{l}{h}}~,\nn\\
\llap{n~{\rm odd}}~:~~\hat{S}_{nn}=\hat{S}_{n{+}1,n{+}1}=
\prod^{\atop 2n-1}_{l=1\atop {\rm step~4}}\usa{\fract{l}{h}}&,&
\hat{S}_{n,n{+}1}=-\!\!%
\prod^{\atop 2n-3}_{l=3\atop {\rm step~4}}\usa{\fract{l}{h}}~.
\label{Sss} \nn
\eea

Otherwise, the
associated kernels are more elaborate. Define
an integer~$t$ by $p=4n{+}t+1$, and then set $g=3(p-1)/2$ and
\eq
\chi_{t}(\theta)=\frac{2g}{t\cosh{\fract{2g}{t}\theta}}~~.
\en
This function has the property that
\[
\chi_{t}{*}f(\theta{+}\fract{i \pi t}{4g})
+\chi_{t}{*}f(\theta{-}\fract{i \pi t}{4g})=f(\theta)~.
\]
Then the kernels in the TBA are given by
\eq
\tphi_{nn}(\theta)=\chi_{t}{*}\tphi_{n,n{-}1}(\theta)\quad,\quad
\tpsi_{nn}(\theta)=\chi_{t}{*}\tpsi_{n,n{-}1}(\theta)~.
\en
(For $p{\le} 6$, $\tphi_{n,n{-}1}$ and $\tpsi_{n,n{-}1}$ are not defined
and we set $\tphi_{nn}{=}\tpsi_{nn}{=}0$.)
Finally, depending on the continued-fraction expansion of $2/g$,
a number of extra kernels are needed. These  are
\bea
&& \tphi_1(\theta)={g\over\cosh g\theta} 
\nn \\
\tphi_2(\theta)={2g\over \cosh 2g\theta}  &&\qquad
\tphi_4(\theta)={8g\cosh{2g\over 3}\theta\over
 3(4\cosh^2{2g\over 3}\theta-3)} \nn\\
\tphi_3(\theta)={{2\over 3}g\over \cosh {2g\over 3}\theta} &&\qquad
\tphi_5(\theta)={8g\cosh{2g\over 3}\theta\over
\sqrt{3}(4\cosh^2{2g\over 3}\theta-1)} 
\label{magkernels}
\eea

This completes the definition of the kernels as needed for the TBA
equations. However, when deriving Y-systems from the TBA (as we did
for these $\phi_{21}$ cases) it is also important to know the precise
locations of the poles in the kernel functions. This is because the
derivation involves complex shifts in the rapidity $\theta$,
and care must be taken when poles cross the contours of convolution
integrals, as extra terms are generated which enter into the Y-systems
in a crucial way.

In most cases these pole locations are easily read off from the
explicit formulae,
but so far we have only given the kernel $\Phi_{KK}$ in terms of an
integral representation (\ref{ker}). Here we show how an alternative
product formula can be obtained.

Let us  define the function $\B(\te)$ such that
\eq
\B(\te+i \rho) \B(\te - i \rho)= {\cal S}_{KB}(\te) 
\label{boostB}
\en
and set
\eq
\Phi_{KK}= -i \ddt \ln 
\B(\te) ~.
\en
Inverting (\ref{boostB}) using Fourier transform we find
\eq
\B(\te)= \exp \left( \int_{-\infty}^{\infty} dk e^{i k \te}
{\tilde{f}_{KB}(k) \over 2 \cosh(k \rho \pi)} \right)
\label{bb}
\en
where   $\tilde{f}_{KB}(k)$ is the Fourier transform  of
\eq
f_{KB}(\te)=  \ln  {\cal S}_{KB}(\te) \,.
\en
In order to get a convergent  product
representation  for $\B(\te)$ let us set 
\eq
\sigma^{\pm}(\te,a)= { \Gamma(1 \mp i(\te/ \pi \mp i a)/2) \over
\Gamma(\mp i(\te/\pi  \pm i a)/2)}
{ \Gamma(1 \mp i(\te/ \pi \mp i (1-a))/2) \over
\Gamma(\mp i(\te/\pi  \pm i(1- a))/2)}
\en
and write
\eq
{\cal S}_{BK}={ \sigma^{+}(\te,a_1) \sigma^{+}(\te,a_2) \over
\sigma^{-}(\te,a_1)
\sigma^{-}(\te,a_2) }
\en
with $a_1=1/2+1/2 \lambda$ and  $a_2=1/6+1/2 \lambda$.
Note  that $\sigma^{+}(\te,a)$ ( $\sigma^{-}(\te,a)$)
 has  only a finite number of zeroes and poles in the upper (lower)
half plane  $\Im m(\te)>0$  ($\Im m(\te)<0$). 
Expanding the cosh function in the  denominator in (\ref{bb}) we can 
formally write
\bea
\B(\te) &=& \prod_{n=0}^{\infty}
 \exp \left( (-1)^n \int_{-\infty}^{\infty} dk
\, e^{i (\te + i \rho \pi (2n+1))k }
(\tilde{\sigma}^{+}(k,a_1)+\tilde{\sigma}^{+}(k,a_2)) \right) \nn \\
& \times  & \prod_{n=0}^{\infty}  \exp \left( (-1)^{(n+1)} \int_{-\infty}^{\infty}
dk \,e^{i (\te - i \rho \pi (2n+1))k }
(\tilde{\sigma}^{-}(k,a_1)+\tilde{\sigma}^{-}(k,a_2)) \right) \nn
\label{bb1}
\eea
and hence
\bea
\B(\te) &=& \prod_{n=0}^{\infty} { \sigma^{+}(\te + i \rho \pi (4n+1),a_1) \over
\sigma^{+}(\te + i \rho \pi (4n+3),a_1)}
{ \sigma^{+}(\te + i \rho \pi (4n+1),a_2) \over
\sigma^{+}(\te + i \rho \pi (4n+3),a_2)} \nn \\
& \times  &
\prod_{n=0}^{\infty}
{ \sigma^{-}(\te - i \rho \pi (4n+3),a_1) \over
\sigma^{-}(\te - i \rho \pi (4n+1),a_1)}
{ \sigma^{-}(\te - i \rho \pi (4n+3),a_2) \over
\sigma^{-}(\te  - i \rho \pi (4n+1),a_2)}
\eea
To derive the Y-systems, the following mass relations were also
important:
\bea
m_B &=& 2 \cos \left( \fract{\pi}{6} \fract{p-5}{p-1} \right )m_K\,; \nn \\
2 \cos \left (\fract{\pi}{6} \fract{p+3}{p-1} \right) m_B &=& 
m_K +2 \cos \left( \fract{\pi}{3} \fract{4}{p-1} \right) m_K \,.
\label{massrel1}
\eea
%
\subsection{General notation for Y-systems}
\label{y21not}
We define
\eq
Y(\te) = e^{\ep(\te)}
\en
and  introduce the shorthand notations
\eq
\CY(\pm n)=Y(\te + i \pi \fract{n}{H})Y(\te - i \pi \fract{n}{H}) \, ,
\en
\eq
\Lambda(\pm n_1, \dots, \pm n_m)= \prod_{j=1}^{m}
\left (1+Y(\te + i \pi  \fract{n_j}{H}) \right)
\left (1+Y(\te - i  \pi \fract{n_j}{H}) \right) \, ,
\en
and
\eq
\CL(\pm n_1, \dots, \pm n_m)= \left[\prod_{j=1}^{m}
\left(1+{ 1 \over Y(\te + i \pi \fract{n_j}{H})} \right)
\left(1+{1 \over Y(\te - i \pi \fract{
n_j}{H})} \right)
\right]^{-1} \, ,
\en
with   $n_i \ge  0$.  Just a single factor appears on the RHS 
for  entries with the $\pm$ omitted, so that, for example,
\eq
\Lambda(0, \pm n_2)=
(1+Y(\te))\left (1+Y(\te + i \pi \fract{n_2}{H}) \right)
\left (1+Y(\te - i \pi \fract{n_2}{H}) \right) \, .
\en
In practice the functions $\varepsilon$, $Y$, ${\cal Y}$ and so on
appear with indices to show which pseudoenergy is involved. In all
of the above definitions,
these indices take the same values in all factors.

We shall also set $H=6|\xi{-}1|$ where, as before,
$c=1-6/\xi(\xi{+}1)$ with $\xi>0$ on the critical branch, and $\xi<0$
on
the tricritical branch. For the unitary
minimal models $\CM_{p,p{+}1}$ this
translates as $H=6(p{-}1)$ for the $\phi_{21}$ 
perturbations, and $H=6(p{+}2)$ for the  $\phi_{12}$ perturbations.

\subsection{TBA equations and Y-systems for $\CM_{p,p+1}+\phi_{21}$}

In the figures below, we supplement the four
families of $\phi_{21}$-related TBA equations and Y-systems with four
sets of graphical representations.
These graphs give a rough idea of
the structure of the   
TBA equations, and also fix the labelling conventions.
On the magnonic nodes,  to any  pair of  numbers $(i,\alpha)$
corresponds, in the TBA
equations, quantities labelled with a lower index $i$ and an upper index
$(\alpha)$: $\ep_{i}^{(\alpha)}$,  
$L_{i}^{(\alpha)}=\ln(1+e^{-\ep_{i}^{(\alpha)}})$.\\

\noindent
{\bf Case A, (p=4n+2 , n $\ge$ 2):}\\
\begin{figure}[ht]
\begin{center}
\resizebox{0.4\linewidth}{!}
{\includegraphics{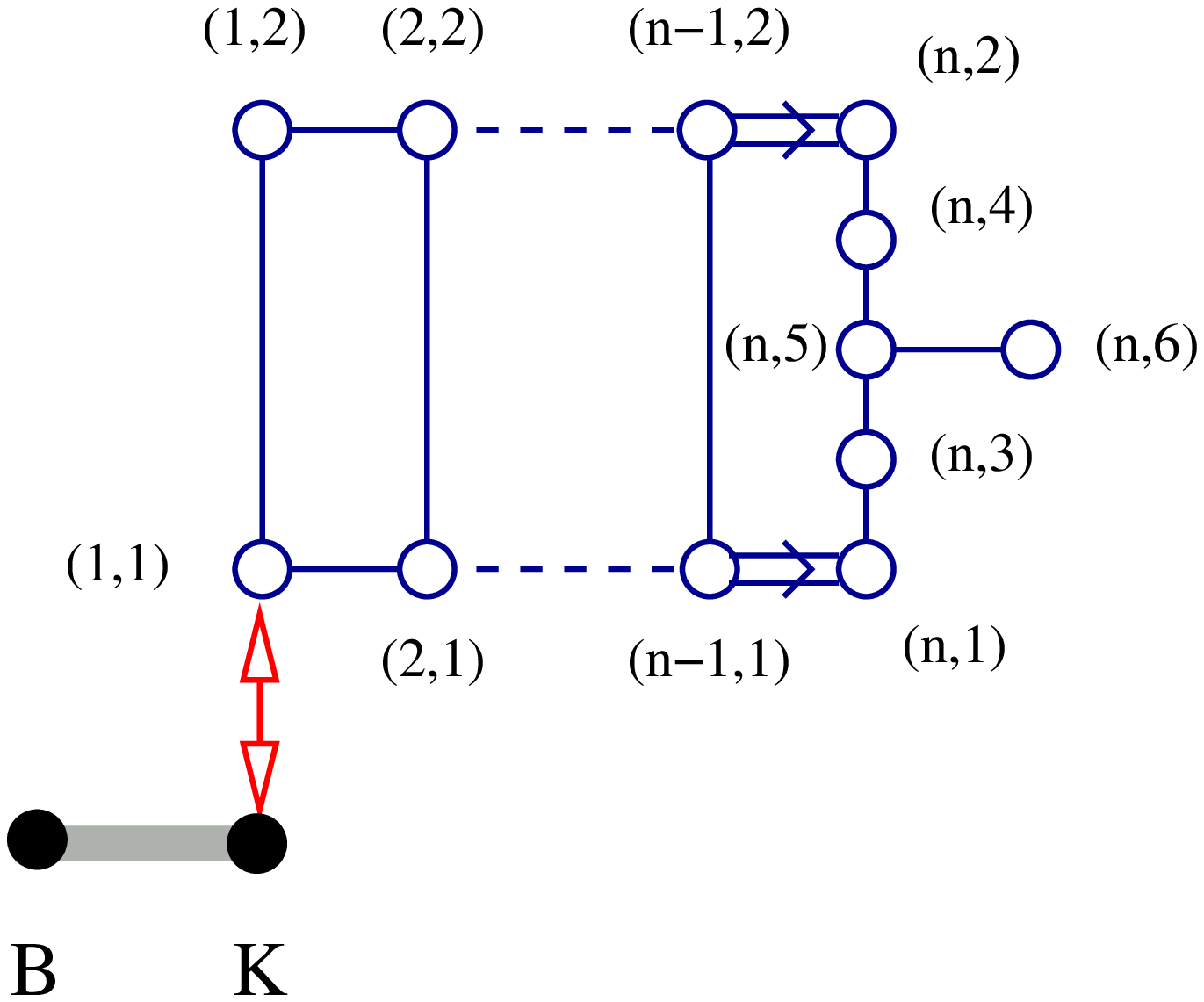}}
\label{fig0F.a}
\end{center}
\end{figure}
\bea
\ep_{B}(\theta) &=& R m_B \cosh \theta  -  \sum_{c \in \{ B,K \}}
\Phi_{Bc}*L_{c}(\theta) \nn \\
\ep_{K}(\theta) &=& R m_K \cosh \theta - \sum_{c \in \{ B,K \}}
\Phi_{Kc}*L_{c}(\theta)
+ \sum_{j=1}^{n}
\tpsi_{1j}*L^{(1)}_{j}(\te) \nn  \\
\ep^{(\alpha)}_i(\theta)&=&\delta_{\alpha 1} \tpsi_{1i}*L_{K}(\te)
-\!\sum^n_{j=1} \lf[\tphi_{ij}{*}L^{(\alpha)}_j(\theta)
-\tpsi_{ij}{*}L^{(\bar\alpha)}_j(\theta)\ri]\nn\\
&&~~~{}-\delta_{in}\,\sum^6_{\beta=1}l_{\alpha\beta}^{[E_6]}
           \tphi_2{*}L^{(\beta)}_n(\theta)\,~~~ 
(i=1,\dots,n;~~\alpha{=}1,2;~~
            \bar\alpha=3{-}\alpha)\,\nn\\
{}~\ep^{(\gamma)}_n(\theta)&=&{~~~}
-\sum^6_{\beta=1}l^{[E_6]}_{\gamma\beta}
  \tphi_2{*}L^{(\beta)}_n(\theta)\, ~~~~~~~~~\quad(\gamma=3,4,5,6) 
\label{TBAzvgen} \nn
\eea
The Y-system is
\bea
\CY_B(\pm (p+3) )&=&\Lambda_K(0,\pm 8)\Lambda_1^{(1)}(\pm 4)
\Lambda_1^{(2)}(0)      \nn \\
\CY_K(\pm(p-5))&=& \Lambda_B(0)  \CL^{(1)}_n(0)
\prod_{i=1}^{n-1} \CL_i^{(1)}(\pm(4n-4i-3))
     \nn \\
\CY_j^{(\alpha)}(\pm 4)&=& 
\left [\CL_K(0) \right]^{\delta_{j  1} \delta_{\alpha 1}}
\CL_j^{(\bar\alpha)}(0) \prod_{k=1}^{n-1}
\left( \Lambda_{k}^{(\alpha)}(0) \right)^{l_{k j}^{[A_{n-1}]}} \nn\\
&\times &
\left[ \Lambda^{(6)}_n(0)
\Lambda^{(4)}_n(0)
\Lambda^{(5)}_n(\pm 1)
\Lambda^{(3)}_n(\pm 2)
\Lambda^{(1)}_n(\pm 3)
\right]^{\delta_{j,n-1} \delta_{\alpha 1}}
\nn \\
&\times &
\left[ \Lambda^{(6)}_n(0)
\Lambda^{(3)}_n(0)
\Lambda^{(5)}_n(\pm 1)
\Lambda^{(4)}_n(\pm 2)
\Lambda^{(2)}_n(\pm 3)
\right]^{\delta_{j,n-1} \delta_{\alpha 2}}     \nn \\
&&\big(\mbox{with }j=1, \dots,n-1;~    \alpha=1,2\,\mbox{ and } 
\bar{\alpha}=3-\alpha \big )
\nn \\
\CY^{(\gamma)}_n(\pm 1)&=& \left[ \Lambda_{n-1}^{(\gamma)} (0) \right
]^{\delta_{\gamma 1}+\delta_{\gamma 2}}
\prod_{\beta=1}^{6} \left[ \CL^{(\beta)}_n(0) \right]^{l^{[E_6]}_{\beta \gamma}}
~~~(\gamma=1,2, \dots, 6) 
\eea
\\

{\bf Case B, (p=4n+3, n $\ge$ 2):}
\begin{figure}[ht]
\begin{center}
\resizebox{0.35\linewidth}{!}
{\includegraphics{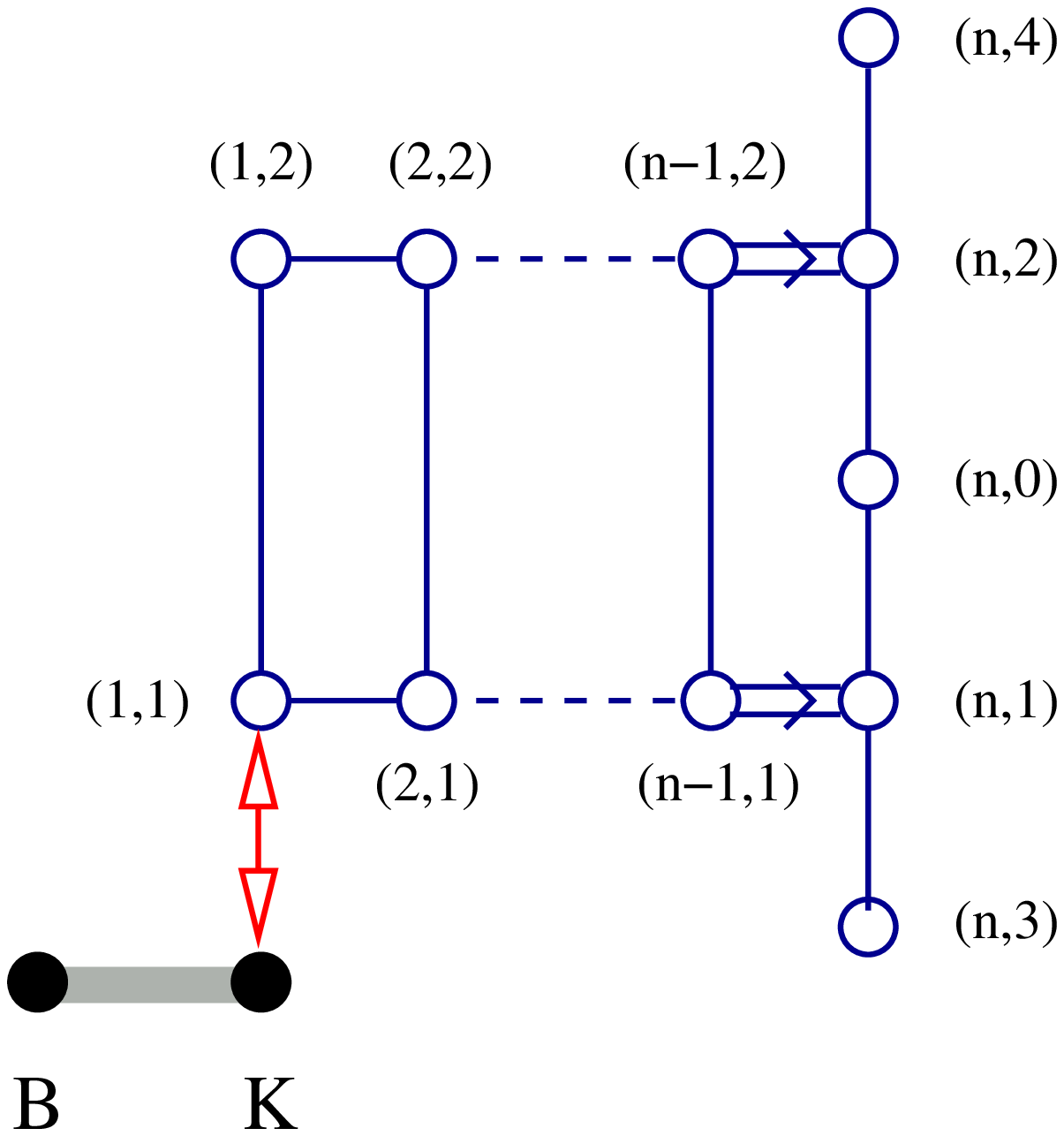}}
\label{fig0F.b}
\end{center}
\end{figure}
\bea
\ep_{B}(\theta) &=& R m_B \cosh \theta  -  \sum_{c \in \{ B,K \}}
\Phi_{Bc}*L_{c}(\theta) \nn \\
\ep_{K}(\theta) &=& R m_K \cosh \theta - \sum_{c \in \{ B,K \}}
\Phi_{Kc}*L_{c}(\theta)
+ \sum_{j=1}^{n}
\tpsi_{1j}*L^{(1)}_{j}(\te)  \nn \\
\ep^{(\alpha)}_i(\theta)&=&\delta_{\alpha 1} \tpsi_{1i}*L_{K}(\te)
-\!\sum^n_{j=1}
\lf[\tphi_{ij}{*}L^{(\alpha)}_j(\theta)
-\tpsi_{ij}{*}L^{(\bar\alpha)}_j(\theta)\ri] \nn\\
&&~~{}-\delta_{in}\!\sum^4_{\beta=0}l_{\alpha\beta}^{[A_5]}
           \tphi_1{*}L^{(\beta)}_n(\theta)\, ~~~~~
(i=1,\dots,n;~~\alpha{=}1,2;~~
            \bar\alpha=3{-}\alpha)\nn\\
\ep^{(\gamma)}_n(\theta)&=&-
\sum^4_{\beta=0}l^{[A_5]}_{\gamma \beta}
     \tphi_1{*}L^{(\beta)}_n(\theta)\,~~~~~\quad(\gamma =0,3,4)
\label{TBAzbn} \\ \nn
\eea
The Y-system is
\bea
\CY_B(\pm(p+3))&=&\Lambda_K(0,\pm 8)\Lambda_1^{(1)}(\pm 4)
\Lambda_1^{(2)}(0)  \nn \\
\CY_K(\pm(p-5))&=& \Lambda_B(0)  \CL^{(1)}_n(0)
\prod_{i=1}^{n-1} \CL_i^{(1)}(\pm(4n-4i-2))  
\nn \\
\CY_j^{(\alpha)}(\pm 4)&=&
\left [\CL_K(0) \right]^{\delta_{j  1} \delta_{\alpha 1}}
\CL_j^{(\bar \alpha)}(0)\prod_{k=1}^{n-1}
\left( \Lambda_{k}^{(\alpha)}(0) \right)^{l_{k j}^{[A_{n-1}]}} \nn\\
&\times &
\left[ \Lambda^{(0)}_n(0)
\Lambda^{(3)}_n(0)
\Lambda^{(1)}_n(\pm 2)
\right]^{\delta_{j,n-1} \delta_{\alpha 1}} \nn \\
& \times &
\left[ \Lambda^{(0)}_n(0)
\Lambda^{(4)}_n(0)
\Lambda^{(2)}_n(\pm 2)
\right]^{\delta_{j,n-1} \delta_{\alpha 2}}
\nn \\
&&\big(\mbox{with } j=1, \dots,n-1;~  \alpha=1,2\,\mbox{ and } 
\bar{\alpha}=3-\alpha \big )
\nn \\
\CY^{(\gamma)}_n(\pm 2)&=& \left[ \Lambda_{n-1}^{(\gamma)}(0)
\right]^{\delta_{\gamma 1} +\delta_{\gamma 2} }
\prod_{\beta=0}^{4}\left[\CL^{(\beta)}_n(0) \right]^{l^{[A_5]}_{\beta \gamma}}
\, ,  (\gamma=0,\dots,4) 
\eea
\\

\noindent
{\bf Case C, (p=4n+4 , n $\ge$ 2):}
\begin{figure}[ht]
\begin{center}
\resizebox{0.4\linewidth}{!}
{\includegraphics{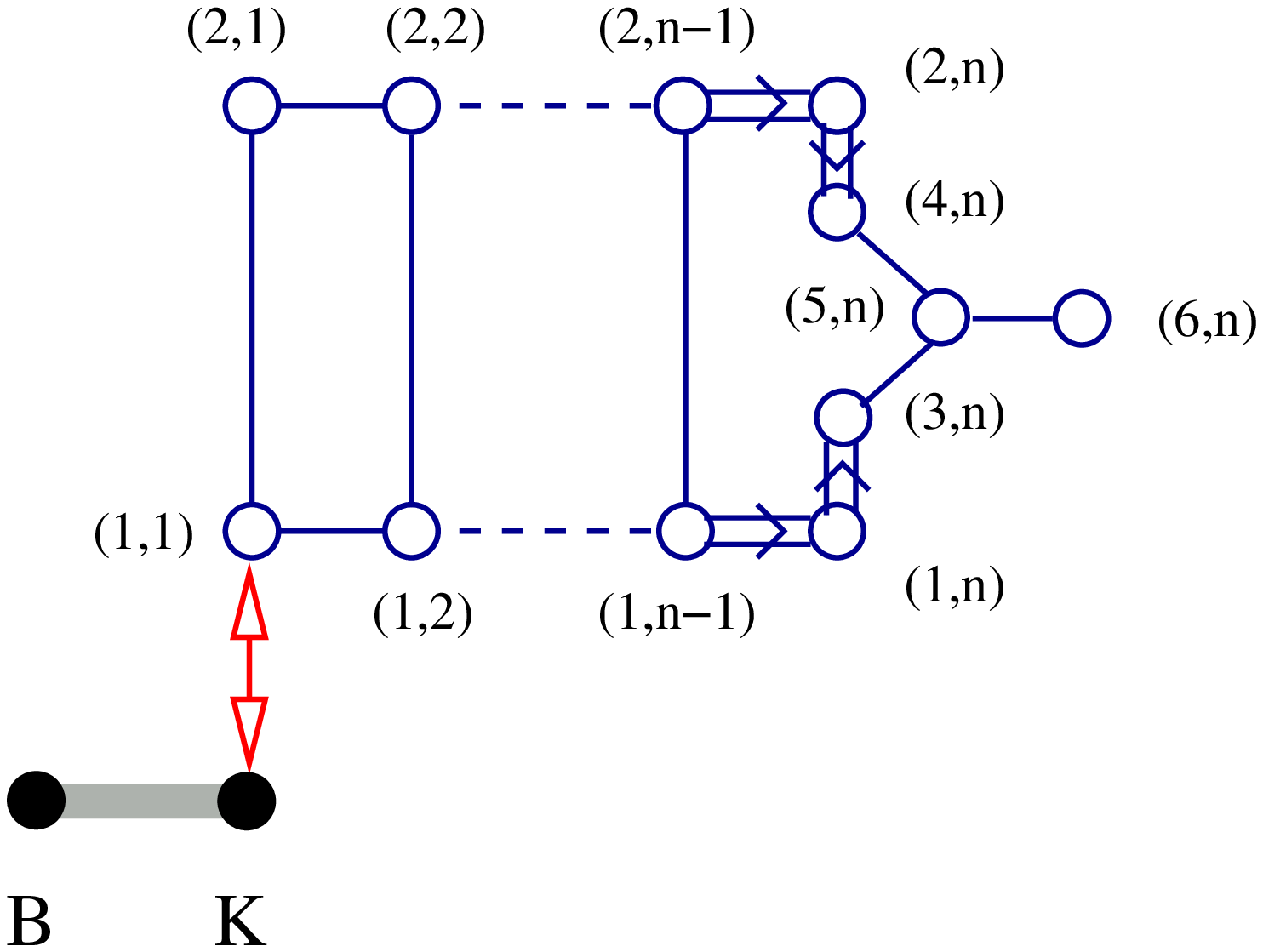}}
\label{fig0F.c}
\end{center}
\end{figure}
\bea
\ep_{B}(\theta) &=& R m_B \cosh \theta  -  \sum_{c \in \{ B,K \}}
\Phi_{Bc}*L_{c}(\theta)~ \nn \\
\ep_{K}(\theta) &=& R m_K \cosh \theta - \sum_{c \in \{ B,K \}}
\Phi_{Kc}*L_{c}(\theta)
+ \sum_{j=1}^{n}
\tpsi_{1j}*L^{(1)}_{j}(\te)  \nn \\
\ep^{(\alpha)}_i(\theta)&=& \delta_{\alpha 1} \tpsi_{1i}*L_{K}(\te)
-\!\sum^n_{j=1} \lf[\tphi_{ij}{*}L^{(\alpha)}_j(\theta)
-\tpsi_{ij}{*}L^{(\bar\alpha)}_j(\theta)\ri]\nn\\
&&~~~{}-\delta_{i n} \delta_{\alpha 1}
\lf[\tphi_3{*}(L^{(4)}_n(\theta)+L^{(6)}_n(\theta))
+\tphi_4{*}L^{(3)}_n(\theta)+\tphi_5{*}L^{(5)}_n(\theta)\ri]\nn\\
&&~~~{}-\delta_{i n} \delta_{\alpha 2}
\lf[\tphi_3{*}(L^{(3)}_n(\theta)+L^{(6)}_n(\theta))
+\tphi_4{*}L^{(4)}_n(\theta)+\tphi_5{*}L^{(5)}_n(\theta)\ri]   \nn\\
\ep^{(3)}_n(\theta)&=&~~\tphi_2{*}(K^{(5)}_n(\theta)-L^{(1)}_n(\theta))  \nn\\
\ep^{(5)}_n(\theta)&=&~~
  \tphi_2{*}(K^{(3)}_n(\theta)+K^{(6)}_n(\theta)+K^{(4)}_n(\theta))  \nn \\
 \ep^{(6)}_n(\theta) &=&~~\tphi_2{*}K^{(5)}_n(\theta) \nn\\
\ep^{(4)}_n(\theta)&=&~~\tphi_2{*}(K^{(5)}_n(\theta)-L^{(2)}_n(\theta)) 
\label{TBAzviigen}
\eea
with $i=1,\dots,n$,~~$\alpha{=}1,2$, $\bar\alpha=3{-}\alpha$, and
$K^{(c)}(\theta)= \ln(1+e^{\ep^{(c)}(\theta)})$ as in (\ref{ccase})
above.\\

The Y-system is
\bea
\CY_B(\pm(p+3))&=&\Lambda_K(0,\pm 8)\Lambda_1^{(1)}(\pm 4)
\Lambda_1^{(2)}(0)  \nn \\
\CY_K(\pm (p-5))&=& \Lambda_B(0)  \CL^{(1)}_n(0)
\prod_{i=1}^{n-1} \CL_i^{(1)}(\pm(4n-4i-1)) 
\nn \\
\CY_j^{(\alpha)}(\pm 4)&=& \left 
[\CL_K(0) \right]^{\delta_{j  1} \delta_{\alpha 1}}
\CL_j^{(\bar \alpha)}(0)
\prod_{k=1}^{n-1}
\left( \Lambda_{k}^{(\alpha)}(0) \right)^{l_{k j}^{[A_{n-1}]}} \nn\\
&\times &
\left[ \Lambda^{(3)}_n(0)
\Lambda^{(1)}_n(\pm 1)
\right]^{\delta_{j,n-1} \delta_{\alpha 1}} \nn \\
& \times &
\left[ \Lambda^{(4)}_n(0)
\Lambda^{(2)}_n(\pm 1)
\right]^{\delta_{j,n-1} \delta_{\alpha 2}} \nn \\
&&\big(\mbox{with } j=1, \dots,n-1; \alpha=1,2\,\mbox{ and } 
\bar{\alpha}=3-\alpha \big )
\nn \\
\CY^{(1)}_n(\pm 3)&=&  \Lambda_{n-1}^{(1)}(0)
\CL^{(4)}_n(0)\CL^{(6)}_n(0)\CL^{(5)}_n(\pm 1) \CL^{(3)}_n(\pm 2)  \nn \\
\CY^{(2)}_n(\pm 3)&=&  \Lambda_{n-1}^{(2)}(0)
\CL^{(3)}_n(0)\CL^{(6)}_n(0)\CL^{(5)}_n(\pm 1) \CL^{(4)}_n(\pm 2)  \nn \\
\CY^{(3)}_n(\pm 1)&=&
\CL^{(1)}_n(0)\Lambda^{(5)}_n(0)  \nn \\
\CY^{(4)}_n(\pm 1)&=&
\CL^{(2)}_n(0)\Lambda^{(5)}_n(0) \nn \\
\CY^{(5)}_n(\pm 1)&=&
\Lambda^{(3)}_n(0)\Lambda^{(6)}_n(0) \Lambda^{(4)}_n(0)  \nn \\
\CY^{(6)}_n(\pm 1)&=&
\Lambda^{(5)}_n(0) 
\eea
\\

\noindent
{\bf Case D, (p=4n+5, n $\ge$ 2):}
\begin{figure}[ht]
\begin{center}
\resizebox{0.35\linewidth}{!}
{\includegraphics{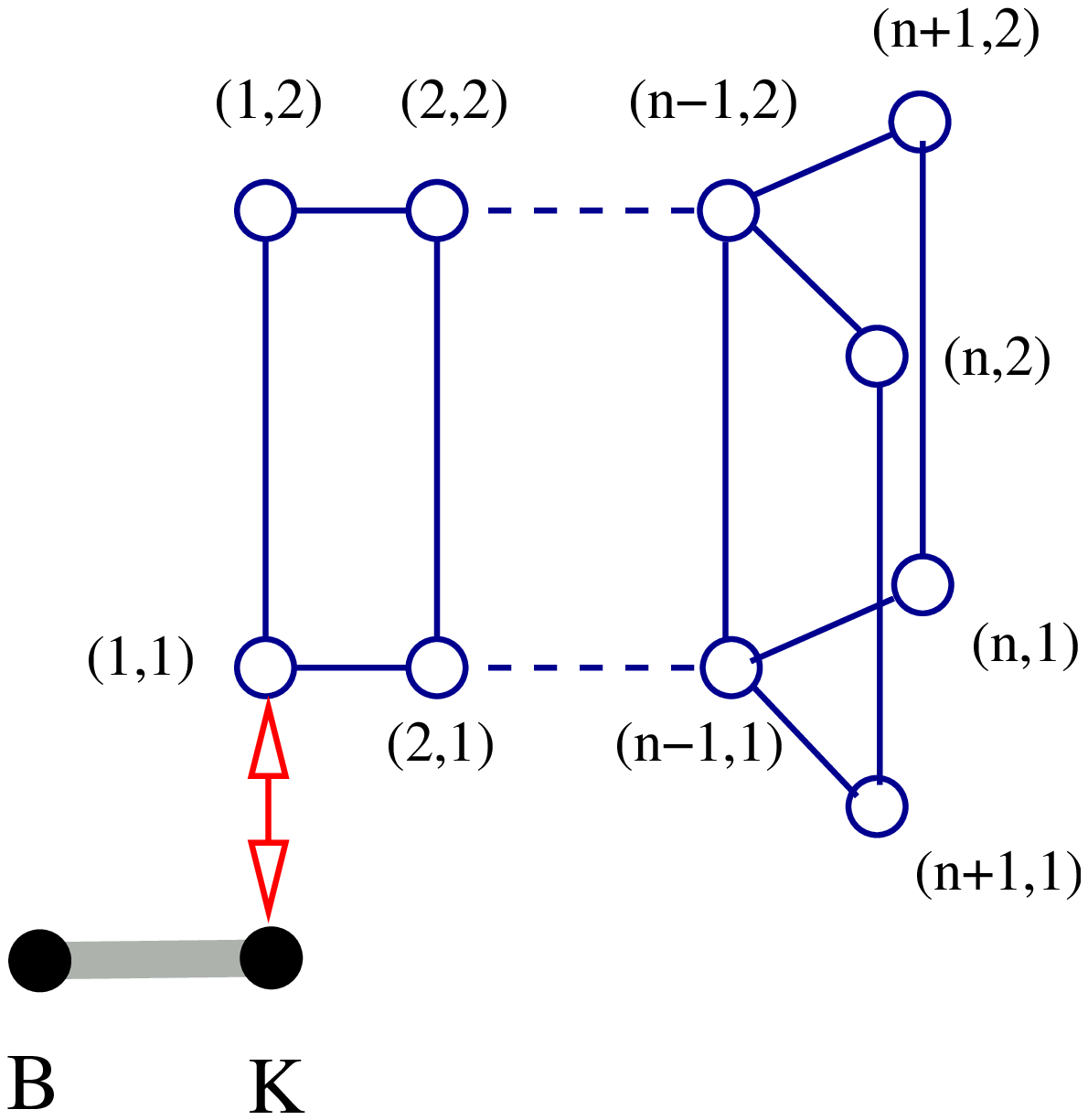}}
\label{fig0F.d}
\end{center}
\end{figure}
\bea
\ep_{B}(\theta) &=& R m_B \cosh \theta  -  \sum_{c \in \{ B,K \}}
\Phi_{Bc}*L_{c}(\theta)  \nn \\
\ep_{K}(\theta) &=& R m_K \cosh \theta - \sum_{c \in \{ B,K \}}
\Phi_{Kc}*L_{c}(\theta)
+ \sum_{j=1}^{n+1}
\tpsi_{1j}*L^{(1)}_{j}(\te)  \nn \\
\ep^{(\alpha)}_i(\theta)&=&
 \delta_{\alpha 1} \tpsi_{1i}*L_{K}(\te)
-\!\!\sum^{n{+}1}_{j=1}
\!\lf[\tphi_{ij}{*}L^{(\alpha)}_j(\theta)
-\tpsi_{ij}{*}L^{(\bar \alpha)}_j(\theta)\ri]
\label{TBAziv}
\eea
with $i=1,\dots,n+1$,~~$\alpha{=}1,2$, and $\bar\alpha=3{-}\alpha$.\\

The Y-system is
\bea
\CY_B(\pm(p+3))&=&\Lambda_K(0,\pm 8)\Lambda_1^{(1)}(\pm 4)
\Lambda_1^{(2)}(0)  \nn \\
\CY_K(\pm(p-5))&=& \Lambda_B(0)  \CL_n^{(1)}(0)
\CL_{n+1}^{(1)}(0)
\prod_{i=1}^{n-1} \CL_i^{(1)}(\pm(4n-4i))  
\nn \\
\CY_j^{(\alpha)}(\pm 4)&=& \left [\CL_K(0) \right]^{\delta_{j 1} 
\delta_{\alpha 1}}
\CL_j^{(\bar\alpha )}(0) \prod_{i=1}^{n+1}
\left[ \Lambda_{i}^{(\alpha)}(0) \right]^{l^{[D_{n+1}]}_{ij}}  \nn \\
&&\big(\mbox{with } j=1, \dots,n+1 \, , \alpha=1,2\,\mbox{ and } 
\bar{\alpha}=3-\alpha \big ) 
\eea

\subsection{Exceptional $\phi_{21}$  Y-systems}
\label{y21}
The well-known~\cite{Zam2} Y-systems for $\CM_{3,4}+\phi_{21}$ and  
$\CM_{5,6}+\phi_{21}$ are
\eq
\CY_K(\pm 6)= 1 \,
\en
and
\eq
\CY_{j}(\pm 8)= \prod_{k=1}^{2} \left[
\Lambda_{k}(0) \right]^{l^{[A_2]}_{kj}}  \,\,\,\,\,(j=1,2) 
\label{aa2}
\en 
respectively, with $\CY_1 = \CY_2=\CY_K$ in the second system.

For $\CM_{4,5}+\phi_{21}$,  a Y-system can be derived from the 
TBA of \cite{BP}. In the current notation, it can be written as
\bea
\CY^{(2)}(\pm 1) &=& \CL^{(1)}(0)  \nn \\
\CY^{(1)}(\pm 1) &=& \CL^{(2)}(0)/\CL_K(0)   \nn \\
\CY_K(\pm 6) &=& {\CY_K(0) \over  \CY^{(2)}(0)^3 }  
{\CL_K(\pm 4) \CL^{(1)}(\pm 1, \pm 5)  \over \CL^{(1)}(\pm 3) \CL_K(\pm 2)} 
\label{M45}
\eea
(The periodicity implied  by this system  is $P=32/H=16/9$,
and the resulting  conformal dimension 
$\Delta=1-1/P=7/16$  matches that of the perturbing operator $\phi_{21}$.) 

The remaining exceptional $\phi_{21}$ Y-systems can be derived from
the TBA equations given in \S\ref{con21} above, and are as follows:

\noindent
$\CM_{6,7}+\phi_{21}$\,:
\bea
\CY_B(\pm 9)&=&\Lambda_K(0,\pm 8)\Lambda^{(1)}(\pm 1,\pm 7)
\Lambda^{(3)}(0,\pm 2,\pm 6) \Lambda^{(5)}(\pm 1,\pm 3,\pm 5) \nn \\ 
&\times& 
\Lambda^{(6)}(0,\pm 4)
\Lambda^{(4)}(\pm 2,\pm 4)
\Lambda^{(2)}(\pm 3)   \nn \\
\CY_K(\pm 1)&=& \Lambda_B(0) \CL^{(1)}(0)      \nn \\
\CY^{(\gamma)}(\pm 1)&=& \left[ \CL_K(0) \right]^{\delta_{\gamma 1}}
\prod_{\beta=1}^{6} \left(\CL^{(\beta)}(0) \right)^{l^{[E_6]}_{\beta \gamma}} 
~~~~(\gamma=1,\dots, 6) 
\eea
\noindent
$\CM_{7,8}+\phi_{21}$\,:
\bea
\CY_B(\pm 10)&=&\Lambda_K(0,\pm 8)\Lambda^{(4)}(0)
\Lambda^{(2)}(\pm 2) \Lambda^{(0)}(0,\pm 4)
\Lambda^{(1)}(\pm 2,\pm 6) \Lambda^{(3)}(\pm 4)      \nn \\
\CY_K(\pm 2)&=& \Lambda_B(0) \CL^{(1)}(0)      \nn \\
\CY^{(\gamma)}(\pm 2)&=& \left[ \CL_K(0) \right]^{\delta_{\gamma 1}}
\prod_{\beta=0}^{4} \left(\CL^{(\beta)}(0) \right)^{l^{[A_5]}_{\beta \gamma}}
~~~~(\gamma=0,\dots, 4) 
\eea
$\CM_{8,9}+\phi_{21}$\,:
\bea
\CY_B(\pm 11)&=&
\Lambda_K(0,\pm 8) \Lambda^{(1)}(\pm 3,\pm 5)
\Lambda^{(2)}(\pm 1) \Lambda^{(3)}(\pm 4)
\Lambda^{(4)}(0)      \nn \\
\CY_K(\pm 3)&=& \Lambda_B(0) \CL^{(1)}(0)      \nn \\
\CY^{(1)}(\pm 3)&=& \CL_K(0) \CL^{(6)}(0)\CL^{(4)}(0) \CL^{(5)}(\pm 1)
\CL^{(3)}(\pm 2)     \nn \\
\CY^{(2)}(\pm 3)&=& \CL^{(6)}(0)\CL^{(3)}(0) \CL^{(5)}(\pm 1)
 \CL^{(4)}(\pm 2)      \nn \\
\CY^{(3)}(\pm 1)&=& \Lambda^{(5)}(0)\CL^{(1)}(0)      \nn \\
\CY^{(4)}(\pm 1)&=& \Lambda^{(5)}(0)\CL^{(2)}(0)      \nn \\
\CY^{(5)}(\pm 1)&=& \Lambda^{(6)}(0)
\Lambda^{(4)}(0)\Lambda^{(3)}(0)      \nn \\
\CY^{(6)}(\pm 1)&=& \Lambda^{(5)}(0)     
\eea
$\CM_{9,10}+\phi_{21}$\,:
\bea
\CY_B(\pm 12)&=&\Lambda_K(0,\pm  8)\Lambda^{(4)}(0)
\Lambda^{(2)}(0) \Lambda^{(1)}(\pm 4)
\Lambda^{(3)}(\pm 4)      \nn \\
\CY_K(\pm 4)&=& \Lambda_B(0) \CL^{(1)}(0) \CL^{(3)}(0)     \nn \\
\CY^{(1)}(\pm 4)&=& \CL_K(0) \CL^{(2)}(0)     \nn \\
\CY^{(3)}(\pm 4)&=& \CL_K(0) \CL^{(4)}(0)     \nn \\
\CY^{(2)}(\pm 4)&=& \CL^{(1)}(0)      \nn \\
\CY^{(4)}(\pm 4)&=& \CL^{(3)}(0) 
\eea

\resection{TBA and Y-systems  ($\phi_{12}$ perturbations)}
\label{a12}
The mass spectrum for the theory $\CM_{p,p+1}+\phi_{12}$ 
for  $p=3,4,5$ and $6$  contains  8, 7, 6 and 6 particle 
types respectively.
In the  region   $p \ge 10$
there are instead only  four particles: two
breathers ( $B_1$ and $B_3$ ) and two kinks ( $K_1$  and $K_2$ ),
with masses
\bea
m_{B_1} &=& 2 m \cos \lf(  \xi_1 \pi +4 \xi_2 \pi  \ri)   \nn \\
m_{B_3} &=& 4 m  \cos \lf(   \xi_1 \pi +4 \xi_2 \pi \ri)
\cos \lf(   \xi_1 \pi   \ri)     \\
m_{K_2} &=& 2 m  \cos \lf(  \xi_2 \pi   \ri)  \nn  \\
m_{K_1} &=&  m \nn
\eea
where $\xi_1=(p-2) /(12+6p)=(p-2)/H$ and $\xi_2=2/ (6+3p)=4/H$.
The  four  masses  also satisfy the following `fusion' relation
\bea
2 \cos\lf(   \xi_2 \pi  \ri) m_{K_1} &=& m_{K_2}   \nn \\
2 \cos\lf(   \xi_2 \pi  \ri) m_{K_2} &=& m_{K_1} + m_{B_3}    \\
2 \cos\lf(   \xi_1 \pi  \ri) m_{B_3}  &=&  m_{B_1}+
2 \cos\lf( \xi_1 \pi -\xi_2  \pi\ri) m_{K_2}    \nn \\
2 \cos\lf( \xi_1 \pi  \ri) m_{B_1} &=& m_{B_3}  \nn
\label{massrel12}  
\eea
We shall introduce a tower of magnonic  pseudoenergies
determined by the ratio
\eq
\trisymb{\rho}= \xi_1 /  \xi_2   = (p-2)/4 \, . 
\en
As for the $\phi_{21}$ perturbations, there  are four distinct families 
of models,
determined by a continued fraction decomposition: 
\bea
\trisymb{A})~~~p=4n+3~~:~~ \trisymb{\rho}
 &=& (n-1)+ {1 \over 4}     \nn    \\
\trisymb{B})~~~p=4n+4~~:~~  \trisymb{\rho} 
 &=& (n-1)+ {1 \over 2}     \nn  \\
\trisymb{C})~~~p=4n+5~~:~~ \trisymb{\rho} 
 &=& (n-1)+ {1 \over{1+{1 \over 3}}} \nn  \\
\trisymb{D})~~~p=4n+6~~:~~\trisymb{\rho} 
 &=&   n         \nn
\label{fourc1}
\eea
\subsection{The kernels}

First we need the kernels involving the breathers and kinks.
These are simply defined as 
\eq
\Phi_{ab}(\te)=\Phi_{ba}(\te)= -i \ddt \ln {\cal S}_{ab}(\te) \, ,~~~ 
(a=B_1,B_3,K_1,K_2;~~b=B_1,B_3)
\en
where the quantities ${\cal S}_{ab}(\te)$ were defined in \S\ref{4part}.
The kink-kink kernels are defined, as  for the 
$\phi_{21}$-related cases
in \S \ref{con21},  to match the mass fusion relation.
\bea
\Phi_{K_2 K_1}=\Phi_{K_1 K_2}(\te) &=&  w*\Phi_{B_3 K_1}(\te) - w(\te) 
\nn \\
\Phi_{K_1 K_1}(\te) &=&  z*\Phi_{K_2 K_1}(\te)  \nn \\
\Phi_{K_2 K_2}(\te) &=&  \Phi_{K_1 K_1}(\te) + \Phi_{B_3 K_1}(\te)  
\eea
with
\eq
 z(\te)= {H \over 8 \cosh( H\te/8)}~~,~~
 w(\te)=  {H \over 4 \sqrt{3}}
{\cosh(H \te/24) \over \cosh(H \te/8)} \, .
\en
Some of the extra  kernels needed are defined in term of
S-matrix elements of
the $A_2$, $A_5$  and  $E_6$  `purely-elastic' scattering theories.
In general these S-matrix elements can be written as \cite{Patrick,BCDSa}
\eq
S_{ij}^{[g]}(\te)= \prod_{\alpha \in A^{[g]}_{ij}} \{ \alpha \} \, ,
\en
and from these we can define the functions
\eq
T_{ij}^{[g]}(\te)= \prod_{\alpha \in A^{[g]}_{ij}} (\alpha ) \, ,
\en
where $g$ is $A_2$, $A_5$  or $E_6$, and $A^{[g]}_{ij}$ is a set of 
rational  numbers depending on $i$, $j$ and $g$, and the blocks $(x)$ and
$\{ x \}$ were defined in (\ref{blocks}). Then the kernels needed  in the 
TBA are defined as
\eq
\tphi_{i j}^{[g]}(\theta) =  -i {d \over d \te} S_{i j}^{[g]}\left ((p+2)
\fract{\theta}{2} \right ) ~~~\, ,~~~
\tpsi_{i j}^{[g]}(\theta)=-i {d \over d \te} T_{i j}^{[g]}\left((p+2)
\fract{\theta}{2} \right) \, , 
\en
and
\eq
\bar{\psi}_{i j}^{[A_2]}(\theta) =
\tpsi_{i j}^{[A_2]}(\theta -i \pi/H) +\tpsi_{i j}^{[A_2]}(\theta+i \pi/H) \, .
\en
\subsection{TBA equations and Y-systems for $\CM_{p,p+1}+\phi_{12}$}
\label{TBAY12}
As in appendix A.3, we illustrate our proposals with a set of graphs,
which give a rough idea of  
the structure of the TBA equations, and also fix the labelling
convention.
Where  convenient,  we  also refer to kink-related 
quantities  using  a single label $K_\alpha$, rather than the pair
$(0,\alpha)$. 
For the Y-systems  we use  the notation defined in  \S\ref{y21not} with 
$H=6(p+2)$.\\

\noindent
{\bf Case $\trisymb{A}$, (p=4n+3, n $\ge$ 1):}
\begin{figure}[ht]
\begin{center}
\resizebox{0.35\linewidth}{!}
{\includegraphics{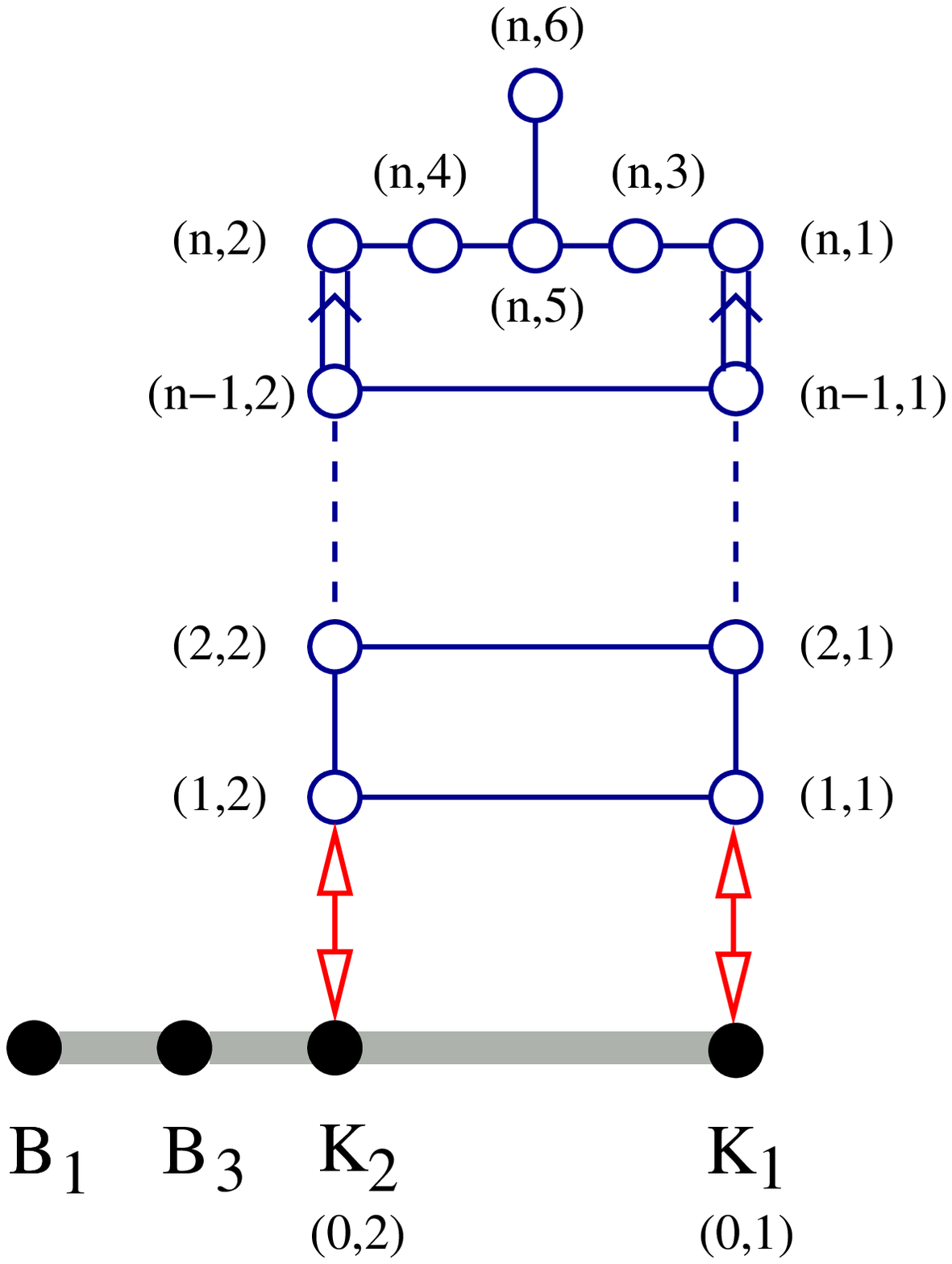}}
\label{figF.a}
\end{center}
\end{figure}
\bea
\ep_{B_a}(\theta) &=& R m_{B_a} \cosh \theta  -
\sum_{c \in \{ B_1,B_2,K_1,K_2 \}}
\Phi_{B_a,c}*L_{c}(\theta)~\,~~~ (a=1,2) \nn \\
\ep_j^{({\alpha})}(\theta) &=& \delta_{j0} \left (
R m_{K_\alpha} \cosh \theta -
\sum_{d \in \{ B_1, B_2 ,K_1,K_2  \}}
\Phi_{K_\alpha,d}*L_{d} (\theta) +
\sum_{\beta=1}^{2}
\tphi^{[A_2]}_{\alpha \beta}*L_j^{(\beta)}(\te)
\right) \nn \\
 &-&
\sum_{\beta=1}^{2}
\left(
\tphi^{[A_2]}_{\alpha \beta}*L_j^{(\beta)}(\te)- \sum_{k=0}^{n-1}
l_{jk}^{[A_{n}]}
\tpsi^{[A_2]}_{\alpha \beta}*L_{k}^{(\beta)}(\te) \right)  \nn \\
&+& \delta_{j,n-1} \sum_{\delta=1}^{6} \tpsi_{\alpha \delta}^{[E_6]}*L^{(\delta)}_n(\te)
~~\, ~~~ (j=0, \dots, n-1, \alpha=1,2)
 \nn \\
\ep^{(\gamma)}_n(\theta) &=&
- \sum_{\delta =1}^{6} \tphi_{\gamma \delta }^{[E_6]}*L^{(\delta)}_n (\theta)
+ \tpsi_{\gamma 1}^{[E_6]}*L_{n-1}^{(1)}(\theta)
+ \tpsi_{\gamma 2}^{[E_6]}*L_{n-1}^{(2)}(\theta) 
\label{TBAzvgen1}
\eea
with  $\gamma=1,\dots, 6$.\\

The Y-system is
\bea
\CY_{B_1}(\pm(p-2)) &=& \Lambda_{B_3}(0) \nn \\
\CY_{B_3}(\pm(p-2)) &=& \Lambda_{B_1}(0) 
\Lambda^{(2)}_n(0)
\prod_{i=0}^{n-1}  \Lambda^{(2)}_i(\pm( 4n-4i-3)) \nn  \\
\CY_{j}^{(\alpha)}(\pm 4) &=&  [\Lambda_{B_3}(0) ]^{\delta_{j 0} 
\delta_{\alpha 2}}
\Lambda_{j}^{(\bar \alpha )}(0)
 \prod_{k=0}^{n-1} \left(  \CL_{k}^{(\alpha)}(0) \right)^{l_{kj}^{[A_{n}]}} 
 \nn \\
&\times&  \left[\CL^{(4)}_n(0)\CL^{(6)}_n(0)\CL^{(5)}_n(\pm 1)\CL^{(3)}_n(\pm 2)
\CL^{(1)}_n(\pm 3)
\right ]^{\delta_{j,n-1} \delta_{\alpha 1}} \nn \\
&\times&  \left[\CL^{(3)}_n(0)\CL^{(6)}_n(0)\CL^{(5)}_n(\pm 1)\CL^{(4)}_n(\pm 2)
\CL^{(2)}_n(\pm 3)
\right ]^{\delta_{j,n-1} \delta_{\alpha 2}} \nn \\
&&\big(\mbox{with } j=0, \dots,n-1;~ \alpha=1,2\,\mbox{ and } 
\bar{\alpha}=3-\alpha \big ) \nn \\ 
\CY^{(\gamma)}_n(\pm 1)&=& \left[ \CL_{n-1}^{(\gamma)}(0) \right]^{
\delta_{\gamma 1} +\delta_{\gamma 2} }
\prod_{\beta=1}^{6} 
\left[ \Lambda^{(\beta)}_n(0) \right]^{l^{[E_6]}_{\beta \gamma}}
~~  (\gamma=1,\dots 6)
\eea
\noindent
{\bf Case $\trisymb{B}$, (p=4n+4, n $\ge$ 1):}
\begin{figure}[ht]
\begin{center}
\resizebox{0.35\linewidth}{!}
{\includegraphics{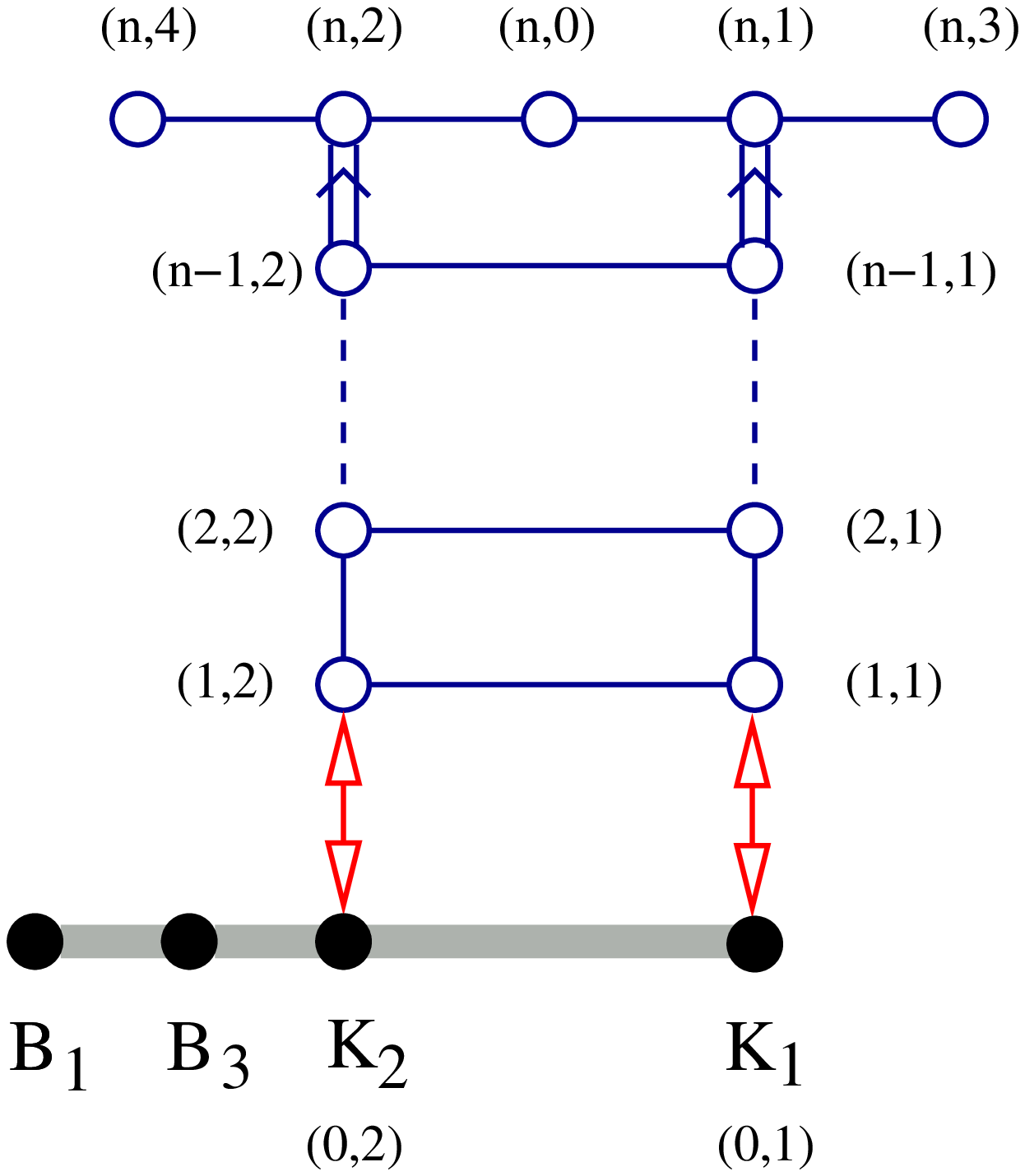}}
\label{figF.b}
\end{center}
\end{figure}
\bea
\ep_{B_a}(\theta) &=& R m_{B_a} \cosh \theta  -
\sum_{c \in \{ B_1,B_2,K_1,K_2 \}}
\Phi_{B_a,c}*L_{c}(\theta)~~~~ (a=1,2) \nn \\
\ep_j^{({\alpha})}(\theta) &=& \delta_{j0} \left (
R m_{K_\alpha} \cosh \theta -
\sum_{d \in \{ B_1, B_2 ,K_1,K_2  \}}
\Phi_{K_\alpha,d}*L_{d} (\theta) +
\sum_{\beta=1}^{2}
\tphi^{[A_2]}_{\alpha \beta}*L_j^{(\beta)}(\te)
\right) \nn \\
 &-&
\sum_{\beta=1}^{2}
\left(
\tphi^{[A_2]}_{\alpha \beta}*L_j^{(\beta)}(\te)- \sum_{k=0}^{n-1}
l_{jk}^{[A_{n}]}
\tpsi^{[A_2]}_{\alpha \beta}*L_{k}^{(\beta)}(\te) \right) \nn \\
&+& 
\delta_{j,n-1} \sum_{\delta=0}^{5} \tpsi_{\delta \alpha}^{[A_5]}*L^{(\delta)}_n(\te)
~~~~~ (j=0, \dots, n-1; \alpha=1,2)
\nn \\
\ep^{(\gamma)}_n(\theta) &=&
- \sum_{\delta=0}^{4} \tphi_{\gamma \delta}^{[A_5]}*L^{(\delta)}_n (\theta)
+ \tpsi_{\gamma 1}^{[A_5]}*L_{n-1}^{(1)}(\theta)
+ \tpsi_{\gamma2}^{[A_5]}*L_{n-1}^{(2)}(\theta)  
\eea
with  $\gamma=0,\dots, 4$.\\

The Y-system is
\bea
\CY_{B_1}(\pm(p-2)) &=& \Lambda_{B_3}(0) \nn \\
\CY_{B_3}(\pm(p-2)) &=& \Lambda_{B_1}(0) 
\Lambda^{(2)}_n(0)
\prod_{i=0}^{n-1}  \Lambda^{(2)}_i(\pm( 4n-4i-2)) \nn  \\
\CY_{j}^{(\alpha)}(\pm 4) &=&  
[\Lambda_{B_3}(0)]^{\delta_{j 0} \delta_{\alpha 2}}
\Lambda_{j}^{(\bar \alpha)}(0) 
 \prod_{k=0}^{n-1} \left(  \CL_{k}^{(\alpha)}(0) \right)^{l_{kj}^{[A_{n}]}} 
\nn \\
&\times&  \left[\CL^{(0)}_n(0)\CL^{(3)}_n(0)\CL^{(1)}_n(\pm 2)
\right ]^{\delta_{j,n-1} \delta_{\alpha 1}} \nn \\
&\times&  \left[\CL^{(0)}_n(0)\CL^{(4)}_n(0)\CL^{(2)}_n(\pm 2 )
\right ]^{\delta_{j,n-1} \delta_{\alpha 2}} \nn \\
&&\big(\mbox{with } j=0, \dots,n-1;~ \alpha=1,2\,\mbox{ and } 
\bar{\alpha}=3-\alpha \big ) \nn \\ 
\CY^{(\gamma)}_n(\pm 2)&=& \left[ \CL_{n-1}^{(\gamma)}(0) \right]^{
\delta_{\gamma 1} +\delta_{\gamma 2}}
\prod_{\beta=0}^{4} 
\left[ \Lambda^{(\beta)}_n(0) \right]^{l^{[A_5]}_{\beta \gamma}}
~~ (\gamma=0,\dots 4)
\eea
\\

\noindent
{\bf Case $\trisymb{C}$, (p=4n+5 , n $\ge$ 1):}
\begin{figure}[ht]
\begin{center}
\resizebox{0.35\linewidth}{!}
{\includegraphics{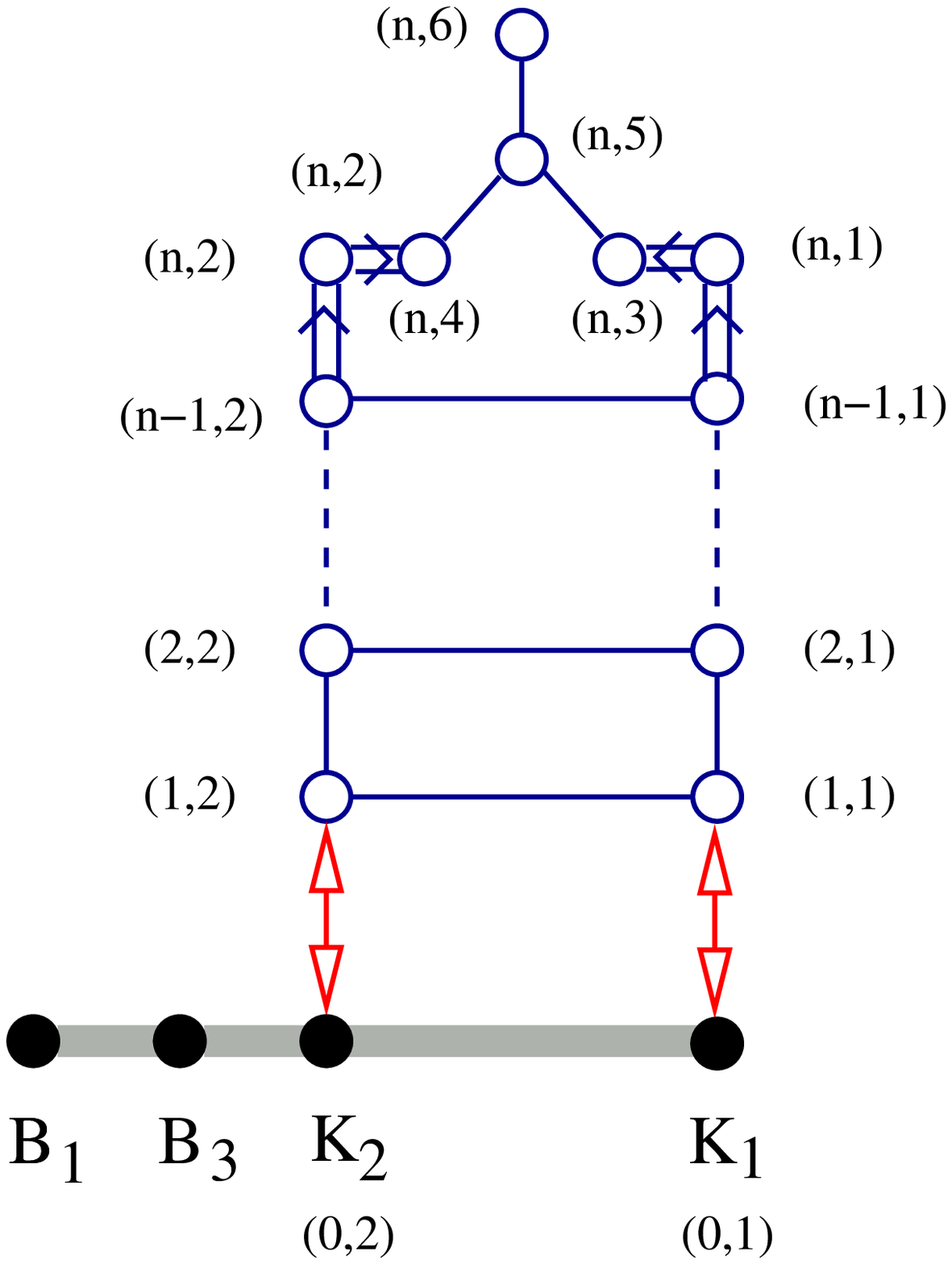}}
\label{figF.c}
\end{center}
\end{figure}
\bea
\ep_{B_a}(\theta) &=& R m_{B_a} \cosh \theta  -
\sum_{c \in \{ B_1,B_2,K_1,K_2 \}}
\Phi_{B_a,c}*L_{c}(\theta)~~~~ (a=1,2) \nn \\
\ep_j^{({\alpha})}(\theta) &=& \delta_{j0} \left (
R m_{K_\alpha} \cosh \theta -
\sum_{d \in \{ B_1, B_2 ,K_1,K_2  \}}
\Phi_{K_\alpha,d}*L_{d} (\theta) +
\sum_{\beta=1}^{2}
\tphi^{[A_2]}_{\alpha \beta}*L_j^{(\beta)}(\te)
\right) \nn \\
 &-&
\sum_{\beta=1}^{2}
\left(
\tphi^{[A_2]}_{\alpha \beta}*L_j^{(\beta)}(\te)- \sum_{k=0}^{n-1}
l_{jk}^{[A_{n}]}
\tpsi^{[A_2]}_{\alpha \beta}*L_{k}^{(\beta)}(\te) \right) \nn \\
&+& \delta_{j,n-1} \sum_{\beta=1}^{2} \left(
\tpsi_{\alpha \beta}^{[A_2]}*L^{(\beta+2)}_n(\te) +
\bar{\psi}_{\alpha \beta}^{[A_2]}*L^{(\beta)}_n(\te)  \right)
~~~~~ (j=0, \dots, n-1; \alpha=1,2)  \nn \\
\ep^{(6)}_n(\theta) &=& -\tphi_2*L^{(5)}_n(\te)  \nn \\
\ep^{(5)}_n(\theta) &=& -\tphi_2*\left( L^{(6)}_n(\te)+ 
L^{(4)}_n(\te)+L^{(5)}_n(\te) \right) \nn \\
\ep^{(3)}_n(\theta) &=& \tphi_2*\left( K^{(1)}_n(\te)-L^{(5)}_n(\te) 
\right)  \nn \\
\ep^{(4)}_n(\theta) &=& 
 \tphi_2*\left( K^{(2)}_n(\te)-L^{(5)}_n(\te) \right)  \nn \\
\ep^{(2)}_n(\theta) &=& \tphi_3*\left( K^{(6)}_n(\te)+ 
K^{(3)}_n(\te)-L_{n-1}^{(2)}(\te) 
\right)+
\tphi_5*K^{(5)}_n(\te)+\tphi_4*K^{(4)}_n(\te)  \nn \\
\ep^{(1)}_n(\theta) &=& \tphi_3*\left( K^{(6)}_n(\te)+ 
K^{(4)}_n(\te)-L_{n-1}^{(1)}(\te) 
\right)+
\tphi_5*K^{(5)}_n(\te)+\tphi_4*K^{(3)}_n(\te)  
\label{TBAzvgen3}
\eea
where $K^{(c)}(\theta)= \ln(1+e^{\ep^{(c)}(\theta)})$, as in (\ref{ccase})
earlier,
and the kernels $\tphi_i$ are defined in 
(\ref{magkernels}) with $g=3(p+2)/2$.\\

The Y-system is
\bea
\CY_{B_1}(\pm(p-2)) &=& \Lambda_{B_3}(0) \nn \\
\CY_{B_3}(\pm(p-2)) &=& \Lambda_{B_1}(0) 
\Lambda^{(2)}_n(0)
\prod_{i=0}^{n-1}  \Lambda^{(2)}_i(\pm( 4n-4i-1)) \nn  \\
\CY_{j}^{(\alpha)}(\pm 4) &=&  [\Lambda_{B_3}(0) ]^{\delta_{j 0} 
\delta_{\alpha 2}}
\Lambda_{j}^{(\bar \alpha)}(0) 
\prod_{k=0}^{n-1} \left(  \CL_{k}^{(\alpha)}(0) \right)^{l_{kj}^{[A_{n}]}} 
\nn \\
&\times&  \left[\CL^{(3)}_n(0) \CL^{(1)}_n(\pm 1)
\right ]^{\delta_{j,n-1} \delta_{\alpha 1}} \nn \\
&\times&  \left[\CL^{(4)}_n(0)\CL^{(2)}_n(\pm 1)\right ]^{\delta_{j,n-1} 
\delta_{\alpha 2}} \nn \\
&&\big(\mbox{with } j=0, \dots,n-1 ;~ \alpha=1,2\,\mbox{ and } 
\bar{\alpha}=3-\alpha \big ) \nn \\ 
\CY^{(1)}_n(\pm 3)&=&  \CL_{n-1}^{(1)}(0) 
\Lambda^{(4)}_n(0)\Lambda^{(6)}_n(0)\Lambda^{(5)}_n(\pm 1) 
\Lambda^{(3)}_n(\pm 2) \nn \\
\CY^{(2)}_n(\pm 3)&=&  \CL_{n-1}^{(2)}(0) 
\Lambda^{(3)}_n(0)\Lambda^{(6)}_n(0)\Lambda^{(5)}_n(\pm 1) 
\Lambda^{(4)}_n(\pm 2) \nn \\
\CY^{(3)}_n(\pm 1)&=&  \Lambda^{(1)}_n(0)\CL^{(5)}_n(0) \nn \\
\CY^{(4)}_n(\pm 1)&=&  \Lambda^{(2)}_n(0)\CL^{(5)}_n(0) \nn \\
\CY^{(5)}_n(\pm 1)&=&  \CL^{(6)}_n(0)\CL^{(3)}_n(0) \CL^{(4)}_n(0)\nn \\
\CY^{(6)}_n(\pm 1)&=&  \CL^{(5)}_n(0)
\eea
\\

\noindent
{\bf Case $\trisymb{D}$, (p=4n+6, n $\ge$ 1):}
\begin{figure}[ht]
\begin{center}
\resizebox{0.35\linewidth}{!}
{\includegraphics{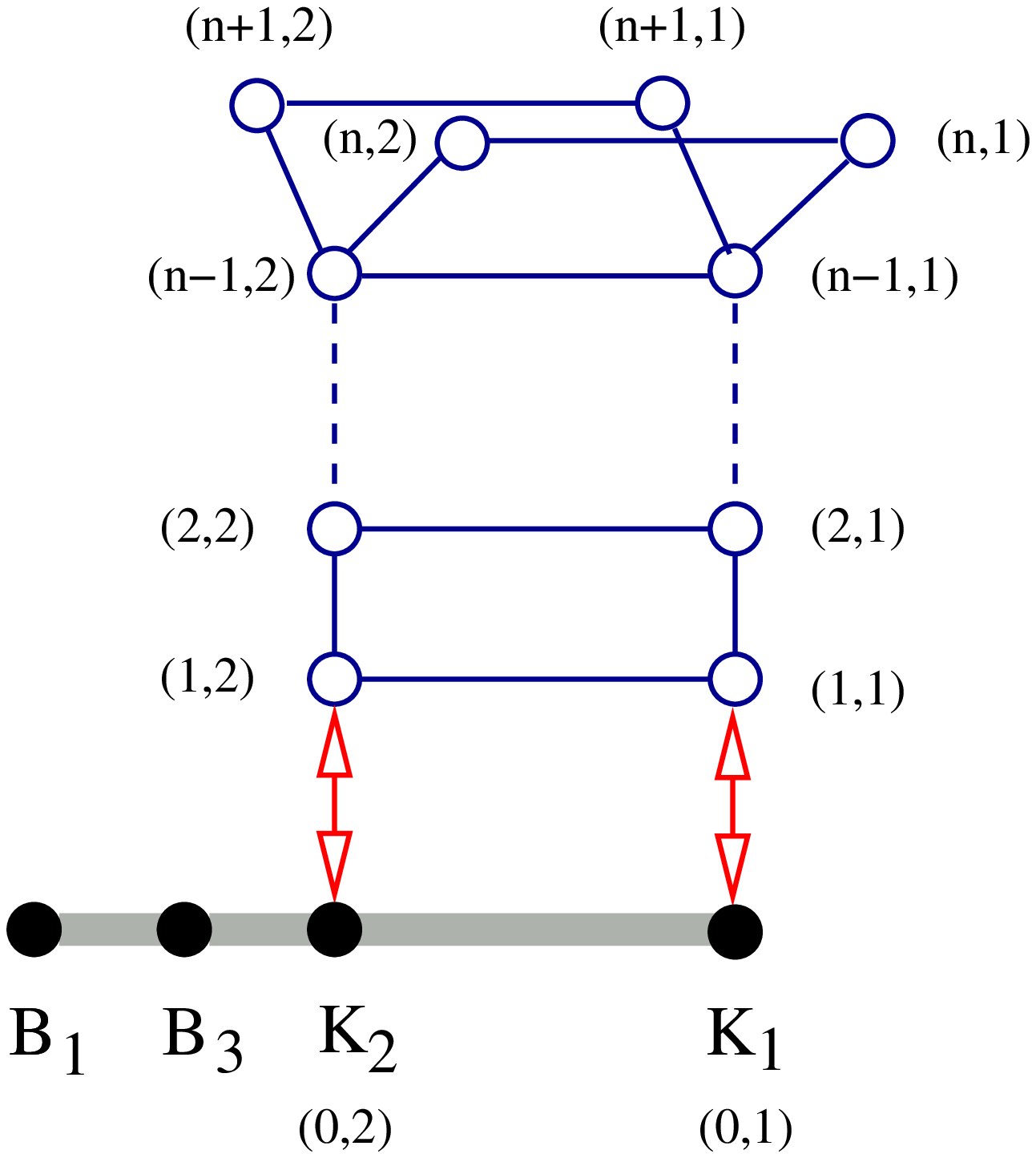}}
\label{figF.d}
\end{center}
\end{figure}
\bea
\ep_{B_a}(\theta) &=& R m_{B_a} \cosh \theta  -
\sum_{c \in \{ B_1,B_2,K_1,K_2 \}}
\Phi_{B_a,c}*L_{c}(\theta)~\, ~~~ (a=1,2) \nn \\
\ep_j^{({\alpha})}(\theta) &=& \delta_{j0} \left (
R m_{K_\alpha} \cosh \theta -
\sum_{d \in \{ B_1, B_2 ,K_1,K_2  \}}
\Phi_{K_\alpha,d}*L_{d} (\theta) +
\sum_{\beta=1}^{2}
\tphi^{[A_2]}_{\alpha \beta}*L_j^{(\beta)}(\te)
\right) \nn \\
 &-&
\sum_{\beta=1}^{2}
\left(
\tphi^{[A_2]}_{\alpha \beta}*L_j^{(\beta)}(\te)-
\sum_{k=0}^{n+1} l^{[D_{n+2}]}_{jk} 
\tpsi^{[A_2]}_{\alpha \beta }*L_{k}^{(\beta)}(\te) \right) 
\label{TBAzvgen4}
\eea
with $j=0, \dots, n+1, \alpha=1,2$.\\

The Y-system is
\bea
\CY_{B_1}(\pm(p-2)) &=& \Lambda_{B_3}(0) \nn \\
\CY_{B_3}(\pm(p-2)) &=& \Lambda_{B_1}(0) 
\Lambda^{(2)}_{n+1}(0)
\prod_{i=0}^{n}  \Lambda^{(2)}_i(\pm(4n-4i)) \nn  \\
\CY_j^{(\alpha)}(\pm 4)&=& \left [\Lambda_{B_3}(0) \right]^{\delta_{j 0} 
\delta_{\alpha 2}}
\Lambda_j^{(\bar \alpha)}(0) \prod_{i=0}^{n+1} 
\left[ \CL_{i}^{(\alpha)}(0) \right]^{l^{[D_{n+2}]}_{ij}} \nn \\
&&\big(\mbox{with } j=0, \dots,n+1 ;~ \alpha=1,2\,\mbox{ and } 
\bar{\alpha}=3-\alpha \big )
\eea 

\subsection{Exceptional $\phi_{12}$  Y-systems}
\label{y12}
The $\CM_{p,p+1}+\phi_{12}$ Y-systems are exceptional for $p=3,4,5$
and $6$. For $p=3,4$ and $6$ they are related to the exceptional Lie
algebras $E_r$ with $r=8,7$ and $6$. These systems are
most conveniently written by departing from the conventions used
elsewhere in this paper, and simply labelling 
the pseudoenegies as $\ep_1(\te),\dots \ep_r(\te)$.     
The Y-systems are then \cite{Zam2}:
\eq
\CY_{j}(\pm(p-2)) = \prod_{k=1}^{r} \Big( \Lambda_{k}(0) \Big)^{l^{[E_r]}_{kj}}
 ~~~~(j=1,\dots,r) 
\label{En} 
\en
where $l^{[E_r]}$  is the incidence matrix
of $E_8$, $E_7$ or $E_6$ for $p=3$, $4$ or $6$
respectively.

The Y-system for $\CM_{5,6}+\phi_{12}$ is new, and
follows from the TBA equations given in
\S\ref{con12} above. It is:
\bea
\CY_{B_1}(\pm 3) &=& \Lambda_{B_3}(0) \nn \\
\CY_{B_3}(\pm 3) &=& \Lambda_{B_1}(0)\Lambda_{B_5}(0) \nn \\
\CY_{B_5}(\pm 3) &=& \Lambda_{B_3}(0) \Lambda_{K_2}(\pm 2) \Lambda_{K_1}(0) 
\Lambda^{(1)}(\pm 1) \Lambda^{(2)}(0) \nn \\
\CY_{B_2}(\pm 3) &=&  \Lambda_{K_1}(\pm 2) \Lambda^{(1)}(\pm 1) \Lambda_{K_2}(0) 
\Lambda^{(2)}(0)  \nn \\ 
\CY_{K_2}(\pm 1) &=& \Lambda_{B_5}(0) \CL^{(1)}(0) \nn \\
\CY_{K_1}(\pm 1) &=& \Lambda_{B_2}(0)\CL^{(1)}(0)  \nn \\
\CY^{(1)}(\pm 1) &=& \CL^{(2)}(0) \CL_{K_2}(0)\CL_{K_1}(0) \nn \\
\CY^{(2)}(\pm 1) &=& \CL^{(1)}(0) 
\label{set}
\eea
%
%

\end{document}